\documentclass[preprint,rmp,aps,floatfix]{revtex4}
\usepackage{amssymb}

\usepackage{graphics}
\usepackage{epsfig}
\usepackage{color}

\begin{document}

\title{Fr\"ohlich polaron and bipolaron: recent developments}
\author{Jozef T.~Devreese}
\email{jozef.devreese@ua.ac.be} \affiliation{Universiteit Antwerpen,
Groenenborgerlaan 171, B-2020 Antwerpen, Belgium}
\author{Alexandre S. ~Alexandrov}
\email{a.s.alexandrov@lboro.ac.uk}
\affiliation{Department of Physics, Loughborough University, Loughborough LE11 3TU,
United Kingdom}

\begin{abstract} {It is remarkable how the Fr\"{o}hlich
 polaron, one of the simplest examples of a {\it Quantum
Field Theoretical} problem, as it basically consists of a single
fermion interacting with a scalar Bose field of ion displacements,
has resisted full analytical or numerical solution at all coupling
since $\sim 1950$, when its Hamiltonian was first written. The field
has been a testing ground for analytical, semi-analytical, and
numerical techniques, such as path integrals,  strong-coupling
perturbation expansion, advanced variational, exact diagonalisation
(ED),  and quantum Monte Carlo (QMC) techniques.  This article
reviews recent developments in the field of continuum and discrete
(lattice) Fr\"ohlich (bi)polarons  starting with the basics and
covering a number of active directions of research.}
\end{abstract}

\maketitle
\tableofcontents

%\date{November 2006}

{\section{INTRODUCTION} \label{sec:intro}

%%%%%%%%%%%%%%%%%%%%%%%%%%%%%%%%%%
%Start of the updated part - S. K.
%%%%%%%%%%%%%%%%%%%%%%%%%%%%%%%%%%

Charge carriers in inorganic and organic matter interact with ion
vibrations. The corresponding electron-phonon interaction (EPI) causes phase
transformations, including superconductivity, and dominates the transport
properties of many metals and semiconductors. EPIs have been shown to be
relevant in cuprate and other high-temperature superconductors through, for
example, isotope substitution experiments \cite%
{Zhao:1995,Zhao:1997,Khasanov:2004}, high resolution angle resolved
photoemission (ARPES) \cite{lan0,Gweon:2004,shen}, a number of earlier
optical \cite{Zamboni:1994,Mih:1990,Calvani:1994}, neutron-scattering \cite%
{Sendyka:1995} and more recent inelastic scattering \cite{rez},
pump-probe \cite{boz,gad} and tunnelling \cite{boz2} measurements.
In colossal magnetoresistance (CMR) manganites, isotope
substitutions \cite{Zhao:1996}, X-ray and neutron scattering
spectroscopies \cite{Campbell:2001,Campbell:2003} and a number of
other experiments also show a significant effect of EPI on the
physical properties (for review see
\cite{tokura}). Therefore it has been suggested that the long-range \cite%
{fro} and/or the molecular-type (e.g. \cite{Jahn:1937}) EPIs play
significant role in high-temperature superconductors (see \cite%
{Bednorz1988,muller:2000,alexandrov:1996,Alexandrov98,devreese:1996,Devreese05},
and references
therein), and in CMR manganites (see \cite%
{millis:1995,alexbra1999,edwards2002} and references therein). Very
recent experimental observations of the optical conductivity in the
Nb-doped SrTiO$_{3}$ \cite{vanMechelen2008} reveal the evidence of
the mid-infrared optical conductivity band provided by the polaron
mechanism like in many other oxides. The effective mass of the
charge carriers is obtained by analyzing the Drude spectral weight.
Defining the mass renormalization of the charge carriers as the
ratio of the total electronic spectral weight and the Drude spectral
weight, a twofold mass enhancement is obtained, which is attributed
in Ref. \cite{vanMechelen2008} to the electron-phonon coupling. The
missing spectral weight is recovered according the sum rule
\cite{SR} in a mid-infrared optical conductivity band. This band
results from the electron-phonon coupling interaction, traditionally
associated with the polaronic nature of the charge carriers. The
effective mass obtained from the optical spectral weights yields an
intermediate electron-phonon coupling strength, $3<\alpha<4$.
Therefore it has been suggested in Ref. \cite{vanMechelen2008} that
the charge transport in the Nb-doped SrTiO$_{3}$ is carried by large
polarons.

When EPI is sufficiently strong, electron Bloch states are affected even in
the normal phase. Phonons are also affected by conduction electrons. In
doped insulators \emph{bare} phonons are well defined in insulating parent
compounds, but microscopic separation of electrons and phonons is not so
straightforward in metals and heavily doped insulators \cite{maximov:1997},
where the Born and Oppenheimer (1927) and density functional \cite%
{kohn:1964,kohn:1965} methods are used. Here we have to start with the first
principle Hamiltonian describing conduction electrons and ions coupled by
the Coulomb forces.

One cannot solve the corresponding Schr\"{o}dinger equation perturbatively
because the Coulomb interaction is strong. The ratio of the characteristic
Coulomb energy to the kinetic energy is $r_{s}=m_{e}e^{2}/(4\pi
n_{e}/3)^{1/3}\approx 1$ for the electron density $n_{e}=ZN=10^{23}cm^{-3}$
(here and further we take the volume of the system as $V=1,$ unless
specified otherwise, and $\hbar=c=k_B=1$). However, one can take advantage
of the small value of the electron to ion mass ratio, $m_{e}/M<10^{-3}$.
Ions are heavy and the amplitudes of their vibrations, $\langle|\mathbf{u}%
|\rangle\simeq \sqrt{1/M\omega _{D}}$, near equilibrium positions are much
smaller than the lattice constant $(a=N^{-1/3})$, $\langle|\mathbf{u}%
|\rangle/a\approx (m_{e}/Mr_{s})^{1/4}\ll 1$. In this estimate we take the
characteristic vibration frequency $\omega _{D}$ of the order of the ion
plasma frequency $\omega_0=\sqrt{4\pi NZ^{2}e^{2}/M}$. Hence one can expand
the Hamiltonian in powers of $|\mathbf{u}|$.

Any further progress requires a simplifying physical idea, which commonly is
to approach the ground state of the many-electron system via a one-electron
picture. In the framework of the local density approximation (LDA), where
the Coulomb electron-electron interaction is replaced by an effective
one-body potential, the Hamiltonian is written \textrm{as }%
\begin{equation}
H=H_{e}+H_{ph}+H_{e-ph}+H_{e-e},
\end{equation}%
\textrm{where }%
\begin{eqnarray}
H_{e} &=&\sum_{\mathbf{k},n,s}\xi _{n\mathbf{k}s}c_{n\mathbf{k}s}^{\dagger
}c_{n\mathbf{k}s}, \\
H_{ph} &=&\sum_{\mathbf{q},\nu }\omega _{\mathbf{q}\nu }(d_{\mathbf{q}\nu
}^{\dagger }d_{\mathbf{q}\nu }+1/2)
\end{eqnarray}%
\textrm{describe} independent Bloch electrons and phonons, created
(annihilated) by $c_{n\mathbf{k}s}^{\dagger }$ ($c_{n\mathbf{k}s}$) and by $%
d_{\mathbf{q}\nu }^{\dagger }$ ($d_{\mathbf{q}\nu }$), respectively, $\xi _{n%
\mathbf{k}s}=E_{n\mathbf{k}s}-\mu $ is the band energy spectrum with respect
to the chemical potential $\mu $, $(\mathbf{k,q})$ are quasi-momenta of
electrons and phonons, respectively, $n$ is the electron band index, $\nu $
is the phonon mode index, and $s$ is the electron spin. The part of the
electron-phonon interaction, which is linear in the phonon operators, is
written as
\begin{equation}
H_{e-ph}={\frac{1}{\sqrt{2N}}}\sum_{\mathbf{k,q},n,n^{\prime },\nu ,s}\gamma
_{nn^{\prime }}(\mathbf{q,k},\nu )\omega _{\mathbf{q}\nu }c_{n^{\prime }%
\mathbf{k}s}^{\dagger }c_{n\mathbf{k-q}s}d_{\mathbf{q}\nu }+H.c.,
\label{EPI2}
\end{equation}%
where $\gamma _{nn^{\prime }}(\mathbf{q,k},\nu )$ is the dimensionless
matrix element. If we restrict the summations over $\mathbf{q}$ and $\mathbf{%
k}$ to the first Brillouin zone of the crystal, then $H_{e-ph}$ should also
include the summation over reciprocal lattice vectors $\mathbf{G}$ of
umklapp scattering contributions where $\mathbf{q}$ is replaced by $\mathbf{%
q+G}$. The terms of $H_{e-ph}$ which are quadratic and of higher orders in
the phonon operators, $d_{\mathbf{q}\nu }$, are usually small. They play a
role only for those phonons which are not coupled with electrons by the
linear interaction, Eq.(\ref{EPI2}).

The electron-electron correlation energy of a homogeneous electron system is
often written as
\begin{equation}
H_{e-e}={\frac{1}{{2}}}\sum_{\mathbf{q}}V_{c}(\mathbf{q})\rho _{\mathbf{q}%
}^{\dagger }\rho _{\mathbf{q}},
\end{equation}
where $V_{c}(\mathbf{q})$ is a matrix element, which is zero for $\mathbf{q}%
=0$ because of electroneutrality and $\rho _{\mathbf{q}}^{\dagger }=\sum_{%
\mathbf{k},s}c_{\mathbf{k}s}^{\dagger }c_{\mathbf{k+q}s}$ is the density
fluctuation operator. $H$ should also include a random potential in doped
semiconductors and amorphous metals, which could affect the EPI matrix
element \cite{Belitz:1994}.

For the purpose of this review we mostly confine our discussions to a single
band approximation with the EPI matrix element $\gamma _{nn}(\mathbf{q,k}%
,\nu )=\gamma(\mathbf{q})$ depending only on the momentum transfer $\mathbf{q%
}$. The approximation allows for qualitative and in many cases quantitative
descriptions of essential polaronic effects in advanced materials.
Nevertheless there are might be degenerate atomic orbitals in solids coupled
to local molecular-type Jahn-Teller distortions, where one has to consider
multi-band electron energy structures.

Quantitative calculations of the matrix element in the whole region
of momenta can be performed from pseudopotentials
\cite{maximov:1997,bar:1999}. On the other hand one can parameterize
EPI rather than to compute it from first principles in many
physically important cases \cite{mahan:1990}. There are three most
important interactions in doped semiconductors, which are polar
coupling to optical phonons (the Fr\"{o}hlich EPI), deformation
potential coupling to acoustical phonons, and the local (Holstein)
EPI with molecular type vibrations in complex lattices. While the
matrix element is ill defined in metals, the bare phonons $\omega
_{\mathbf{q}\nu }$ and the electron band structure
$E_{n\mathbf{k}s}$ are well defined in doped semiconductors, which
have their parent dielectric compounds. Here the effect of carriers
on the crystal field and on the dynamic matrix is small while the
carrier density is much less than the atomic one. Hence one can use
the band structure and the crystal field of parent insulators to
calculate the matrix element in doped semiconductors. The
interaction constant $\gamma (\mathbf{q})$ has different
$q$-dependence for different
phonon branches. In the long wavelength limit ($q\ll \pi /a$), $\gamma (%
\mathbf{q})\propto q^{n}$, where $n=-1,0$ and $n=-1/2$ for polar optical,
molecular ($\omega _{\mathbf{q}}=\omega _{0})$) and acoustic $(\omega _{%
\mathbf{q}}\propto q)$ phonons, respectively. Not only $q$ dependence is
known but also the absolute values of $\gamma (\mathbf{q})$ are well
parameterized in this limit. For example in polar semiconductors $|\gamma (%
\mathbf{q})|^{2}=4\pi e^{2}/\kappa \omega _{0}q^{2}$, where $\kappa
=(\varepsilon ^{-1}-\varepsilon _{0}^{-1})^{-1}$, and $\varepsilon $ and $%
\varepsilon _{0}$ are high-frequency and static dielectric constants,
respectively. If the crystal lacks an inversion center to be piezoelectric,
there is EPI with piezoelectric (acoustic) phonons with an anysotropic
matrix element, which also contribute to a polaron effect and a Coulomb-like
attraction of two polarons \cite{mahan:1972}.

To get a better insight into physical constraints of the above approximation
let us transform the Bloch states to the real space or Wannier states using
the canonical linear transformation of the electron operators, $%
c_{i}=N^{-1}\sum_{\mathbf{k}}e^{i\mathbf{k\cdot m}}c_{\mathbf{k}s}$, where $%
i=(\mathbf{m},s)$ includes both site $\mathbf{m}$ and spin quantum numbers.
In this site (Wannier) representation the electron kinetic energy takes the
following form
\begin{equation}
H_{e}=\sum_{i,j} t(\mathbf{m-n})\delta _{ss^{\prime }} c_{i}^{\dagger }c_{j},
\end{equation}
where $t(\mathbf{m})=N^{-1}\sum_{\mathbf{k}}E_{\mathbf{k}}e^{i\mathbf{k\cdot
m}}$ is the ``bare'' hopping integral, $j=(\mathbf{n},s^{\prime })$, and $E_{%
\mathbf{k}}$ is the Bloch band dispersion in the rigid lattice.

The electron-phonon interaction and the Coulomb correlations acquire simple
forms in the Wannier representation, if their matrix elements in the
momentum representation depend only on the momentum transfer $\mathbf{q}$
(here we follow \cite{alexandrov:1995}),
\begin{equation}
H_{e-ph}=\sum_{\mathbf{q},i}\omega _{\mathbf{q} }\hat{n}_{i}\left[ u_{i}(%
\mathbf{q,})d_{\mathbf{q} }+H.c.\right] ,
\end{equation}
\begin{equation}
H_{e-e}={\frac{1}{{2}}}\sum_{i\neq j}V_{c}(\mathbf{m-n})\hat{n}_{i}\hat{n}%
_{j},
\end{equation}
where
\begin{equation}
u_{i}(\mathbf{q})={\frac{1}{\sqrt{2N}}}\gamma (\mathbf{q} )e^{i\mathbf{%
q\cdot m}}
\end{equation}
and
\begin{equation}
V_{c}(\mathbf{m})={\frac{1}{{N}}}\sum_{\mathbf{q}}V_{c}(\mathbf{q})e^{i%
\mathbf{q\cdot m}},
\end{equation}
are the matrix elements of the electron-phonon and Coulomb interactions,
respectively, in the Wannier representation for electrons, and $\hat{n}%
_{i}=c_{i}^{\dagger }c_{i}$ is the density operator.

We see that taking the interaction matrix element depending only on the
momentum transfer one neglects the terms in the electron-phonon and Coulomb
interactions, which are proportional to the overlap integrals of the Wannier
orbitals on different sites. This approximation is justified for narrow band
materials, where the electron bandwidth is less than the characteristic
magnitude of the crystal field potential. In the Wannier representation the
Hamiltonian is
\begin{eqnarray}
H &=&\sum_{i,j} t(\mathbf{m-n})\delta _{ss^{\prime }} c_{i}^{\dagger
}c_{j}+\sum_{\mathbf{q},i}\omega _{\mathbf{q}}\hat{n}_{i}\left[ u_{i}(%
\mathbf{q} )d_{\mathbf{q} }+H.c.\right]  \nonumber \\
&&+{\frac{1}{{2}}}\sum_{i\neq j}V_{c}(\mathbf{m-n})\hat{n}_{i}\hat{n}%
_{j}+\sum_{\mathbf{q}}\omega _{\mathbf{q}}(d_{\mathbf{q} }^{\dagger }d_{%
\mathbf{q}}+1/2).  \label{hamiltonian}
\end{eqnarray}
One can transform it further using the site-representation also for phonons.
The site representation of $H_{e-ph}$ is particularly convenient for the
interaction with dispersionless modes, when $\omega _{\mathbf{q}}=\omega_0$
and the phonon polarization vector $\mathbf{e}_{\mathbf{q}}=\mathbf{e}$ are
roughly $q$-independent. Introducing the phonon site-operators $d_{\mathbf{n
}}=N^{-1}\sum_{\mathbf{q}}e^{i\mathbf{q\cdot n}}d_{\mathbf{q}}$ one obtains
in this case,
\begin{equation}
H_{e-ph}=\omega_0 \sum_{\mathbf{n,m},s }g(\mathbf{m-n})(\mathbf{e}\cdot
\mathbf{e}_{\mathbf{m-n}})\hat{n}_{\mathbf{m}s}(d_{\mathbf{n}}^{\dagger }+d_{%
\mathbf{n} }),
\end{equation}
where $g(\mathbf{m})$ is the dimensionless \emph{force} acting between the
electron on site\textbf{\ }$\mathbf{m}$ and the displacement of ion $\mathbf{%
n}$, proportional to the Fourier transform of $\gamma(\mathbf{q})$, and $%
\mathbf{e}_{\mathbf{m-n}}\equiv (\mathbf{m-n})/|\mathbf{m-n}|$ is the unit
vector in the direction from the electron on site $\mathbf{m}$ to the ion $%
\mathbf{n.}$ The real space representation is particularly convenient in
parameterizing EPI in complex lattices. Atomic orbitals of an ion
adiabatically follow its motion. Therefore the electron does not interact
with the displacement of the ion, whose orbitals it occupies, that is $%
g(0)=0 $.

\section{ Continuum polaron}

\label{sec:single} If characteristic phonon frequencies are sufficiently
low, the local deformation of ions, caused by electron itself, creates a
potential well, which traps the electron even in a perfect crystal lattice.
This \textit{self-trapping }phenomenon was predicted by Landau \cite{land}
more than 70 years ago. It was studied in greater detail \cite%
{pek,fro,Feynman,ras,devreese:1996} in the effective mass approximation for
the electron placed in a continuum polarizible (or deformable) medium, which
leads to a so-called \textit{large} or \textit{continuum} polaron. Large
polaron wave functions and corresponding lattice distortions spread over
many lattice sites. The self-trapping is never complete in the perfect
lattice. Due to finite phonon frequencies ion polarisations can follow
polaron motion if the motion is sufficiently slow. Hence, large polarons
with a low kinetic energy propagate through the lattice as free electrons
but with an enhanced effective mass.

When the polaron binding energy $E_{p}$ is larger than the \textrm{%
halfbandwidth} $D$ of the electron band, all states in the Bloch bands are
``dressed''\ by phonons. In this
strong-coupling regime, $\lambda =E_{p}/D>1,$ the finite bandwidth becomes
important, so the continuum approximation cannot be applied. In this case
the carriers are described as \textit{small} or discrete (lattice) polarons.
The main features of small polarons were understood a long time ago \cite%
{tja,yam,sew,hola,holb,lan,eag}. The first identification of small polarons
in solids was made for {non-stoichiometric} uranium dioxide in Refs. \cite%
{D1963,NDD1963}. Large and small polarons were discussed in a number of
review papers and textbooks, for example \cite%
{appel,fir,boettger:1985,mitra1987,alexandrov:1994,salje:1995,alexandrov:1995,devreese:1996,itoh:2001,rashba:2005,mahan:1990}%
.

In many models of EPI the ground-state polaron energy is an analytical
function of the coupling constant for any dimensionality of space \cite%
{peeters,lowen0,lowen,fehskebook,jim}. There is no abrupt (nonanalytical)
phase transition of the ground state as the electron-phonon coupling
increases. It is instead a crossover from Bloch states of band electrons or
large polarons propagating with almost bare mass in a rigid lattice to
heavily dressed Bloch states of small polarons propagating at low
temperatures with an exponentially enhanced effective mass. The ground-state
wave function of any polaron is delocalized for any coupling strength. This
result holds for both finite-site models and infinite-site models \cite%
{lowen0}.

\subsection{Pekar's polaron}

First let us briefly discuss a single electron interacting with the lattice
deformation in the continuum approximation, as studied by Pekar (1946). In
his model a free electron interacts with the dielectric polarisable
continuum, described by the static $\varepsilon _{0}$ and the optical (high
frequency) $\varepsilon $ dielectric constants. This is the case for
carriers interacting with optical phonons in ionic crystals under the
condition that the size of the self-trapped state is large compared with the
lattice constant so that the lattice discreteness is irrelevant.

\subsubsection{Ground state}

Describing the ionic crystal as a polarisable dielectric continuum one
should keep in mind that only the ionic part of the total polarisation
contributes to the polaron state. The interaction of a carrier with valence
electrons responsible for the optical properties is taken into account via
the Hartree-Fock periodic potential and included in the band mass $m$.
Therefore only ion displacements contribute to the self-trapping. Following
Pekar we minimise the sum $E(\psi )$ of the electron kinetic energy and the
potential energy due to the self-induced polarisation field (here we follow
\cite{alexandrov:1995}),
\begin{equation}
E(\psi )=\int d\mathbf{r}\left[ \psi ^{\ast }(\mathbf{r})\left( -{\frac{%
\nabla ^{2}}{{2m}}}\right) \psi (\mathbf{r})-\mathbf{P}(\mathbf{r})\cdot
\mathbf{D}(\mathbf{r})\right]
\end{equation}%
where
\begin{equation}
\mathbf{D}(\mathbf{r})=e\mathbf{\nabla }\int d\mathbf{r^{\prime }}{\frac{%
|\psi (\mathbf{r^{\prime }})|^{2}}{{|\mathbf{r}-\mathbf{r^{\prime }}|}}}
\end{equation}%
is the electric field of the electron in the state with the wave function $%
\psi (\mathbf{r})$ and $\mathbf{P}$ is the ionic part of the lattice
polarisation. Minimising $E(\psi )$ with respect to $\psi ^{\ast }(\mathbf{r}%
)$ at \emph{fixed} $\mathbf{P}$ and $\int d\mathbf{r}|\psi (\mathbf{r}%
)|^{2}=1$ one arrives at the equation of motion,
\begin{equation}
\left( -{\frac{\nabla ^{2}}{{2m}}}-e\int d\mathbf{r^{\prime }}\mathbf{P}(%
\mathbf{r^{\prime }})\cdot \mathbf{\nabla ^{\prime }}{\frac{1}{{|\mathbf{%
r^{\prime }}-\mathbf{r}|}}}\right) \psi (\mathbf{r})=E_{0}\psi (\mathbf{r}),
\end{equation}%
where $E_{0}$ is the polaron ground state energy. The ionic part of the
total polarisation is obtained using the definition of the susceptibilities $%
\chi _{0}$ and $\chi $, $\mathbf{P}=(\chi _{0}-\chi )\mathbf{D}$. The
dielectric susceptibilities $\chi _{0}$ and $\chi $ are related to the
static and high frequency dielectric constants, respectively, $(\chi
_{0}=(\varepsilon _{0}-1)/4\pi \varepsilon _{0}$, and $\chi =(\varepsilon
-1)/4\pi \varepsilon )$, so that $\mathbf{P}={\frac{\mathbf{D}}{{4\pi \kappa
}}}$. Then the equation of motion becomes
\begin{equation}
\left( -{\frac{\nabla ^{2}}{{2m}}}-{\frac{e^{2}}{{4\pi \kappa }}}\int d%
\mathbf{r^{\prime }}\int d\mathbf{r^{\prime \prime }}|\psi (\mathbf{%
r^{\prime \prime }})|^{2}\mathbf{\nabla ^{\prime }}{\frac{1}{{|\mathbf{%
r^{\prime }}-\mathbf{r^{\prime \prime }}|}}}\cdot \mathbf{\nabla ^{\prime }}{%
\frac{1}{{|\mathbf{r^{\prime }}-\mathbf{r}|}}}\right) \psi (\mathbf{r}%
)=E_{0}\psi (\mathbf{r}).
\end{equation}%
Differentiating by parts with the use of $\nabla ^{2}r^{-1}=-4\pi \delta (%
\mathbf{r})$ one obtains
\begin{equation}
\left( -{\frac{\nabla ^{2}}{{2m}}}-{\frac{e^{2}}{{\kappa }}}\int d\mathbf{%
r^{\prime }}{\frac{|\psi (\mathbf{r^{\prime }})|^{2}}{{|\mathbf{r^{\prime }}-%
\mathbf{r}|}}}\right) \psi (\mathbf{r})=E_{0}\psi (\mathbf{r}).
\end{equation}%
The solution of this nonlinear integro-differential equation can be found
using a variational minimisation of the functional
\begin{equation}
J(\psi )={\frac{1}{{2m}}}\int d\mathbf{r}|\mathbf{\nabla }\psi (\mathbf{r}%
)|^{2}-{\frac{1}{{2ma_{B}}}}\int d\mathbf{r}d\mathbf{r^{\prime }}{\frac{%
|\psi (\mathbf{r})|^{2}|\psi (\mathbf{r^{\prime }})|^{2}}{{|\mathbf{%
r^{\prime }}-\mathbf{r}|}}},
\end{equation}%
where $a_{B}=\kappa /me^{2}$ is the effective Bohr radius. The simplest
choice of the normalised trial function is $\psi (\mathbf{r})=Ae^{-r/r_{p}}$
with $A=(\pi r_{p}^{3})^{-1/2}$. Substituting the trial function into the
functional yields $J(\psi )=T+{\frac{1}{{2}}}U$, where the kinetic energy is
$T=1/2mr_{p}^{2}$, and the potential energy $U=-5/8ma_{B}r_{p}$. Minimising
the functional with respect to $r_{p}$ yields the polaron radius, $%
r_{p}=16a_{B}/5$, and the ground state energy $E_{0}=T+U$ as $%
E_{0}=-0.146/ma_{B}^{2}$. This can be compared with the ground state energy
of the hydrogen atom $-0.5m_{e}e^{4}$, where $m_{e}$ is the free electron
mass. Their ratio is $0.3m/(m_{e}\kappa ^{2})$. In typical polar solids $%
\kappa \gtrsim 4$, so that the continuum polaron binding energy is about $%
0.25eV$ or less, if $m\simeq m_{e}$. The potential energy in the ground
state is $U=-4T=4E_{0}/3$.

The lowest photon energy $\omega_{min}$ to excite the polaron into the bare
electron band is $\omega_{min}=|E_{0}|$. The ion configuration does not
change in the photoexcitation process of the polaron. However, a lower
activation energy $W_{T}$ is necessary, if the self-trapped state disappears
together with the polarisation well due to thermal fluctuations, $%
W_{T}=|E_{0}|-U_{d}$, where $U_{d}$ is the deformation energy. In ionic
crystals
\begin{equation}
U_{d}={\frac{1}{{2}}}\int d\mathbf{r} \mathbf{P}(\mathbf{r})\cdot \mathbf{D}(%
\mathbf{r}),
\end{equation}
which for the ground state is $U_{d}=2|E_{0}|/3$. Therefore the thermal
activation energy is $W_{T}=|E_{0}|/3$. The ratio of four characteristic
energies for the continuum Pekar's polaron is $W_{T}:U_{d}:%
\omega_{min}:|U|=1:2:3:4$ \cite{pekarmon}.

Using Pekar's choice,
\begin{equation}
\psi(r)=A(1+ r/r_{p} +\beta r^{2}) e^{- r/r_{p}}  \label{trial}
\end{equation}
one obtains $A=0.12 /r_{p}^{3/2}$, $\beta=0.45/r_{p}^{2}$, the polaron
radius $r_{p}=1.51 a_{B}$ and a better estimate for the ground state energy,
$E_{0}=-0.164/ma_{B}^{2}$ as compared to the result of the simplest
exponential choice.

\subsubsection{Effective mass of Pekar's polaron}

The lattice polarization is responsible for the polaron mass enhancement
\cite{LP48}. Within the continuum approximation the evolution of the lattice
polarisation $\mathbf{P}(\mathbf{r},t)$ is described by a harmonic
oscillator subjected to an external force $\sim \mathbf{\ D}/\kappa$:
\begin{equation}
\omega_0^{-2}{\frac{\partial^{2}\mathbf{P}(\mathbf{r},t)}{{\partial t^{2}}}}
+\mathbf{P}(\mathbf{r},t)={\frac{\mathbf{D}(\mathbf{r},t)}{{4\pi \kappa}}},
\label{pol0}
\end{equation}
where $\omega_0$ is the optical phonon frequency. If during the
characteristic time of the lattice relaxation $\simeq\omega_0^{-1}$ the
polaron moves a distance much less than the polaron radius, the polarisation
practically follows the polaron motion. Hence for a slow motion with the
velocity $v \ll \omega_0 a_{B}$ the first term in Eq.(\ref{pol0}) is a small
perturbation, so that
\begin{equation}
\mathbf{P}(\mathbf{r},t)\approx{\frac{1}{{4\pi \kappa}}} \left(\mathbf{D}(%
\mathbf{r},t)- \omega_0^{-2}{\frac{\partial^{2}\mathbf{D}(\mathbf{r},t)}{{%
\partial t^{2}}}}\right).
\end{equation}
The total energy of the crystal with an extra electron,
\begin{equation}
E=E({\psi})+2\pi \kappa \int d\mathbf{r}\left[\mathbf{P(r,t)}^{2}
+\omega_0^{-2}\left({\frac{\partial\mathbf{P}(\mathbf{r},t)}{{\partial t}}}%
\right)^{2}\right],
\end{equation}
is determined in such a way that it gives Eq.(\ref{pol0}) when it is
minimised with respect to $\mathbf{P}$. We note that the first term of the
lattice contribution to $E$ is the deformation energy $U_{d}$, discussed in
the previous section. The lattice part of the total energy depends on the
polaron velocity and contributes to the effective mass. Replacing the static
wave function $\psi(\mathbf{r})$ in all expressions for $\psi(\mathbf{r}-%
\mathbf{v}t)$ and neglecting a contribution to the total energy of higher
orders than $v^{2}$ one obtains
\begin{equation}
E=E_{0}+U_{d}+{\frac{m^{*}v^{2}}{{2}}},
\end{equation}
where
\begin{equation}
m^{*}=-{\frac{1}{{12\pi \omega_0^{2}\kappa}}}\int d\mathbf{r} \mathbf{D}(%
\mathbf{r})\cdot \nabla^{2}\mathbf{D}(\mathbf{r})
\end{equation}
is the polaron mass. Using the equation $\mathbf{\nabla} \cdot \mathbf{D}%
=-4\pi e |\psi(\mathbf{r})|^{2}$ yields
\begin{equation}
m^{*}={\frac{4\pi e^{2}}{{3 \omega_0^{2} \kappa}}} \int d\mathbf{r}|\psi(%
\mathbf{r})|^{4}.
\end{equation}
Calculating the integral with the trial function Eq.(\ref{trial}) one
obtains
\begin{equation}
m^{*}\thickapprox 0.02 \alpha^{4} m,
\end{equation}
where $\alpha$ is the dimensionless constant, defined as $%
\alpha=(e^{2}/\kappa)\sqrt{m/2\omega_0}$.

Concluding the discussion of the Pekar's polarons let us specify conditions
of its existence. The polaron radius should be large compared with the
lattice constant, $r_{p}\gg a$ to justify the effective mass approximation
for the electron. Hence the value of the coupling constant $\alpha$ should
not be very large, $\alpha \ll (D/z\omega_0)^{1/2}$, where $D\simeq
z/2ma^{2} $ is the bare half-bandwidth, and $z$ is the lattice coordination
number. On the other hand the classical approximation for the lattice
polarisation is justified if the number of phonons taking part in the
polaron cloud is large. This number is of the order of $U_{d}/\omega_0$. The
total energy of the immobile polaron and the deformed lattice is $%
E=-0.109\alpha^{2}\omega_0$ and $U_{d}=0.218\alpha^{2}\omega_0$,
respectively. Then $\alpha$ is bounded from below by the condition $%
U_{d}/\omega_0 \gg 1$ as $\alpha^{2}\gg 5$. The typical adiabatic ratio $%
D/\omega_0$ is of the order of $10$ to $100$. In fact in many transition
metal oxides with narrow bands and high optical phonon frequencies this
ratio is about $10$ or even less, which makes the continuum strong-coupling
polaron hard to be realised in oxides and related ionic compounds with light
ions \cite{alexandrov:1995}.

\subsection{Weak-coupling Fr\"ohlich polaron}

Fr\"{o}hlich (1954) applied the second quantisation form of the
electron-lattice interaction \cite{frohlich0} to describe the large polaron
in the weak-coupling regime, $\alpha <1$, where the quantum nature of
lattice polarisation becomes important,
\begin{equation}
H=-{\frac{\nabla ^{2}}{{2m}}}+\sum_{\mathbf{q}}\left( V_{\mathbf{q}}d_{%
\mathbf{q}}e^{i\mathbf{q}\cdot \mathbf{r}}+h.c.\right) +\sum_{\mathbf{q}%
}\omega _{\mathbf{q}}(d_{\mathbf{q}}^{\dagger }d_{\mathbf{q}}+1/2).
\label{frohlichH}
\end{equation}%
The quantum states of the noninteracting electron and phonons are specified
by the electron momentum $\mathbf{k}$ and the phonon occupation number $%
\langle d_{\mathbf{q}}^{\dagger }d_{\mathbf{q}}\rangle \equiv n_{\mathbf{q}%
}=0,1,2,...\infty $. At zero temperature the unperturbed state is the
vacuum, $|0\rangle $, of phonons and the electron plane wave
\begin{equation}
|\mathbf{k},0\rangle =e^{i\mathbf{k}\cdot \mathbf{r}}|0\rangle .
\label{ground}
\end{equation}%
While the coupling is weak one can apply the perturbation theory. The
interaction couples the state Eq.(\ref{ground}) with the energy $k^{2}/2m$
and the state with the energy $(\mathbf{k}-\mathbf{q})^{2}/2m+\omega _{0}$
of a single phonon with momentum $\mathbf{q}$ and the electron with momentum
$\mathbf{k}-\mathbf{q}$, $|\mathbf{k}-\mathbf{q},1_{\mathbf{q}}\rangle =e^{i(%
\mathbf{k}-\mathbf{q})\cdot \mathbf{r}}|1_{\mathbf{q}}\rangle $. The
corresponding matrix element is $\langle \mathbf{k}-\mathbf{q},1_{\mathbf{q}%
}|H_{e-ph}|\mathbf{k},0\rangle =V_{\mathbf{q}}^{\ast }.$ There is no
diagonal matrix elements of $H_{e-ph}$, so the second order energy $\tilde{E}%
_{\mathbf{k}}$ is
\begin{equation}
\tilde{E}_{\mathbf{k}}={\frac{k^{2}}{{2m}}}-\sum_{\mathbf{q}}{\frac{|V_{%
\mathbf{q}}|^{2}}{{(\mathbf{k}-\mathbf{q})^{2}/2m+\omega _{0}-k^{2}/2m}}}.
\end{equation}%
There is no imaginary part of $\tilde{E}_{\mathbf{k}}$ for a slow electron
with $k<q_{p},$ where
\begin{equation}
q_{p}=\min (m\omega _{0}/q+q/2)=(2m\omega _{0})^{1/2},
\end{equation}%
which means that the momentum is conserved. Evaluating the integrals one
arrives at
\begin{equation}
\tilde{E}_{\mathbf{k}}={\frac{k^{2}}{{2m}}}-{\frac{\alpha \omega _{0}q_{p}}{{%
k}}}\arcsin \left( {\frac{k}{{q_{p}}}}\right) ,
\end{equation}%
which for a very slow motion $k\ll q_{p}$ yields
\begin{equation}
\tilde{E}_{\mathbf{k}}\simeq -\alpha \omega _{0}+{\frac{k^{2}}{{2m^{\ast }}}}%
.
\end{equation}%
Here the first term is minus the polaron binding energy, $-E_{p}$. The
effective mass of the polaron is enhanced as
\begin{equation}
m^{\ast }={\frac{m}{{1-\alpha /6}}}\simeq m(1+\alpha /6)  \label{massF}
\end{equation}%
due to a phonon ``cloud''\ accompanying the
slow polaron. The number of virtual phonons $N_{ph}$ in the cloud is given
by taking the expectation value of the phonon number operator, $%
N_{ph}=\langle \sum_{\mathbf{q}}d_{\mathbf{q}}^{\dagger }d_{\mathbf{q}%
}\rangle ,$ where $bra$ and $ket$ refer to the perturbed state,
\begin{equation}
|\rangle =|0\rangle +\sum_{\mathbf{q^{\prime }}}{\frac{V_{\mathbf{q^{\prime }%
}}^{\ast }}{{\ k^{2}/2m-(\mathbf{k}-\mathbf{q^{\prime }})^{2}/2m-\omega _{0}}%
}}|1_{\mathbf{q^{\prime }}}\rangle .
\end{equation}%
For the polaron at rest ($\mathbf{k}=0$) one obtains $N_{ph}=\alpha /2$.
Hence the Fr\"{o}hlich coupling constant, $\alpha $, measures the cloud
`thickness'. One can also calculate the lattice charge density induced by
the electron. The electrostatic potential $e\phi (\mathbf{r})$ is given by
the average of the interaction term of the Hamiltonian
\begin{equation}
e\phi (\mathbf{r})=\langle \sum_{\mathbf{q}}V_{\mathbf{q}}e^{i\mathbf{q}%
\cdot \mathbf{r}}d_{\mathbf{q}}+H.c.\rangle ,
\end{equation}%
and the charge density $\rho (\mathbf{r})$ is related to the electrostatic
potential by Poisson's equation $\nabla \phi =-4\pi \rho $. Using these
equations one obtains
\begin{equation}
\rho (\mathbf{r})=-{\frac{1}{{2\pi e}}}\sum_{\mathbf{q}}{\frac{q^{2}|V_{%
\mathbf{q}}|^{2}\cos (\mathbf{q}\cdot \mathbf{r})}{{\omega _{0}+q^{2}/2m}}},
\end{equation}%
which yields
\begin{equation}
\rho (\mathbf{r})=-{\frac{eq_{p}^{3}}{{4\pi \kappa }}}{\frac{e^{-q_{p}r}}{{%
q_{p}r}}}.
\end{equation}%
The mean extension of the phonon cloud, which can be taken as the radius of
the weak coupling polaron, is $r_{p}=q_{p}^{-1}$, and the total induced
charge is $Q=\int d\mathbf{r}\rho (\mathbf{r})=-e/\kappa $.

\subsection{Lee-Low-Pines transformation}

One can put the Fr\"ohlich result on a variational basis by applying
the Lee-Low-Pines (LLP)  transformation \cite{LLP}, which removes
the
electron coordinate, followed by the displacement transformation \cite%
{tja,leepines,gurari}. The latter serves to account for that part of the
lattice polarisation which follows the electron instantaneously. The
remaining part of the polarisation field turns out to be small, if the
coupling constant is not extremely large. In the opposite extreme limit,
which is Pekar's strong-coupling regime discussed above, one can construct
the perturbation theory by an expansion in descending powers of $\alpha$
\cite{bogol,Allcock1,bogoljr}. Alternatively one can apply Feynman's
path-integral formalism to remove the phonon field at the expense of a
non-instantaneous interaction of electron with itself (\ref{Ground_Feynman},
and also \cite{schultz}).

The transformation can be written as
\begin{equation}
|\tilde N\rangle=\exp(S) |N \rangle,
\end{equation}
 where $S$ is an anti-Hermitian operator: $S^+ = - S$. In our case $|N\rangle$ is a single-electron multiphonon wave
function. The transformed eigenstate, $|\tilde {N}\rangle$ satisfies
the Schr\"odinger equation, $\tilde{H}|\tilde N>=E|\tilde N\rangle$,
with the transformed Hamiltonian
\begin{equation}
\tilde {H}=\exp(S) H \exp(-S).  \label{trans0}
\end{equation}
If all operators are transformed according to Eq.(\ref{trans0}) the physical
averages remain unchanged. LLP transformation eliminating the electron
coordinate in the Hamiltonian is defined as
\begin{equation}
S_{LLP}=i\sum_{\mathbf{q}}(\mathbf{q}\cdot \mathbf{r})d^{\dagger}_{\mathbf{q}%
}d_{\mathbf{q}}.
\end{equation}
The transformed Hamiltonian is obtained as
\begin{equation}
\tilde{H}={\frac{1}{{2m}}}(-i\nabla-\sum_{\mathbf{q}}\mathbf{q}d^{\dagger}_{%
\mathbf{q}}d_{\mathbf{q}})^{2}+\sum_{\mathbf{q}}(V_{\mathbf{q}}d_{\mathbf{q}%
}+H.c.)+\omega_0\sum_{\mathbf{q}}(d^{\dagger}_{\mathbf{q}}d_{\mathbf{q}%
}+1/2).
\end{equation}
The electron coordinate is absent in $\tilde{H}$. Hence the eigenstates $|%
\tilde{N}\rangle$ are classified with the momentum $\mathbf{\ K}$, which is
the conserving total momentum of the system, $|\tilde{N}\rangle=e^{i\mathbf{K%
}\cdot \mathbf{r}}|\tilde{N}_{ph}\rangle,$ where $|\tilde{N}_{ph}\rangle$ is
an eigenstate of phonons. The number of virtual phonons is not small in the
intermediate coupling regime. Therefore one cannot apply the perturbation
theory for $\tilde{H}$. However, one can remove the essential part of the
interaction term in the Hamiltonian by the displacement canonical
transformation,
\begin{equation}
S=\sum_{\mathbf{q}}f(\mathbf{q})d_{\mathbf{q}}-H.c.,
\end{equation}
where $c$-number $f(\mathbf{q})$ is determined by minimisation of the ground
state energy. Assuming that the transformed ground state is the phonon
vacuum $e^{S}|\tilde{N}_{ph}\rangle=|0\rangle,$ one obtains the energy $E_{%
\mathbf{K}}$ as
\begin{equation}
E_{\mathbf{K}}={\frac{(1-\eta^{2})K^{2}}{{2m}}}-{\frac{\alpha \omega_0 q_{p}%
}{{K(1-\eta)}}}\sin^{-1} \left({\frac{K(1-\eta)}{{q_{p}}}}\right),
\end{equation}
where
\begin{equation}
\eta(1-\eta)^{2}={\frac{\alpha q_{p}^{3}}{{2K^{3}}}}\left({\frac{(1-\eta) K
}{\sqrt{q_{p}^{2}-(1-\eta)^{2}K^{2}}}}-\sin^{-1}{\frac{(1-\eta)K}{{q_{p}}}}%
\right).
\end{equation}
Only the term independent of $K$ needs to be retained in $\eta$ for a slow
polaron with $K\ll q_{p}$,
\begin{equation}
\eta= {\frac{\alpha/6}{{1+\alpha/6}}}.
\end{equation}
Then the energy up to the second order in $K$ is given by
\begin{equation}
E_{\mathbf{K}}=-\alpha \omega_0 +{\frac{K^{2}}{{2m^{*}}}},
\end{equation}
where the polaron mass is $m^{*}=m(1+\alpha/6)$ as in Eq.(\ref{massF}). Lee,
Low and Pines evaluated also the corrections due to off-diagonal parts of
the transformed Hamiltonian and found that they are small.

\subsection{All-coupling continuum polaron}

\label{Ground_Feynman}

\subsubsection{Feynman theory}

Feynman developed a superior all-coupling continuum-polaron theory using his
path-integral formalism\thinspace \cite{Feynman}. He got the idea to
formulate the polaron problem into the Lagrangian form of quantum mechanics
and then eliminate the field oscillators, ``\ldots in exact
analogy to Q.~E.~D. \ldots (resulting in) \ldots a sum over all trajectories
\ldots''. The resulting path integral (here limited to the
ground-state properties) is of the form:
\begin{eqnarray}
\langle 0,\beta |0,0\rangle \! &=&\!\int \!\mathcal{D}\mathbf{r}(\tau )e^{S},
\label{eq_8a} \\
S &=&\exp \left[ -\int_{0}^{\beta }\frac{\mathbf{{\dot{r}}}^{2}}{2}d\tau
\!+\!\frac{\alpha }{2^{3/2}}\int_{0}^{\beta }\int_{0}^{\beta }\frac{%
e^{-|\tau -\sigma |}}{|\mathbf{r}(\tau )-\mathbf{r}(\sigma )|}d\tau d\sigma %
\right] ,  \label{eq_8b}
\end{eqnarray}%
where $\beta =1/T$. \textrm{Here, the Feynman units are used: }$\hbar =1$%
\textrm{, }$m=1$\textrm{, }$\omega _{0}=1$. Eq. (\ref{eq_8a}) gives the
\textit{amplitude} that an electron found at a point in space at time zero
will appear at the same point at the imaginary time $\beta $. The
interaction term in the action function $S$ may be interpreted as indicating
that at ``time''\ $\tau $, the electron
behaves as if it were in a potential
\begin{equation}
\frac{\alpha }{2^{3/2}}\int_{0}^{\beta }\frac{e^{-|\tau -\sigma |}}{|\mathbf{%
r}(\tau )-\mathbf{r}(\sigma )|}d\sigma   \label{eq_9a}
\end{equation}%
which results from the electrostatic interaction of the electron with the
mean charge density of its ``previous''\
positions, weighted with the function $e^{-|\tau -\sigma |}.$ This path
integral (\ref{eq_8a}) with (\ref{eq_8b}) has a great intuitive appeal: it
shows the polaron problem as an equivalent one-particle problem in which the
interaction, non-local in time or ``retarded'', occurs between the electron and itself.

Subsequently Feynman introduced a variational principle for path integrals
to study the polaron. He then simulated the interaction between the electron
and the polarization modes by a harmonic interaction (with force constant $k$%
) between a hypothetical (``fictitious'') particle with mass $M$ and the
electron. Within his model, the action function $S$ (\ref{eq_8b}) is
imitated by a quadratic trial action (non-local in time):
\begin{equation}
S_{0}=\exp \left[ -\int_{0}^{\beta }\frac{\mathbf{{\dot{r}}}^{2}}{2}d\tau
\!+\!\frac{C}{2}\int_{0}^{\beta }\int_{0}^{\beta }\left[ \mathbf{r}(\tau )-%
\mathbf{r}(\sigma )\right] ^{2}e^{-w|\tau -\sigma |}d\tau d\sigma \right] ,
\label{eq_8c}
\end{equation}%
where the interaction potential (\ref{eq_9a}) is replaced by a parabolic
potential
\begin{equation}
\frac{C}{2}\int_{0}^{\beta }\left[ \mathbf{r}(\tau )-\mathbf{r}(\sigma )%
\right] ^{2}e^{-w|\tau -\sigma |}d\sigma  \label{eq_9b}
\end{equation}%
with the weight function $e^{-w|\tau -\sigma |}$. The variational parameters
$C$ and $w$ in Eq. (\ref{eq_9b}) are adjusted in order to partly compensate
for the error of exploiting the trial potential (\ref{eq_9b}) instead of the
true potential (\ref{eq_9a}). Following the Feynman approach, an upper bound
for the polaron ground-state energy can be written down as%
\begin{equation}
E=E_{0}-\lim_{\beta\rightarrow\infty}\frac{1}{\beta}\left\langle
S-S_{0}\right\rangle _{0},  \label{sr11}
\end{equation}
where $S$ is the exact action functional of the polaron problem, while $%
S_{0} $ is the trial action functional, which corresponds to the above model
system, $E_{0}$ is the ground-state energy of the model system, and%
\begin{equation}
\left\langle F\right\rangle _{0}\equiv\frac{\int Fe^{S_{0}}\mathcal{D}%
\mathbf{r}\left( t\right) }{\int e^{S_{0}}\mathcal{D}\mathbf{r}\left(
t\right) }.  \label{sr12}
\end{equation}
The parameters of the model system $C$ and $w$ are found from the
variational condition that they provide a mimimum to the upper bound for the
ground state energy of Eq. (\ref{sr11}). (For the details of the
calculation, see subsection \ref{scaling}.) At nonzero temperatures, the
best values of the model parameters can be determined from a variational
principle for the free energy \cite{FeynmanSM}, see Refs. \cite%
{osaka1959,KP1957}.

\begin{figure}[tbh]
\begin{center}
\newpage \includegraphics[width=0.7\textwidth]{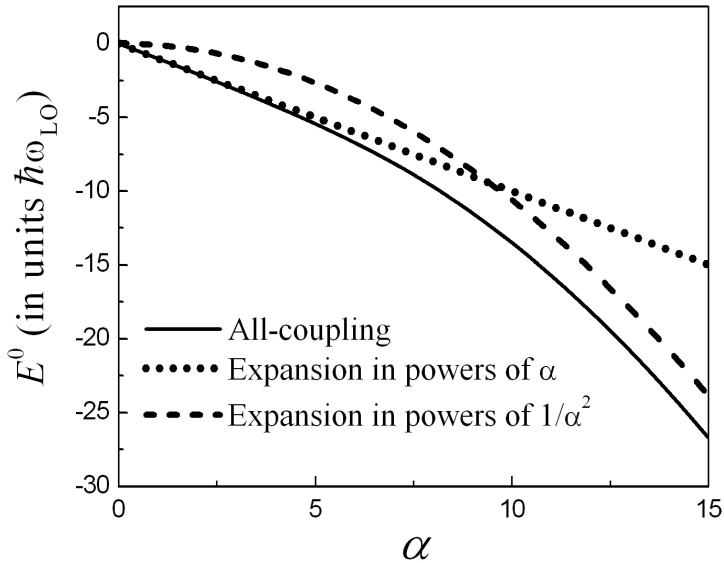}
\end{center}
\caption{Feynman-polaron energy as a function of $\protect\alpha$: the
all-coupling theory.}
\label{fig_Feynman_Interpolation}
\end{figure}

Eq.(\ref{sr11}) constitutes an upper bound for the polaron self-energy at
all $\alpha$, which in Fig. \ref{fig_Feynman_Interpolation} is compared with
the results of weak- and strong-coupling expansions. The weak-coupling
expansions of Feynman for the ground-state energy and the effective mass of
the polaron are:
\begin{equation}
\frac{E_{0}}{\omega_{0}}=-\alpha-0.0123\alpha^{2}-0.00064\alpha^{3}-\ldots%
\>(\alpha\rightarrow0),  \label{eq_7a}
\end{equation}%
\begin{equation}
\frac{m^{\ast}}{m}=1+\frac{\alpha}{6}+0.025\alpha^{2}+\ldots
\>(\alpha\rightarrow0).  \label{eq_7b}
\end{equation}
In the strong-coupling limit Feynman found for the ground-state energy:
\begin{equation}
{\frac{E_{0}}{\omega_{0}}\equiv\frac{E_{3D}(\alpha)}{\omega_{0}}%
=-0.106\alpha^{2}-2.83-\ldots\hspace*{0.3cm}(\alpha\rightarrow\infty)}
\label{eq_9a1}
\end{equation}
and for the polaron mass:
\begin{equation}
\frac{m^{\ast}}{m}\equiv\frac{m_{3D}^{\ast}(\alpha)}{m}=0.0202\alpha
^{4}+\ldots\hspace*{0.3cm}(\alpha\rightarrow\infty).  \label{eq_9b1}
\end{equation}

Becker \emph{et al.} (1983), using a Monte Carlo calculation, derived the
ground-state energy of a polaron as $E_{0}=\lim_{\beta\rightarrow\infty}%
\Delta F,$ where $\Delta F=F_{\beta}-F_{\beta}^{0}$ with $F_{\beta}$ the
free energy per polaron and $F_{\beta}^{0}=\left[ 3/\left( 2\beta\right) %
\right] \ln\left( 2\pi\beta\right) $ the free energy per electron. The value
$\beta\omega_{0}=25$, used for the actual computation in Ref. \cite%
{bgsprb285735}, corresponds to $T/T_{D}=0.04$ ($T_{D}=\hbar \omega_{\text{LO}%
}/k_{B};$ $\omega_{\text{LO}}\equiv\omega_0$ is the longitudinal (LO)
optical phonon frequency in conventional units). The authors of Ref. \cite%
{bgsprb285735} actually calculated the \textit{free energy} $\Delta F$,
rather than the polaron \textit{ground-state energy}. To investigate the
importance of temperature effects on $\Delta F,$ \cite{pdprb316826}
considered the polaron energy as obtained by Osaka (1959), who generalized
the Feynman polaron theory to nonzero temperatures:%
\begin{eqnarray}
\frac{\Delta F}{\omega_{0}} & =\frac{3}{\beta}\ln\left( \frac{w}{v}\frac{%
\sinh\frac{\beta_{0}v}{2}}{\sinh\frac{\beta_{0}w}{2}}\right) -\frac {3}{4}%
\frac{v^{2}-w^{2}}{v}\left( \coth\frac{\beta_{0}v}{2}-\frac{2}{\beta_{0}v}%
\right)  \nonumber \\
& -\frac{\alpha}{\sqrt{2\pi}}\left[ 1+n\left( \omega_{0}\right) \right]
\int_{0}^{\beta_{0}}du\frac{e^{-u}}{\sqrt{D\left( u\right) }},  \label{e1}
\end{eqnarray}
where $\beta_{0}=\beta\omega_{0},$ $n\left( \omega\right) =1/\left( e^{\beta
\omega}-1\right) ,$ and%
\begin{equation}
D\left( u\right) =\frac{w^{2}}{v^{2}}\frac{u}{2}\left( 1-\frac{u}{\beta _{0}}%
\right) +\frac{v^{2}-w^{2}}{2v^{3}}\left( 1-e^{-vu}-4n\left( v\right)
\sinh^{2}\frac{vu}{2}\right) .  \label{e2}
\end{equation}
This result is variational, with variational parameters $v$ and $w,$ and
gives an upper bound to the exact polaron free energy. $-\Delta F$ increases
with increasing temperature and the effect of temperature on $\Delta F$
increases with increasing $\alpha$. The values for the free energy obtained
analytically from the Feynman polaron model are\textit{\ lower }than the
published MC results for $\alpha\lesssim2$ and $\alpha\geq4$ (but lie within
the 1\% error of the Monte Carlo results). Since the Feynman result for the
polaron free energy is an upper bound to the exact result, we conclude that
for $\alpha\lesssim2$ and $\alpha\geq4$ the results of the Feynman model are
closer to the exact result than the MC results of \cite{bgsprb285735}.

\begin{table}[htbp]
\caption{Comparison between the free energy of the Feynman polaron theory, $%
-(\Delta F)_{\mathrm{F}}$, and the Monte Carlo results of Ref.~\protect\cite%
{bgsprb285735}, $-(\Delta F)_{\mathrm{MC}}$, for $T/T_D=0.04$. The relative
difference is defined as $\Delta=100\times[(\Delta F)_{\mathrm{F}}-(\Delta
F)_{\mathrm{MC}}]/(\Delta F)_{\mathrm{MC}}$. (From \protect\cite{pdprb316826}%
) }
\label{tablePD316826}\centering
\newpage
\begin{tabular}[t]{|c|c|c|c|}
\hline\hline
$\alpha$ & $-\left( \Delta F\right)_{\text{F}}$ & $-\left( \Delta F\right)_{%
\text{MC}}$ & $\Delta$ (\%) \\ \hline
$0.5$ & $0.50860$ & $0.505$ & $0.71$ \\ \hline
$1.0$ & $1.02429$ & $1.020$ & $0.42$ \\ \hline
$1.5$ & $1.54776$ & $1.545$ & $0.18$ \\ \hline
$2.0$ & $2.07979$ & $2.080$ & $-0.010$ \\ \hline
$2.5$ & $2.62137$ & $2.627$ & $-0.21$ \\ \hline
$3.0$ & $3.17365$ & $3.184$ & $-0.32$ \\ \hline
$3.5$ & $3.73814$ & $3.747$ & $-0.24$ \\ \hline
$4.0$ & $4.31670$ & $4.314$ & $0.063$ \\ \hline\hline
\end{tabular}%
\end{table}

\subsubsection{Diagrammatic Monte-Carlo algorithm}

Mishchenko \emph{et al.} (2000) performed a study of the Fr\"{o}hlich
polaron on the basis of the Diagrammatic Quantum Monte Carlo (DQMC) method
\cite{Prokofev}. This method is based on the direct summation of Feynman
diagrams for Green's functions in momentum space. The basic object of their
investigation is the Matsu\-bara Green's function of the polaron in the
momentum ($\mathbf{k}$)--imaginary time ($\tau $) representation
\begin{eqnarray}
G(\mathbf{k},\tau ) &=&\langle \mathrm{vac}|c_{\mathbf{k}}(\tau )c_{\mathbf{k%
}}^{\dagger }(0)|\mathrm{vac}\rangle ,\tau \geq 0,  \label{G} \\
c_{\mathbf{k}}(\tau ) &=&\exp (H\tau )c_{\mathbf{k}}\exp (-H\tau ).
\end{eqnarray}

In terms of a complete set $\{|\nu \rangle \}$ of eigenstates of the Fr\"{o}%
hlich polaron Hamiltonian $H$, so that for a given $\mathbf{k}$ the relation
$H|\nu (\mathbf{k})\rangle =E_{\nu }(\mathbf{k})|\nu (\mathbf{(}k)\rangle $
and $H|\mathrm{vac}\rangle =E_{0}|\mathrm{vac}\rangle $, the expansion of
the Green's function (\ref{G})
\begin{equation}
G(\mathbf{k},\tau )=\sum_{\nu }|\langle \nu |c_{\mathbf{k}}^{\dagger }|%
\mathrm{vac}\rangle |^{2}\exp \{-[E_{\nu }(\mathbf{k})-E_{0}]\tau \}
\label{G1}
\end{equation}%
follows straightforwardly. For the calculations discussed in what follows, $%
E_{0}=0$.

The spectral function (Lehmann function) $g_{\mathbf{k}}(\omega )$ is
defined through the representation of the Green's function (\ref{G1}) in the
form
\begin{eqnarray}
G(\mathbf{k},\tau ) &=&\int_{0}^{\infty }d\Omega \quad g_{\mathbf{k}}(\Omega
),  \label{G2} \\
g_{\mathbf{k}}(\Omega ) &=&\sum_{\nu }\delta \lbrack \Omega -E_{\nu }(%
\mathbf{k})]|\langle \nu |c_{\mathbf{k}}^{\dagger }|\mathrm{vac}\rangle
|^{2}.  \label{G3}
\end{eqnarray}%
This spectral function is normalized, $\int_{0}^{\infty }g_{\mathbf{k}%
}(\Omega )d\Omega =1$. It can be interpreted as the probability that a
polaron has momentum $\mathbf{k}$ and energy $\Omega $. The significance of
the spectral function (\ref{G3}) is determined by the fact that it has poles
(sharp peaks) at frequencies, which correspond to stable (metastable)
quasiparticle states.

If, for a given $\mathbf{k}$, there is a stable state at energy $E(\mathbf{k}%
)$, the spectral function takes the form
\begin{equation}
g_{\mathbf{k}}(\Omega )=Z_{0}^{(\mathbf{k})}\delta \lbrack \Omega -E(\mathbf{%
k})]...,  \label{G4}
\end{equation}%
where $Z_{0}^{(\mathbf{k})}$ is the weight of the bare-electron state. The
energy $E_{\mathrm{gs}}(\mathbf{k})$ and the weight $Z_{0,\mathrm{gs}}^{(%
\mathbf{k})}$ for the polaron ground state can be extracted from the Green's
function behaviour at long times:
\begin{equation}
G(\mathbf{k},\tau \gg \omega _{0}^{-1})\rightarrow Z_{0}^{(\mathbf{k})}\exp
[-E(\mathbf{k})\tau ].  \label{G5}
\end{equation}%
Similarly to Eq. (\ref{G}), the $N$-phonon Green's function is defined:
\begin{eqnarray}
G_{N}(\mathbf{k},\tau ;\mathbf{q}_{1},...,\mathbf{q}_{N}) &=&\langle \mathrm{%
vac}|d_{\mathbf{q}_{N}}(\tau )...d_{\mathbf{q}_{1}}(\tau )c_{\mathbf{p}%
}(\tau )c_{\mathbf{p}}^{\dagger }(0)d_{\mathbf{q}_{1}}^{\dagger }(0)...d_{%
\mathbf{q}_{N}}^{\dagger }(0)|\mathrm{vac}\rangle ,\tau \geq 0,  \label{G6}
\\
\mathbf{p} &=&\mathbf{k}-\sum_{j-1}^{N}\mathbf{q}_{j}.  \nonumber
\end{eqnarray}%
From the asymptotic properties of the Green's functions (\ref{G6}) at long
times, the characteristics of the polaron ground state are found. In
particular, the weight of the $N$-phonon state for the polaron ground state
is given by
\begin{equation}
G_{N}(\mathbf{k},\tau \gg \omega _{0}^{-1};\mathbf{q}_{1},...,\mathbf{q}%
_{N})\rightarrow Z_{N}^{(\mathbf{k})}(\mathbf{q}_{1},...,\mathbf{q}_{N})\exp
[-E(\mathbf{k})\tau ].  \label{G7}
\end{equation}%
A standard diagrammatic expansion of the above described Green's functions
generates a series of Feynman diagrams. The following function is further
introduced:
\begin{equation}
P(\mathbf{k},\tau )=G(\mathbf{k},\tau )+\sum_{N=1}^{\infty }\int d\mathbf{q}%
_{1}...d\mathbf{q}_{N}\tilde{G}_{N}(\mathbf{k},\tau ;\mathbf{q}_{1},...,%
\mathbf{q}_{N}),  \label{G8}
\end{equation}%
where $\tilde{G}_{N}$ are irreducible $N$-phonon Green's functions (which do
not contain disconnected phonon propagators). From Eqs. (\ref{G5}), (\ref{G7}%
) and the completeness condition for the non-degenerate ground state
\begin{equation}
Z_{0}^{(\mathbf{k})}+\sum_{N=1}^{\infty }\int d\mathbf{q}_{1}...d\mathbf{q}%
_{N}Z_{N}^{(\mathbf{k})}(\mathbf{q}_{1},...,\mathbf{q}_{N})=1  \label{G9}
\end{equation}%
it follows that the polaron ground state energy is determined by the
asymptotic behaviour of the function (\ref{G8}):
\begin{equation}
P(\mathbf{k},\tau \gg \omega _{0}^{-1})\rightarrow \exp [-E(\mathbf{k})\tau
].  \label{G10}
\end{equation}

The function $P(\mathbf{k},\tau)$ is an infinite series of integrals
containing an ever increasing number of integration variables. The essence
of the DQMC method is to construct a process, which generates \textit{%
continuum random variables} $(\mathbf{k},\tau)$ with a distribution function
that coincides exactly with $P(\mathbf{k},\tau)$. Taking into account Eq. (%
\ref{G8}) and the diagrammatic rules, $P(\mathbf{k},\tau)$ is identified
with the distribution function
\begin{equation}
Q(\{y\})=\sum_{m=0}^{\infty} \sum_{\xi_m}\int dx_1...dx_m F_m(\xi_m, \{y\},
x_1,...,x_m),  \label{G11}
\end{equation}
where the external variables $\{y\}$ include $\mathbf{k},\tau, \alpha$ and $%
N $, while the internal variables describe the topology of the diagram
(labelled with $\xi_m$), times assigned to electron-phonon vertices and
momenta of phonon propagators. The diagrammatic Monte Carlo process is a
numeric procedure, which samples various diagrams in parameter space and
collects statistics for $Q(\{y\})$ according to the Metropolis algorithm
\cite{Metropolis} is such a way that --- when the process is repeated a
large number of times --- the result converges to the exact answer. The
distribution function given by the convergent series (\ref{G11}) is
simulated within the process of sequential stochastic generation of diagrams
described by functions $F_m(\xi_m, \{y\}, x_1,...,x_m)$. Further, using Eq. (%
\ref{G2}), the spectral function $g_{\mathbf{k}}(\omega)$ is obtained
applying a stochastic optimization technique. We refer to the review \cite%
{MN2006} for further details on the DQMC and stochastic optimization, where
information on the excited states of the polaron is also derived by the
analytic continuation of the imaginary time Green's functions to real
frequencies.

DQMC confirms that for $\alpha \gtrsim 1$ the bare-electron state in the
polaron wave function is no longer the dominant contribution and
perturbation theory is not adequate. The bare-electron weight $Z_0^{0}$ for
the polaron ground state, as a function of the polaron coupling constant,
rapidly vanishes for $\alpha \gtrsim 3$. It is suggested in \cite%
{Mishchenko2000} that in the interval $3 \lesssim \alpha \lesssim 10$ the
polaron ground states smoothly transforms between weak- and strong-coupling
limits.

Below (see sections \ref{section-A1}, \ref{method_OAAC}, \ref{sum_rules})
earlier analytical studies and results on Fr\"ohlich polarons are compared
with DQMC results. It would be beneficial, to have an independent numerical
check of the DQMC results. The comparison of the DQMC results for the
``low-energy'' ($\Omega < 0$) part of the spectral density \footnote{%
It is worth of noticing, that the spectral density is not identical to the
optical absorption coefficient, which is discussed below in sections \ref%
{OptWeak}, \ref{section-A1} and \ref{method_OAAC}.} for the Fr\"ohlich
polaron at $\alpha=0.05$ \cite{Mishchenko2000} demonstrates perfect
agreement with the perturbation theory result:
\begin{equation}
g_{0}(\Omega<0) = {\frac{\alpha }{2 }}\delta[\Omega+\alpha]
\end{equation}
($m$ and $\omega_{0}$ are set equal to unity). The DQMC results for the
``high-energy'' ($\Omega > 0$) part of the spectral density significantly
differ from the perturbation-theory curve. This is attributed to the fact
that for the Fr\"ohlich polaron the perturbation-theory expression for
\begin{equation}
g_{0}(\Omega>0) = {\frac{\alpha}{\pi}} {\frac{\theta(\Omega-1) }{\Omega^2
\sqrt{\Omega-1}}}
\end{equation}
diverges as $\Omega \to 1$ and, consequently, the perturbation approach is
no longer adequate. The main difference between the DQMC spectrum of the
Fr\"ohlich polaron and the perturbation-theory result is the broad peak in
the spectral density at $\Omega \sim 3.5$. This peak appears for
significantly larger values of the coupling constant and its weight grows
with $\alpha$, see Fig. \ref{SpectralDensity1}. As shown in the inset, near
the threshold, $\Omega=1$, the spectral density behaves as $\sqrt{\Omega-1}$.

\begin{figure}[tbh]
\begin{center}
\includegraphics[width=0.5\textwidth]{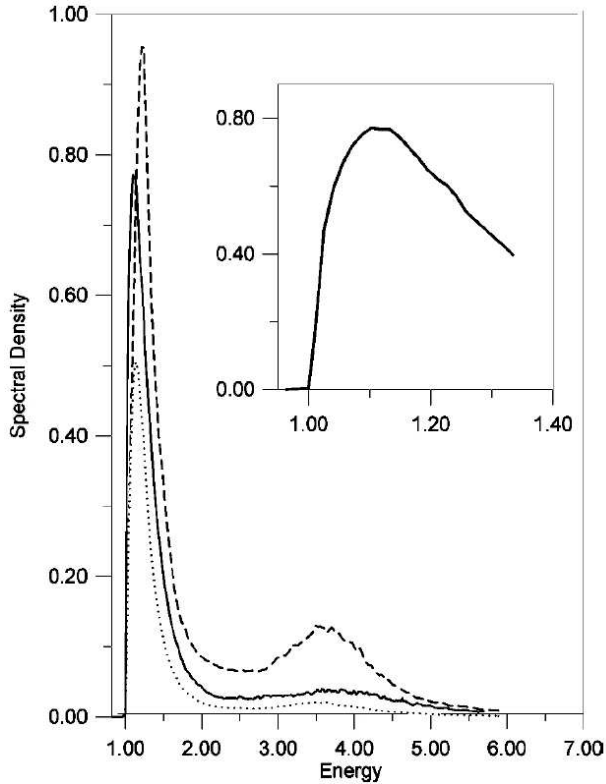}
\end{center}
\caption{The spectral density of the Fr\"ohlich polaron for various values
of the coupling constant: $\protect\alpha=0.5$ (dotted line), $\protect\alpha%
=1$ (solid line), and $\protect\alpha=2$ (dashed line) with energy counted
from the ground-state energy of the polaron. The initial fragment of the
spectral density for $\protect\alpha=1$ is shown in the inset (from
\protect\cite{Mishchenko2000}).}
\label{SpectralDensity1}
\end{figure}

Next to Fr\"{o}hlich polarons, other polaron models have been extensively
investigated using the Monte Carlo approach. In particular Alexandrov (1996)
proposed a long-range \textit{discrete} Fr\"{o}hlich interaction as a model
of the interaction between a hole and the oxygen ions in cuprate
superconductors. The essence of the model is that a charge carrier moves
from site to site on a discrete lattice (or chain in 1D) and interacts with
all the ions, which reside at the lattice sites. Numerically rigorous
polaron characteristics (ground-state energy, number of phonons in the
polaron cloud, effective mass and isotope exponent) for such a lattice
polaron, valid for arbitrary EPI, were obtained using a path-integral
continuous-time quantum Monte Carlo technique (CTQMC) \cite%
{alexandrov:1999,jim,spencer2005}. This study leads to a ``mobile small Fr\"{o}hlich''\ polaron (\ref{lightSFP}).

Over the years the Feynman model for the Fr\"ohlich polaron has remained the
most successful approach to the problem. The analysis of an exactly solvable
(``symmetrical'') 1D-polaron model \cite%
{DThesis,DE64,DE1968}, Monte Carlo schemes \cite{Mishchenko2000,TPC} and,
recently, a unifying variational approach \cite{DeFilippis2003,catbook}
demonstrate the remarkable accuracy of Feynman's path-integral method. Using
the variational wave-functions, which combine both the Landau-Pekar and the
Feynman approximations, \cite{catbook} found in the $\alpha \rightarrow
\infty $ limit:
\begin{equation}
E_0=\left( -0.108507\alpha ^{2}-2.67\right) \omega_{0}.
\end{equation}%
which is slightly lower than the variational Feynman's estimate, Eq. (\ref%
{eq_9a}), and at small values of $\alpha $,
\begin{equation}
E_0=-\alpha \omega_{0}-0.0123\alpha ^{2} \omega_{0},~~\alpha \rightarrow 0~,
\end{equation}
i.e. the same result, as Eq. (\ref{eq_7a}). In the latter limit the correct
result for the electron self-energy is
\begin{equation}
E=-\alpha \omega_{0}-0.0159\alpha ^{2} \omega_{0},
\end{equation}%
as found by Grosjean (1957) and confirmed by H\"ohler and Mullensiefen
(1959), and R\"oseler (1968).

Within the path-integral approach, Feynman \textit{et al}. studied the
mobility of polarons \cite{FHIP,TF70}. Subsequently the path-integral
approach to the polaron problem was generalized and developed to become a
powerful tool to study optical absorption and cyclotron resonance \cite%
{DSG72,PD86}.

\section{Response of continuum polarons}

\subsection{Mobility}

The mobility of large polarons was studied within various theoretical
approaches. Fr\"ohlich \cite{Frohlich1937,fro} pointed out the typical
behavior of the large-polaron mobility
\begin{equation}
\mu\propto\exp(\omega_{0}\beta),
\end{equation}
which is characteristic for weak coupling. Within the weak-coupling regime,
the mobility of the polaron was then derived, e. g., using the Boltzmann
equation in \cite{HS1953,Osaka1961} and starting from the LLP-transformation
in Ref. \cite{LP1955}. In \cite{LK1964} it was shown that for weak coupling:
\begin{equation}
\mu =\frac{e}{2\alpha}\left[\exp(\omega_{0}\beta)-1\right]\left[1-\frac{%
\alpha}{6}+O(\alpha^2)\right].
\end{equation}
A nonperturbative analysis was based on the Feynman polaron theory, where
the mobility $\mu$ of the polaron using the path-integral formalism was
derived by Feynman et al. (1962) (FHIP) as a static limit, starting from a
frequency-dependent impedance function. Details of the FHIP theory are given
in \cite{Platzman1963}. An approximate expression for the impedance function
of a Fr\"{o}hlich polaron at all frequencies, temperatures and coupling
strengths was obtained in Refs. \cite{FHIP,Platzman1963} within the
path-integral technique. Assuming the crystal to be isotropic, an
alternating electric field $\mathbf{E}=E_{0}\mathbf{e}_{x}\exp \left(
i\Omega t\right) $ induces a current in the $x$-direction
\begin{equation}
j(\Omega )=\left[ z(\Omega )\right] ^{-1}E_{0}\exp \left( i\Omega t\right) .
\label{F1}
\end{equation}%
The complex function $z(\Omega )$ is called the impedance function. The
electric field is considered sufficiently weak, so that linear-response
theory can be applied. The frequency-depending mobility $\mu(\Omega)$ is
defined as
\begin{equation}
\mu(\Omega )=\mathrm{Re}\left[ z(\Omega )\right] ^{-1}.  \label{F1a}
\end{equation}%
Taking the electric charge as unity, one arrives at $j=\left\langle \dot{x}%
\right\rangle $, where $\left\langle x\right\rangle $ is the expectation
value of the electron displacement in the direction of the applied field: $%
\left\langle x\right\rangle =E/i\Omega z(\Omega ).$ In terms of time
dependent variables,
\begin{equation}
\left\langle x(\tau )\right\rangle =-\int\limits_{-\infty }^{\tau }iG(\tau
-\sigma )E(\sigma )d\sigma ,  \label{F2}
\end{equation}%
where $G(\tau )$ has the inverse Fourier transform
\begin{equation}
G(\Omega )=\int\limits_{-\infty }^{\infty }G(\tau )\exp \left( -i\Omega \tau
\right) d\tau =\left[ \Omega z(\Omega )\right] ^{-1}.  \label{F3}
\end{equation}
The expected displacement at time $\tau $ is
\begin{equation}
\left\langle x(\tau )\right\rangle =\text{Tr}\left[ xU(\tau ,a)\rho
_{a}U^{\prime -1}(\tau ,a)\right] .  \label{F4}
\end{equation}%
Here, $\rho _{a}$ is the density matrix of the system at some time $a$, long
before the field is turned on, and
\begin{equation}
U(\tau,a)=\text{T}\exp \left\{-i\int\limits_{a}^{\tau }\left[ H_{s}-x_{s}E(s)%
\right] ds\right\}  \label{F5a}
\end{equation}%
is the unitary operator of the development of a state in time with the
complete Hamiltonian $H-xE,$ where $H$ is the Fr\"{o}hlich polaron
Hamiltonian, and T is the time ordering operator \cite{Feynman1951}. Primed
operators are ordered antichronologically:
\begin{equation}
U^{\prime -1}(\tau,a)=\text{T}^\prime \exp \left\{i\int\limits_{a}^{\tau } %
\left[H_{s}^{\prime }-x_{s}^{\prime }E^{\prime }(s)\right] ds\right\} .
\label{F5b}
\end{equation}%
$G(\tau -\sigma )$ can be represented as the second derivative
\begin{equation}
G(\tau -\sigma )=\frac{1}{2}\left. \left( \frac{\partial ^{2}g}{\partial
\eta \partial \varepsilon }\right) \right\vert _{\varepsilon =\eta =0}
\label{F6}
\end{equation}%
of%
\begin{equation}
g=\text{Tr}\left[ U(b,a)\rho _{a}U^{\prime -1}(b,a)\right] ,\quad
b\rightarrow \infty ,\quad a\rightarrow -\infty  \label{F7a}
\end{equation}%
with $E(s)=\varepsilon \delta (s-\sigma )+\eta \delta (s-\tau ),E^{\prime
}(s)=\varepsilon \delta (s-\sigma )-\eta \delta (s-\tau ).$ The initial
state is chosen at a definite temperature $T$, $\rho _{a}\propto \exp
(-\beta H)$. If the time $a$ is sufficiently far in the past, FHIP assume
that only the phonon subsystem was in thermal equilibrium at temperature $T$%
. The energy of the single electron and of the electron-phonon coupling are
infinitesimal (of order $1/V$) with respect to that of the phonons. With
this choice of the initial distribution, the phonon coordinates can be
eliminated from the expression (\ref{F7a}), and the entire expression is
reduced to a double path integral over the electron coordinates only:
\begin{equation}
g=\int \int \exp \left( i\Phi \right) \mathcal{D}\mathbf{r}(t)\mathcal{D}%
\mathbf{r}^{\prime }(t),  \label{F8a}
\end{equation}%
where (taking $m=1$)

\begin{eqnarray}
\Phi &=&\int\limits_{-\infty }^{\infty }\left[ \frac{\mathbf{\dot{r}}^{2}}{2}%
-\frac{\mathbf{\dot{r}}^{\prime 2}}{2}\right]dt - \int\limits_{-\infty
}^{\infty }\left[ \mathbf{E}(t)\cdot \mathbf{r}(t)-\mathbf{E}^{\prime
}(t)\cdot \mathbf{r}^{\prime }(t)\right] dt  \label{F8b} \\
&&+\frac{i\alpha }{2^{3/2}}\int\limits_{-\infty }^{\infty
}\int\limits_{-\infty }^{\infty }\left[ \frac{\exp (-i\left\vert
t-s\right\vert )+2P(\beta )\cos (t-s)}{\left\vert \mathbf{r}(t)-\mathbf{r}%
(s)\right\vert }+\frac{\exp (+i\left\vert t-s\right\vert )+2P(\beta )\cos
(t-s)}{\left\vert \mathbf{r}^{\prime }(t)-\mathbf{r}^{\prime }(s)\right\vert
}\right.  \nonumber \\
&&\left. -\frac{2\left[ \exp (-i\left(t-s\right))+ 2P(\beta )\cos (t-s)%
\right] }{\left\vert \mathbf{r}^{\prime }(t)-\mathbf{r}(s)\right\vert }%
\right] dtds,  \nonumber
\end{eqnarray}
where $P(\beta )=\left[ e^{\beta }-1\right] ^{-1}.$ The double path integral
in (\ref{F8a}) is over paths which satisfy the boundary condition $\mathbf{r}%
(t)\mathcal{-}\mathbf{r}^{\prime }(t)=0$ at times $t$ approaching $\pm
\infty .$ The expression (\ref{F8a}) with (\ref{F8b}) is supposed to be
exact \cite{FHIP}. Clearly to provide analytical solutions at all $\alpha $
presumably is impossible.

Following Feynman's idea to describe the ground state energy properties of a
polaron by introducing a parabolic ``retarded'' interaction of the electron
with itself (see subsection \ref{Ground_Feynman}), it is assumed in \cite%
{FHIP} that the dynamical behavior of the polaron can be described
approximately by replacing $\Phi $ of Eq. (\ref{F8a}) by a parabolic
(retarded) expression
\begin{eqnarray}
\Phi _{0} &=&\int\limits_{-\infty }^{\infty }\left[ \frac{\mathbf{\dot{r}}%
^{2}}{2}-\frac{\mathbf{\dot{r}}^{\prime 2}}{2}\right] dt
-\int\limits_{-\infty }^{\infty }\left[ \mathbf{E}(t)\cdot \mathbf{r(}t%
\mathbf{)}-\mathbf{E}^{\prime }(t)\cdot \mathbf{r}^{\prime }\mathbf{(}t%
\mathbf{)}\right] dt  \label{F8c} \\
&&-\frac{iC}{2}\int\limits_{-\infty }^{\infty }\int\limits_{-\infty
}^{\infty }\left\{\left[ \mathbf{r}(t)-\mathbf{r}(s)\right] ^{2}\left[
e^{-iw\left\vert t-s\right\vert }+2P(\beta w)\cos w(t-s)\right] \right.
\nonumber \\
&&+\left[ \mathbf{r}^{\prime }(t)-\mathbf{r}^{\prime }(s)\right] ^{2}\left[
e^{+iw\left\vert t-s\right\vert }+2P(\beta w)\cos (t-s)\right]  \nonumber \\
&&\left. -2\left[ \mathbf{r}^{\prime }(t)-\mathbf{r}(s)\right] ^{2}\left[
e^{-iw\left(t-s\right)}+2P(\beta w)\cos w(t-s)\right] \right\} dtds.
\nonumber
\end{eqnarray}%
The parameters $C$ and $w$ are to be determined so as to approximate $\Phi $
as closely as possible. However, no variational principle is known for the
mobility. At zero temperature, $C$ and $w$ are fixed at the values, which
provide the best upper bound for the polaron ground state energy (\ref{sr11}%
). At finite temperatures ($T\neq 0$) the parameters $C$ and $w$ are
determined from the variational principle for the polaron free energy \cite%
{FeynmanSM}. This way of selection of the model parameters $C$ and $w$ is
based on the supposition, that ``the comparison Lagrangian, which gives a
good fit to the ground-state energy at zero temperature, will also give the
dynamical behaviour of the system'' \cite{FHIP}.

The analytical calculation of path integrals in (\ref{F8a}) in \cite{FHIP}
was performed to the first term in an expansion of $\exp \left[ i\left( \Phi
-\Phi _{0}\right) \right] $:
\begin{eqnarray}
g &=&\int \int \exp \left( i\Phi _{0}\right) \exp \left[ i\left( \Phi -\Phi
_{0}\right) \right] \mathcal{D}\mathbf{r}(t)\mathcal{D}\mathbf{r}^{\prime
}(t)\approx g_{0}+g_{1},  \label{F9a} \\
g_{0} &=&\int \int \exp \left( i\Phi _{0}\right) \mathcal{D}\mathbf{r}(t)%
\mathcal{D}\mathbf{r}^{\prime }(t),  \label{F9b} \\
g_{1} &=&i\int \int \exp \left( i\Phi _{0}\right) \left( \Phi -\Phi
_{0}\right) \mathcal{D}\mathbf{r}(t)\mathcal{D}\mathbf{r}^{\prime }(t).
\label{F9c}
\end{eqnarray}%
Using (\ref{F3}) and (\ref{F6}), one finds from(\ref{F9a})
\begin{equation}
G(\Omega )\approx G_{0}(\Omega )+G_{1}(\Omega ).  \label{F10}
\end{equation}%
The first term in the rhs of Eq. (\ref{F10}) is
\begin{equation}
G_{0}(\Omega )=iY_{0}(\Omega ),\quad Y_{0}(\Omega )=-\frac{\Omega ^{2}-w^{2}%
}{(\Omega -i\varepsilon )^{2}[(\Omega -i\varepsilon )^{2}-v^{2}]},\quad
\varepsilon \rightarrow +0  \label{F11}
\end{equation}%
with $v^{2}=w^{2}+4C/w$. The second term in the rhs of Eq. (\ref{F10}) is%
\begin{eqnarray}
G_{1}(\Omega ) &=&-iY_{0}^{2}(\Omega )\left[ \chi (\Omega )-\frac{4C}{w}%
\frac{\Omega ^{2}}{\Omega ^{2}-w^{2}}\right] ,  \label{F12a} \\
\chi (\Omega ) &=&\int\limits_{0}^{\infty }\left( 1-e^{i\Omega u}\right)
\mathrm{Im}S(u)du,  \label{F12b} \\
S(u) &=&\frac{2\alpha }{3\sqrt{\pi }}\frac{e^{iu}+2P(\beta )\cos u}{\left[
D(u)\right] ^{3/2}},  \label{F12c} \\
D(u) &=&\frac{w^{2}}{v^{2}}\left\{\frac{v^{2}-w^{2}}{w^{2}v}\left[
1-e^{ivu}+4P(\beta v)\sin ^{2}\left( \frac{vu}{2}\right) \right] \right.
\label{F12d} \\
&&\left. -iu+\frac{u^{2}}{\beta }\right\} .  \nonumber
\end{eqnarray}

From (\ref{F3}) and (\ref{F10}), the impedance results in the form
\begin{equation}
\Omega z(\Omega )\approx \left[ G_{0}(\Omega )+G_{1}(\Omega )\right] ^{-1}.
\label{F13a}
\end{equation}%
Feynman \emph{et al.}(1962) suggested to use this expression expanded to
first order in $G_{1}(\Omega )$:%
\begin{equation}
\Omega z(\Omega )\approx \left[ G_{0}(\Omega )\right] ^{-1}-[G_{0}(\Omega
)]^{-2}G_{1}(\Omega ).  \label{F13b}
\end{equation}
Substitution of (\ref{F11}) and (\ref{F12a}) into Eq.(\ref{F13b}) leads to
the final expression (Ref. \cite{FHIP}) for the impedance function of the
Fr\"ohlich polaron:
\begin{equation}
\Omega z(\Omega )= i\left[ \Omega^2-\chi(\Omega) \right].  \label{F15}
\end{equation}

An alternative derivation of the impedance function of the Fr\"{o}hlich
polaron, based on simple operator techniques, was presented in \cite{PD1983}%
. The FHIP result was worked out in \cite{PD1984} in detail to get a
physical insight into the scattering processes incorporated in the FHIP
approximation. For sufficiently low temperature the FHIP polaron mobility
takes the form \cite{FHIP}
\begin{equation}
\mu=\left( \frac{w}{v}\right) ^{3} \frac{3e}{4m \omega ^{2}_{0} \alpha\beta}%
\mbox{e}^{\omega_{0}\beta} \exp\{(v^{2}-w^{2})/w^{2}v\}\ ,  \label{eq:P24-1}
\end{equation}
where $v$ and $w$ are variational functions of $\alpha$ obtained from the
Feynman polaron model.

Using the Boltzmann equation for the Feynman polaron model, Kadanoff (1963)
found the mobility, which for low temperatures can be represented as
follows:
\begin{equation}
\mu=\left( \frac{w}{v}\right) ^{3} \frac{e}{2m \omega_{0}\alpha} \mbox{e}%
^{\omega_{0}\beta} \exp\{(v^{2}-w^{2})/w^{2}v\}\ ,  \label{eq:P24-1-kadanoff}
\end{equation}
The weak-coupling perturbation expansion of the low-temperature polaron
mobility as found using the Green's function technique \cite{LK1964} has
confirmed that the mobility derived from the Boltzmann equation is
asymptotically exact for weak coupling ($\alpha \ll1$) and at low
temperatures ($T \ll\omega_{0}$). As first noticed in \cite{Kadanoff}, the
mobility of Eq. (\ref{eq:P24-1}) differs by the factor of $3/(2\beta
\omega_{0})$ from that derived using the polaron Boltzmann equation as given
by Eq. (\ref{eq:P24-1-kadanoff}). \footnote{%
In the asymptotic limit of weak electron-phonon coupling and low
temperature, the FHIP polaron mobility of Eq. (\ref{eq:P24-1}) differs by
the same factor of $3/(2\beta\omega_{0})$ from the earlier result~\cite%
{HS1953,LP1955,Osaka1961}, which, as pointed out in Ref. \cite{FHIP} and in
later publications \cite{Kadanoff,LK1964,PD1984}, is correct for $\beta\gg1$.%
} As follows from this comparison, the result of ~\cite{FHIP} is not valid
when $T\to 0$. As emphasized in ~\cite{FHIP} and later confirmed, in
particular, in ~\cite{PDpss1983,PD1984} the above discrepancy can be
attributed to an interchange of two limits in calculating the impedance. In
FHIP, for weak electron-phonon coupling, one takes $\lim_{\Omega\to0}
\lim_{\alpha\to0}$, whereas $\lim_{\alpha\to0} \lim_{\Omega\to0}$ should be
calculated. It turns out that for the asymptotically correct result the
mobility at low temperatures is mainly limited by the absorption of phonons,
while in the theory of FHIP it is the emission of phonons which gives the
dominant contribution as $T$ goes to zero~\cite{PDpss1983}.

The analysis based on the Boltzmann equation takes into account the phonon
emission processes whenever the energy of the polaron is above the emission
threshold. The independent-collision model, which underlies the
Boltzmann-equation approach, however, fails in the ``strong coupling
regime'' of the Fr\"ohlich polaron, when the thermal mean free path becomes
less than the de Broglie wavelength; in this case, the Boltzmann equation
cannot be expected to be adequate \cite{FHIP,HB1999}.

Experimental work on alkali halides and silver halides indicates that the
mobility obtained from Eq.~(\ref{eq:P24-1}) describes the experimental
results accurately (see \cite{Brown1963,Brown1972} and references therein).
Measurements of mobility as a function of temperature for photoexcited
electrons in cubic $n$-type Bi$_{12}$SiO$_{20}$ are explained well in terms
of large polarons within the Feynman approach \cite{HB1999}. The
experimental findings on electron transport in crystalline TiO$_{2}$ (rutile
phase) probed by THz time-domain spectroscopy were quantitatively
interpreted within the Feynman model \cite{Hendry2004}. One of the reasons
for the agreement between theory based on Eq. (\ref{eq:P24-1}) and
experiment is that in the path-integral approximation to the polaron
mobility, a Maxwellian distribution for the \textit{electron velocities} is
assumed, when applying the adiabatic switching on of the Fr\"{o}hlich
interaction. Although such a distribution is not inherent in the Fr\"{o}%
hlich interaction, its incorporation tends to favor agreement with
experiment because other mechanisms (interaction with acoustic phonons etc.)
cause a Gaussian distribution.

\subsection{Optical absorption at weak coupling}

\label{OptWeak} At zero temperature and in the weak-coupling limit, the
optical absorption {of a Fr\"{o}hlich polaron} is due to the elementary
polaron scattering process with the absorption of incoming photon and
emission of an outgoing phonon. In the weak-coupling limit ($\alpha \ll 1$)
the polaron absorption coefficient {for a many-polaron gas} was first
obtained by Gurevich \emph{et al.} (1962). Their optical-absorption
coefficient is equivalent to a particular case of \cite{TDPRB01} with the
dynamic structure factor $S(\mathbf{q},\omega )$ corresponding to the
Hartree-Fock approximation. In \cite{TDPRB01} the optical absorption
coefficient of a many-polaron gas was shown to be given, to order $\alpha $,
by {\
\begin{equation}
\mathrm{Re}[\sigma (\Omega )]=n_{0}e^{2}\frac{2}{3}\alpha \frac{1}{2\pi
\Omega ^{3}}\int_{0}^{\infty }dqq^{2}S(\mathbf{q},\Omega -\omega _{0}),
\label{opticabsTD}
\end{equation}%
where $n_{0}$ is the density of charge carriers. }

In the zero-temperature limit, starting from the Kubo formula the optical
conductivity {of a \textit{single} Fr\"ohlich polaron} can be represented in
the form%
\begin{equation}
\sigma(\Omega) =i\frac{e^{2}}{m\left( \Omega+i\delta\right) } +\frac{e^{2}}{%
m^{2}}\frac{1}{\left( \Omega+i\delta\right) ^{3}}\int_{0}^{\infty}e^{-\delta
t}\left( e^{i\Omega t}-1\right) \sum_{\mathbf{q,q}^{\prime}}q_{x}q_{x}^{%
\prime}\left\langle \Psi_{0}|[\hat{B}_{\mathbf{q}}(t), \hat{B}_{-\mathbf{%
q^{\prime }}}(0)]|\Psi_{0}\right\rangle dt,  \label{FF5}
\end{equation}
{where $\delta =+0$, $\hat{B}_{\mathbf{q}}(t)= [V_{\mathbf{q}}d_{\mathbf{q}%
}(t) +V_{-\mathbf{q}}^{\ast }d_{-\mathbf{q}}^{\dagger}(t)] e^{i\mathbf{%
q\cdot r}(t)}$, and} $\left\vert \Psi_{0}\right\rangle $ is the ground-state
wave function of the electron-phonon system. Within the weak coupling
approximation, the following analytic expression for the real part of the
polaron optical conductivity results from Eq. (\ref{FF5}):
\begin{equation}
\mathrm{Re}\sigma\left( \Omega\right) =\frac{\pi e^{2}}{2m^{\ast}}%
\delta\left( \Omega\right) +\frac{2e^{2}}{3m}\frac{\omega_{0}\alpha}{%
\Omega^{3}}\sqrt{\Omega-\omega_{0}}\Theta\left( \Omega-\omega_{0}\right) ,
\label{Resigma}
\end{equation}
{where $\Theta(x)=1$ if $x>0$, and zero otherwise.} The spectrum of the real
part of the polaron optical conductivity (\ref{Resigma}) is represented in
Fig. \ref{weakabscoef}.

\begin{figure}[tbh]
\begin{center}
\includegraphics[width=.6\textwidth]{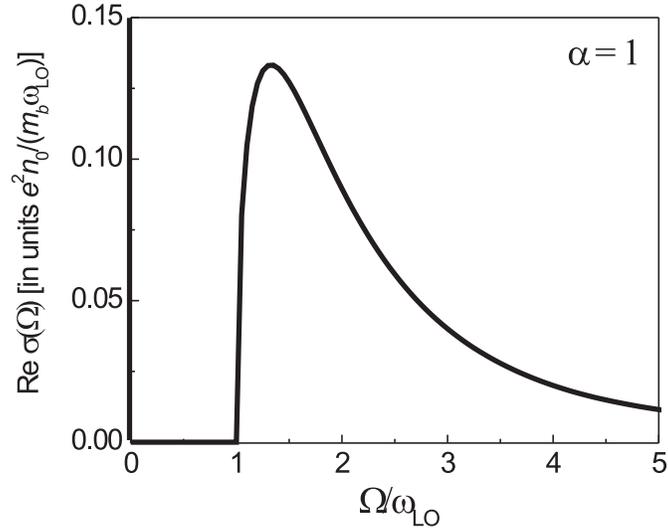}
\end{center}
\caption{Polaron optical conductivity for $\protect\alpha = 1$ in the
weak-coupling approximation, according to \protect\cite{Devreese72} A $%
\protect\delta$-like central peak (at $\Omega=0$) is schematically shown by
a vertical line. {(Reprinted with permission from \protect\cite%
{Devreese2005_V}. \copyright 2006, Societ\`{a} Italiana di Fisica.)}}
\label{weakabscoef}
\end{figure}

According to Eq. (\ref{Resigma}), the absorption coefficient of light with
frequency $\Omega >0$ by free polarons for $\alpha \rightarrow 0$ takes the
form
\begin{equation}
\Gamma (\Omega )=\frac{4\pi }{nc}\frac{2n_{p}e^{2}\alpha \omega _{0}^{2}}{%
3m\Omega ^{3}}(\Omega /\omega _{0}-1)^{1/2}\quad \Theta \left( \Omega
-\omega _{0}\right) ,  \label{DHL24}
\end{equation}%
\textrm{where }$c$\textrm{\ is the velocity of light in vacuum,} $n$ is the
refractive index of the medium, {$n_{p}$} is the concentration of polarons.
A simple derivation \cite{DHL1971} using a canonical transformation method
gives the absorption coefficient of free polarons, which coincides with the
result (\ref{DHL24}). {The} step function in (\ref{DHL24}) reflects the fact
that at zero temperature the absorption of light accompanied by the emission
of a phonon can occur only if the energy of the incident photon is larger
than that of a phonon ($\Omega >\omega _{0}$). In the weak-coupling limit,
according to Eq. (\ref{DHL24}), the absorption spectrum consists of a
``one-phonon line''. At nonzero
temperature, the absorption of a photon can be accompanied not only by
emission, but also by absorption of one or more phonons. Similarity between
the temperature dependence of several features of the experimental infrared
absorption spectra in high-T$_{c}$ superconductors and the temperature
dependence predicted for the optical absorption of a single Fr\"{o}hlich
polaron \cite{DSG72,Devreese72} has been revealed in Ref. \cite{DT1998}.

Experimentally, this one-phonon line has been observed for free polarons in
the infrared absorption spectra of CdO-films \cite{Finkenrath}, which is a
weakly polar material with $\alpha \approx 0.74$. The polaron absorption
band is observed in the spectral region between 6 and 20~$\mu$m (above the
LO phonon frequency). The difference between the one-polaron theoretical
absorption and experiment in the wavelength region where polaron absorption
dominates the spectrum has been explained as due to many-polaron effects
\cite{TDPRB01}.

\subsection{Optical absorption at strong coupling}

\label{section-A1}

The structure of the large polaron excitation spectrum constituted a central
question at the early stages of the development of polaron theory. The
exactly solvable polaron model of Ref. \cite{DThesis} was used to
demonstrate the existence of the so-called ``relaxed excited
states'' of large polarons \cite{DE64}. In Ref. \cite%
{DThesis}, an exactly solvable (``symmetric'') 1D-polaron model was introduced and analysed.
The further study of this model was performed in Refs. \cite{DE64,DE1968}.
The model consists of an electron interacting with two oscillators
possessing opposite wave vectors:\textbf{\ }$\mathbf{q}$ and $\mathbf{-q}$.
The parity operator, which changes $d_{\mathbf{q}}$ and $d_{-\mathbf{q}}$
(and also $d_{\mathbf{q}}^{\dag}$ and $d_{-\mathbf{q}}^{\dag}$), commutes
with the Hamiltonian of the system. Hence, the polaron states are classified
into even and odd states with eigenvalues of the parity operator $+$1 and $-$%
1, respectively. For the lowest even and odd states, the phonon distribution
functions $W_{N}$ are plotted in Fig. \ref{NPolarons}, upper panel, for some
values of the effective coupling constant $\lambda$ of this
``symmetric'' model. The value of the
parameter $\varkappa=[ q^{2}/ m\omega_{0}]^{1/2}$ for these graphs was taken
to be 1, while the total polaron momentum $\mathbf{P}=0$. In the
weak-coupling case ($\lambda\approx0.6$) $W_{N}$ is a decaying function of $%
N $. When increasing $\lambda$, $W_{N}$ acquires a maximum, e.~g. at $N=8$
for the lowest even state at $\lambda=5.0625$. The phonon distribution
function $W_{N}$ has the same character for the lowest even and the lowest
odd states at all values of the number of virtual phonons in the ground
state (as distinct from the higher states). This led to the conclusion that
the lowest odd state is an internal excited state of the polaron.

\begin{figure}[h]
\begin{center}
\includegraphics[width=0.7\textwidth]{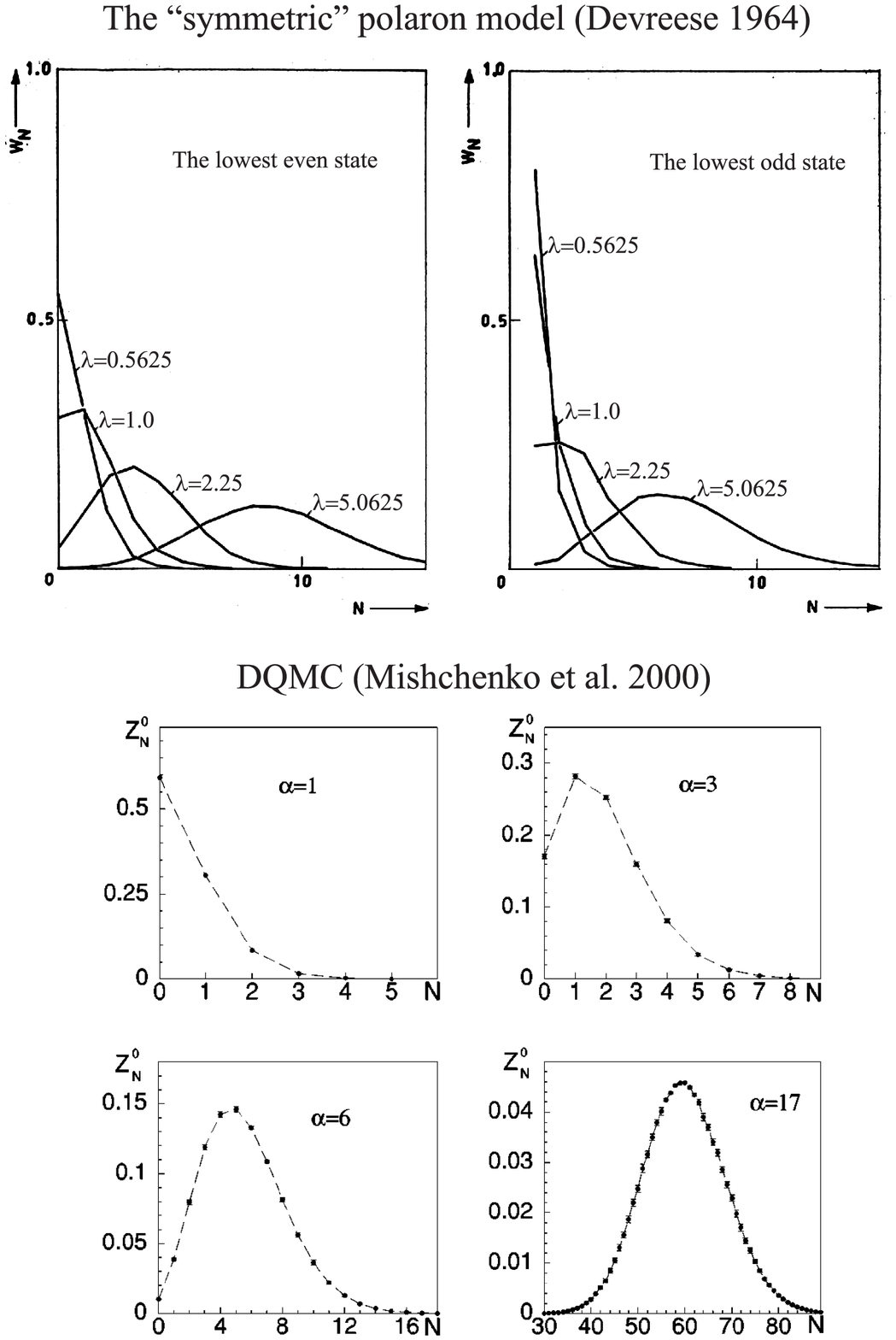}
\end{center}
\caption{\textit{Upper panel}: The phonon distribution functions $W_{N}$ in
the ``symmetric''\ polaron model for various
values of the effective coupling constant $\protect\lambda$ at $\varkappa=1,%
\mathbf{P}=0$ (from \protect\cite{DThesis}). \textit{Lower panel}:
Distribution of multiphonon states in the polaron cloud within DQMC method
for various values of $\protect\alpha$ (from \protect\cite{Mishchenko2000}).
(Reprinted with permission from \protect\cite{Devreese2005_V}. \copyright %
2006, Societ\`{a} Italiana di Fisica.)}
\label{NPolarons}
\end{figure}

In Ref. \cite{Mishchenko2000}, the structure of the Fr\"{o}hlich polaron
cloud was investigated using the DQMC method. Contributions of $N$-phonon
states to the polaron ground state were calculated as a function of $N$ for
a few values of the coupling constant $\alpha$, see Fig.\ref{NPolarons},
lower panel. As follows from Fig.\ref{NPolarons}, the evolution from the
weak-coupling case ($\alpha=1$) into the strong-coupling regime ($\alpha=17$%
) was studied. The evolution of the shape and the scale of the distribution
of the $N$-phonon states with increasing $\alpha$ as derived for a Fr\"{o}%
hlich polaron within DQMC method \cite{Mishchenko2000} is \textit{in notable
agreement} with the results obtained within the ``symmetric'' 1D polaron model \cite{DThesis,DE64,DE1968}.

The insight gained as a result of those investigations concerning the
structure of the excited polaron states, was subsequently used to develop a
theory of the optical absorption spectra of polarons. The first work was
limited to the strong coupling limit \cite{KED69}, {where} the impact of the
internal degrees of freedom of polarons on their optical properties {was
revealed}. The optical absorption of light by free Fr\"ohlich polarons was
treated in Ref.~\cite{KED69} using the polaron states obtained within the
adiabatic strong-coupling approximation. It was argued in Ref.~\cite{KED69},
that for sufficiently large $\alpha$ ($\alpha \gtrsim 3$), the (first)
relaxed excited state (RES) of a polaron is a relatively stable state, which
{gives rise to a ``resonance'' in the polaron optical absorption spectrum}.
The following scenario of a transition, which leads to a \textit{%
``zero-phonon''\ peak} in the absorption by
a strong-coupling polaron, {was} then suggested. If the frequency of the
incoming photon is equal to $\Omega_{\mathrm{RES}}=0.065\alpha^{2}%
\omega_{0}, $ the electron jumps from the ground state (which, at large
coupling, is well-characterised by ``$s$''-symmetry for the electron) to an
excited state (``$2p$''), while the lattice polarization in the final state
is adapted to the ``$2p$'' electronic state of the polaron. In Ref. \cite%
{KED69} considering the decay of the RES with emission of one real phonon,
it is argued, that the ``zero-phonon''\ peak
can be described using the Wigner-Weisskopf formula which is valid when the
linewidth of that peak is much smaller than $\omega_{0}.$

For photon energies larger than $\Omega_{\mathrm{RES}}+\omega_{\mathrm{LO}},$
a transition of the polaron towards the \textit{first scattering state},
belonging to the RES, becomes possible. The final state of the optical
absorption process then consists of a polaron in its lowest RES plus a free
phonon. A ``one-phonon sideband''\ then
appears in the polaron absorption spectrum. This process is called \textit{%
one-phonon sideband absorption}. The one-, two-, ... $K$-, ... phonon
sidebands of the zero-phonon peak give rise to a broad structure in the
absorption spectrum. It turns out that the \textit{first moment} of the
phonon sidebands corresponds to the Franck-Condon (FC) frequency $\Omega_{%
\mathrm{FC}}=0.141\alpha^{2}\omega_{0}.$

To summarise, following \cite{KED69}, the polaron optical absorption
spectrum at strong coupling is characterised by the following features (at $%
T=0$):

\begin{enumerate}
\item[a)] An absorption peak (``zero-phonon
line'') appears, which corresponds to a transition from the
ground state to the first RES at $\Omega_{\mathrm{RES}}$.

\item[b)] For $\Omega>\Omega_{\mathrm{RES}}+\omega_{0}$, a phonon sideband
structure arises. This sideband structure peaks around $\Omega _{\mathrm{FC}%
} $. {Even when the zero-phonon line becomes weak, and most oscillator
strength is in the LO-phonon sidebands, the zero-phonon line continues to
determine the onset of the phonon sideband structure.}
\end{enumerate}

The basic qualitative strong coupling behaviour predicted in Ref. \cite%
{KED69}, namely, zero-phonon (RES) line with a broader sideband at the
high-frequency side, was confirmed by later studies, as discussed below.

\subsection{Optical absorption at arbitrary coupling}

\label{method_OAAC}

The optical absorption of the Fr\"ohlich polaron was calculated in 1972 \cite%
{DSG72,Devreese72} (``DSG'') for the Feynman polaron model using path
integrals. Until recently DSG combined with \cite{KED69} constituted the
basic picture for the optical absorption of the Fr\"ohlich polaron. \cite%
{PD1983} rederived the DSG-result using the memory function formalism (MFF).
The DSG-approach is successful at {weak} electron-phonon coupling and is
able to identify the excitations at intermediate electron-phonon coupling {($%
3 \lesssim \alpha \lesssim 6$). In the strong coupling limit DSG still gives
an accurate first moment for the polaron optical absorption but does not
reproduce the broad phonon sideband structure (cf. \cite{KED69} and \cite%
{Goovaerts73}).} A comparison of the DSG results with the optical
conductivity spectra given by recently developed ``approximation-free''
numerical \cite{Mishchenko2003} and approximate analytical \cite%
{DeFilippis2003,PRL2006} approaches {was carried out } recently \cite%
{PRL2006}, {see also the review articles \cite{catbook} and \cite{MN2006}}.

The polaron absorption coefficient $\Gamma (\Omega )$ of light with
frequency $\Omega $ at arbitrary coupling was first derived in
Ref.\thinspace \cite{DSG72}. \textrm{It was represented in the form }%
\begin{equation}
\Gamma (\Omega )=-\frac{4\pi }{nc}\frac{e^{2}}{m}\frac{\mathrm{Im}\Sigma
(\Omega )}{\left[ \Omega -\mathrm{Re}\Sigma (\Omega )\right] ^{2}+\left[
\mathrm{Im}\Sigma (\Omega )\right] ^{2}}\ .  \label{eq:P24-2}
\end{equation}%
\textrm{This general expression} was the starting point for a derivation of
the theoretical optical absorption spectrum of a single Fr\"{o}hlich polaron
at \textit{all electron-phonon coupling strengths} by \cite{DSG72}. $\Sigma
(\Omega )$ is the so-called ``memory
function'', which contains the dynamics of the polaron and
depends on $\Omega $, $\alpha $, temperature {and applied external fields}.
The key contribution of \cite{DSG72} was to introduce $\Gamma (\Omega )$ in
the form (\ref{eq:P24-2}) and to calculate $\mathrm{Re}\Sigma (\Omega )$,
which is essentially a (technically not trivial) Kramers--Kronig transform
of the more simple function $\mathrm{Im}\Sigma (\Omega )$. {Only} the
function $\mathrm{Im}\Sigma (\Omega )$ had been derived for the Feynman
polaron \cite{FHIP} {to} {study} the polaron mobility $\mu $ using the
impedance function {(\ref{F1a}):} $\mu ^{-1}=\lim\limits_{\Omega \rightarrow
0}\left( \mathrm{Im}\Sigma (\Omega )/\Omega \right) $.

The {basic} nature of the Fr\"ohlich polaron excitations was clearly
revealed through this polaron optical absorption given by Eq. (\ref{eq:P24-2}%
). It was demonstrated \cite{DSG72} that the FC states for Fr\"ohlich
polarons are nothing else but a superposition of phonon sidebands {and} a
relatively large value of the electron-phonon coupling strength ($\alpha >
5.9$) is needed to stabilise the relaxed excited state of the polaron. It
was, further, revealed that at weaker coupling only ``scattering states'' of the polaron play a significant role
in the optical absorption \cite{DSG72,DDG71}.

In the weak coupling limit, the absorption spectrum (\ref{eq:P24-2}) of the
polaron is determined by the absorption of radiation energy, which is
re-emitted in the form of LO phonons. As $\alpha$ increases between
approximately 3 and 6, a resonance with increasing stability appears in the
optical absorption of the Fr\"ohlich polaron of Ref. \cite{DSG72} (see Fig. %
\ref{fig_3}). The RES peak in the optical absorption spectrum also has a
phonon sideband-structure, whose average transition frequency can be related
to a FC-type transition. Furthermore, at zero temperature, the optical
absorption spectrum of one polaron exhibits also a zero-frequency
``central peak''\ [$\propto\delta(\Omega)$].
For nonzero temperature, this ``central
peak''\ smears out and gives rise to an ``anomalous''\ Drude-type low-frequency component of the
optical absorption spectrum.
\begin{figure}[b]
\begin{center}
\includegraphics[width=0.9\textwidth]{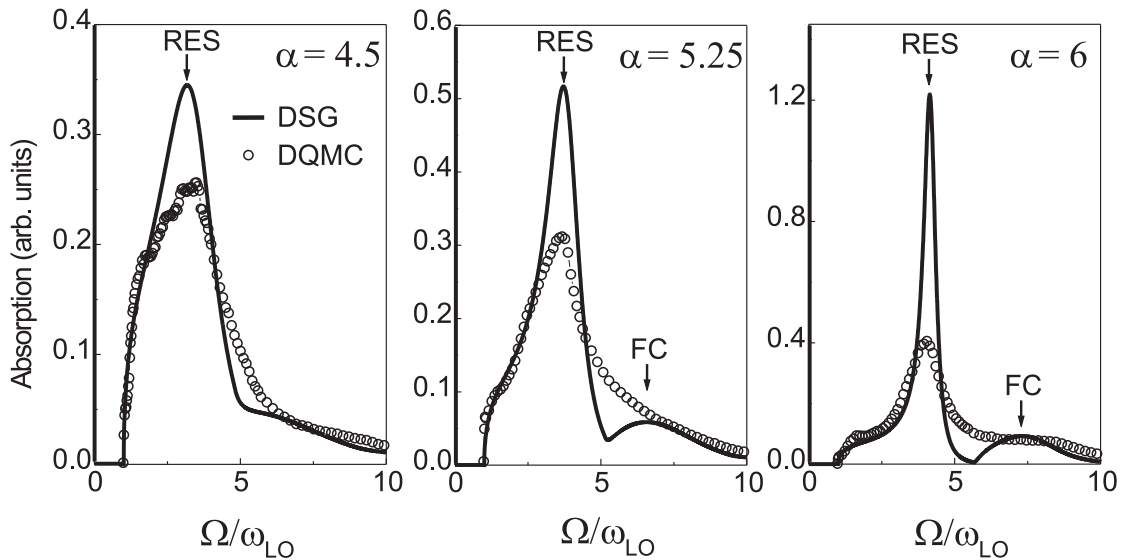}
\end{center}
\caption{Optical absorption spectrum of a Fr\"ohlich polaron for {$\protect%
\alpha=4.5, \protect\alpha=5.25$ and $\protect\alpha = 6$} {after Ref.
\protect\cite{DSG72} (DSG)}. The RES peak is very intense compared with the
FC peak. The $\protect\delta$-like central peaks (at $\Omega=0$) are
schematically shown by vertical lines. {The DQMC results of Ref.
\protect\cite{Mishchenko2003} are shown with open circles.} }
\label{fig_3}
\end{figure}

For $\alpha > 6.5$ the polaron optical absorption gradually develops the
structure \textit{qualitatively} proposed in \cite{KED69}: a broad LO-phonon
sideband structure appears with the zero-phonon (``RES'') transition as
onset. Ref. \cite{DSG72} does not predict the \textit{broad} LO-phonon
sidebands at large coupling constant, although it still gives an accurate
first Stieltjes moment of the optical absorption spectrum.

Based on \cite{DSG72}, it was argued that it is rather Holstein polarons
that determine the optical properties of the charge carriers in oxides like
SrTiO$_{3}$, BaTiO$_{3}$ \cite{HD75}, while {Fr\"ohlich} weak-coupling
polarons could be identified e.~g. in CdO \cite{DHL1971}. The Fr\"{o}hlich
coupling constants of polar semiconductors and ionic crystals are generally
too small to allow for a static ``RES''. In
Ref. \cite{Eagles1995} the RES-peaks of Ref. \cite{DSG72} were invoked to
explain the optical absorption spectrum of Pr$_{2}$NiO$_{4.22}$. The
RES-like resonances in $\Gamma (\Omega )$, Eq. (\ref{eq:P24-2}), due to the
zero's of $\Omega -\mathrm{Re}\Sigma (\Omega )$, can effectively be
displaced to smaller polaron coupling by applying an external magnetic field
$B$, in which case the contribution for what is formally a ``RES-type resonance''\ {arises at} $\Omega -\omega _{c}-%
\mathrm{Re}\Sigma (\Omega )=0$ ($\omega _{c}=eB/m$ is the cyclotron
frequency). Resonances in the magnetoabsorption governed by this
contribution have been {clearly} observed and analysed {in many solids and
structures}, see {e.~g. \cite%
{PD86,Hodby1987,SPD93,devreese:1996,miu97,Devreese03} and references therein.%
}

Evidence for the polaron character of charge carriers in AgBr and AgCl was
obtained through high-precision cyclotron resonance experiments in external
magnetic fields up to 16~T. The all-coupling magneto-absorption calculated
in \cite{PD86} leads to the best quantitative agreement between theory and
experiment for AgBr and AgCl. This quantitative interpretation of the
cyclotron resonance experiment in AgBr and AgCl \cite{Hodby1987} by the
theory of \cite{PD86} provided one of the most convincing and clearest
demonstrations of {Fr\"ohlich} polaron features in solids. The energy
spectra of polaronic systems {such} as shallow donors (``bound polarons''),
e. g., the D$_0$ and D$^-$ centres, constitute the most complete and
detailed polaron spectroscopy realised in the literature \cite{SPD93}.

The numerical calculations of the optical conductivity for the Fr\"{o}hlich
polaron performed within the DQMC method \cite{Mishchenko2003} confirm the
analytical results derived in Ref. \cite{DSG72} for $\alpha\lesssim 3.$ In
the intermediate coupling regime $3<\alpha<6,$ the low-energy behaviour and
the position of the RES-peak in the optical conductivity spectrum of \cite%
{Mishchenko2003} follow closely the prediction of Ref. \cite{DSG72}. There
are some minor quantitative differences between the two approaches in the
intermediate coupling regime: the dominant (``RES'') peak is less intense in the Monte-Carlo numerical
simulations and the second (``FC'') peak
develops less prominently. The following qualitative differences exist
between the two approaches: in \cite{Mishchenko2003}, the dominant peak
broadens {for $\alpha \gtrsim 6$} and the second peak does not develop, {but
gives} rise to a flat shoulder in the optical conductivity spectrum at {$%
\alpha \approx 6$}. As $\alpha$ increases beyond $\alpha \approx 6$, the DSG
results for the OC do not produce the \textit{broad} phonon sideband
spectrum of the RES-transition that was \textit{qualitatively} predicted in
Ref. \cite{KED69} and obtained with DQMC.

An instructive comparison between the positions of the main peak in the
optical absorption spectra of {Fr\"ohlich} polarons obtained within the DSG
and DQMC approaches has been performed {recently} \cite{DK2006}. The
main-peak positions, obtained within DSG, have been found in good agreement
with the results of DQMC for all considered values of $\alpha$. At large $%
\alpha$ the positions of the main peak in the DSG spectra are remarkably
close to those given by DQMC. The difference between the DSG and DQMC
results is relatively larger at $\alpha$~=~8 and for $\alpha$~=~9.5, but
even for those values of the coupling constant the agreement is {rather}
good.

It is suggested that the RES-peak at $\alpha\approx 6$ in the DSG-treatment,
as $\alpha$ increases, gradually transforms onto a FC-peak. As stated above
and in Ref. \cite{DSG72}, DSG predicts a much too narrow FC-peak in the
strong coupling limit, but still at the ``correct'' frequency. The DSG
spectrum also satisfies the \textit{zero} and \textit{first} moment sum
rules at all $\alpha$ as discussed below.

\subsection{Main-peak line and strong-coupling expansion}

In order to describe the OC main peak line width at intermediate
electron-phonon coupling, the DSG approach was modified \cite{PRL2006} to
include additional dissipation processes, the strength of which is fixed by
an exact sum rule \cite{catbook}.

Within the memory function formalism (MFF) \cite{Mori,gotze} the interaction
of the charge carriers with the free phonon oscillations is expressed in
terms of the electron density-density correlation function,
\begin{equation}
\chi (\mathbf{q},t)=-i\Theta (t)\left\langle \exp \left[ i\mathbf{q}\cdot
\mathbf{r}(t)\right] \exp \left[ -i\mathbf{q}\cdot \mathbf{r}(0)\right]
\right\rangle ,
\end{equation}%
which is evaluated in a direct way \cite{PD1983} using the {Feynman polaron}
model, where the electron is coupled via a harmonic force to a fictitious
particle that simulates the phonon degrees of freedom. Within this procedure
the electron density-density correlation function takes the form:
\begin{equation}
\chi _{m}(\mathbf{q},t)=-i\Theta (t)\exp \left[ -iq^{2}t/2M\right] \exp %
\left[ -q^{2}R(1-e^{-ivt})/2M\right] ,
\end{equation}%
where $R=(M-1)/v$, $M$ (the total mass of electron and fictitious particle),
and $v$ are determined variationally within the path integral approach \cite%
{Feynman}. The associated spectral function $A_{m}(\mathbf{q},\omega )=-2%
\mathrm{Im}\chi _{m}(\mathbf{q},\omega )$ is a series of $\delta $ functions
centered at $q^{2}/2M+nv$ ($n$ is integer). Here $q^{2}/2M$ represents the
energy of the center of mass of electron and fictitious particle, and $v$ is
the energy gap between the levels of the relative motion. To include
dissipation \cite{PRL2006}, a finite lifetime {was introduced} for the
states of the relative motion, which can be considered as the result of the
residual EPI not included into the {Feynman variational} model. To this end,
in $\chi _{m}(\mathbf{q},t)$ the factor $\exp \left[ -ivt\right] $ was
replaced with $(1+it/\tau )^{-v\tau }$ which leads to the replacement of $%
\delta $ functions by Gamma functions with mean value and variance given
respectively by $q^{2}/2M+nv$ and $nv/\tau $. The parameter of dissipation $%
\tau $ is determined by the third sum rule for $A(\mathbf{q},\omega )$,
which is additional to the first two sum rules that are already satisfied in
the DSG model without damping. As expected, $\tau $ turns out to be of the
order of $\omega _{0}^{-1}$. If broadening of the oscillator levels is
neglected, the DSG results \cite{DSG72,PD1983} are recovered.

Starting from the Kubo formula, the strong-coupling polaron optical
conductivity can be evaluated using the strong-coupling expansion
(SCE), Refs. {\cite{PRL2006,catbook}}. In Refs.
\cite{DK2006,Devreese07a,DevreeseL} SCE has been extended. In order
to apply the extended SCE for the polaron OC, a scaling
transformation of the coordinates and moments of the electron-phonon
system is made following Allcock \cite{Allcock1}, $\mathbf{r} = \alpha^{-1}%
\mathbf{x}$, $\mathbf{p} = -i\alpha \partial/\partial\mathbf{x}$, and $%
\mathbf{q} = \alpha \mathbf{\tilde {q}}$. This transformation allows us to
see explicitly the order of magnitude of different terms in the Hamiltonian.
Expressed in terms of the new variables, the Hamiltonian can be written as a
sum of two terms, which are of different orders in powers of $\alpha$, $%
H=H_{1}+H_{2}$, where $H_{1}\sim\alpha^{2}$ is the leading term, and $%
H_{2}\sim\alpha^{-2}$ is the kinetic energy of the vibrating ions. The next
step is the Born-Oppenheimer approximation \cite{Allcock1}, which neglects
the non-adiabatic transitions between different polaron levels in
calculating the dipole-dipole correlation function of the Kubo formula \cite%
{DK2006,DevreeseL}.

Figure \ref{fig_SCE} {shows} the polaron OC spectra for {different} values
of $\alpha $ calculated numerically using {the extended SCE with} different
approximations. The OC spectra calculated within the {extended} SCE
approach, taking into account both the Jahn-Teller effect and the
corrections of order $\alpha ^{0}$, are shown by the solid curves. The OC
obtained {with} the leading-term strong-coupling approximation taking into
account the Jahn-Teller effect and {with} the leading term of the
Landau-Pekar adiabatic approximation are plotted {as} dashed and dash-dotted
curves, respectively. The full circles show the DQMC data \cite%
{Mishchenko2003,PRL2006}.
\begin{figure}[t]
\begin{center}
\newpage \includegraphics[width=0.8\textwidth]{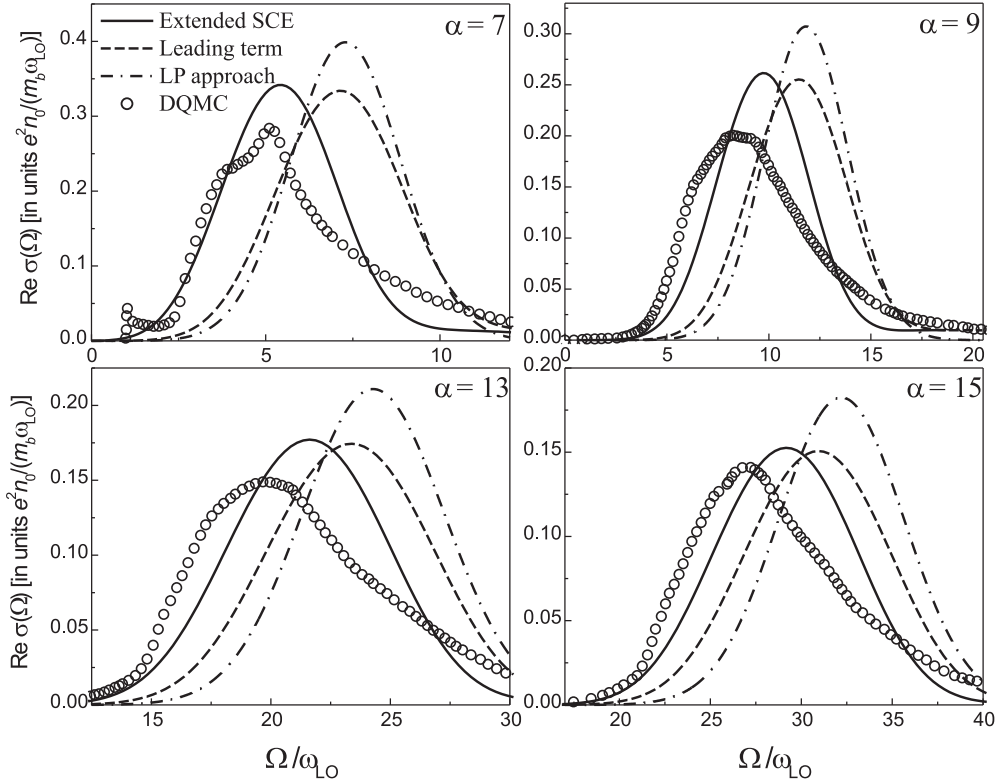}
\end{center}
\caption{The polaron OC calculated within {the extended} {SCE} taking into
account corrections of order $\protect\alpha ^{0}$ (solid curve), the OC
calculated within the leading-term strong-coupling approximation {(dashed
curve)}, {with} the leading term of the Landau-Pekar (LP) adiabatic
approximation (dash-dotted curve), and the numerical DQMC data (open
circles) for $\protect\alpha =7,9,13$ and $15$. (From Ref. \protect\cite%
{DK2006}.) }
\label{fig_SCE}
\end{figure}

{The polaron OC band of Fig. \ref{fig_SCE} obtained within the extended SCE
generalises the Gaussian-like polaron OC band (as given e.g. by Eq. (3) of
Ref.\cite{PRL2006}) thanks to (i) the use of the numerically exact
strong-coupling polaron wave functions \cite{Miyake} and (ii) the
incorporation of both static and dynamic Jahn-Teller effects. The polaron OC
broad structure calculated within the extended SCE consists of a series of
LO-phonon sidebands and provides a realisation --- with all LO-phonons
involved for a given $\alpha$ --- of the strong-coupling scheme proposed in
\cite{KED69}.}

As seen from Fig. \ref{fig_SCE}, the polaron OC spectra calculated within
the asymptotically exact strong-coupling approach are shifted {towards}
lower frequencies {as compared with} the OC spectra calculated within the LP
approximation. This shift is due to {the use of} the numerically exact (in
the strong-coupling limit) energy levels and wave functions of the internal {%
excited} polaron states, as well as the numerically exact self-consistent
adiabatic polaron potential. Furthermore, the {inclusion} of the corrections
of order $\alpha ^{0}$ leads to {a} shift of the OC spectra to lower
frequencies with respect to the OC spectra calculated within the
leading-term approximation. The value of this shift $\Delta \Omega
_{n,0}/\omega_{\mathrm{L}O}\approx -1.8$ obtained in the present
calculation, is close to the LP value $\Delta \Omega _{n,0}^{\left(
LP\right)}/\omega_0 =4\ln 2-1\approx 1.7726$ (cf. \cite%
{Allcock1956,FeynmanSM}). The distinction between the OC spectra calculated
with and without the Jahn-Teller effect is very small.

Starting from $\alpha \approx 9$ {towards} larger values of $\alpha $, the
agreement between the {extended} SCE polaron OC spectra and the numerical
DQMC data becomes gradually better, {consistent with the fact that} the {%
extended} SCE for the polaron OC is asymptotically correct in the
strong-coupling limit. The results of the {extended} SCE are qualitatively
consistent with the interpretation advanced in Ref. \cite{KED69}. In \cite%
{KED69} only the 1-LO-phonon sideband was taken into account, while in Ref.
\cite{Goovaerts73} 2-LO-phonon emission was included. The extended SCE
carries on the program started in Ref. \cite{KED69}. The spectra in Fig. \ref%
{fig_SCE}, in the strong coupling approximation, consist of LO-phonon
sidebands to the RES (which itself has negligible oscillator strength in
this limit, similar to the optical absorption for some colour centres in
alkali halides). These LO-phonon sidebands form a broad FC-structure.

\subsection{Comparison between optical conductivity spectra obtained by
different methods}

\label{comparison}

A comparison between the optical conductivity spectra obtained with the DQMC
method, extended MFF, SCE and DSG for different values of $\alpha $ is shown
in Figs. \ref{figa_graph}, \ref{figb_graph}, taken from Ref.\thinspace \cite%
{PRL2006}. The key results of the comparison are the following.

\begin{figure}[b]
\centering
\newpage \includegraphics[width=1.0\textwidth]{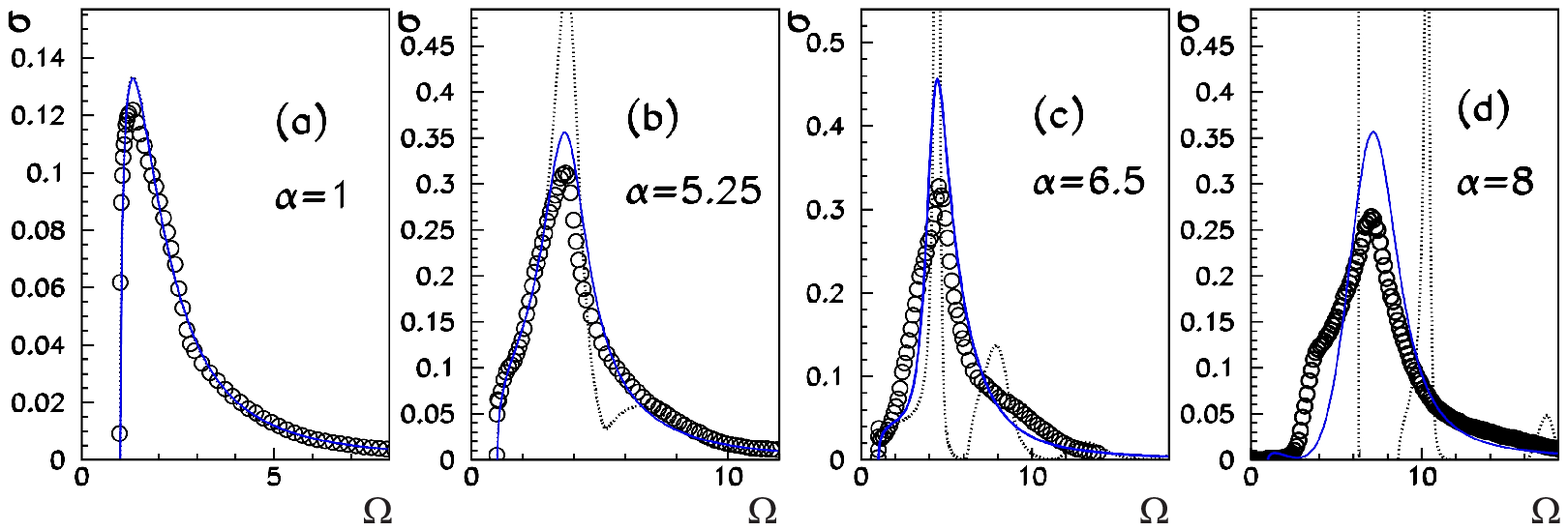}
\caption{{Comparison of the optical conductivity calculated with the DQMC
method (circles), extended MFF (solid line) and DSG \protect\cite%
{DSG72,PD1983} (dotted line), for four different values of $\protect\alpha $%
. The arrow indicates the lower-frequency feature in the DQMC data. ({%
Reprinted with permission from} Ref.\thinspace \protect\cite{PRL2006}. {%
\copyright 2006 by the American Physical Society}.)}}
\label{figa_graph}
\end{figure}

\begin{figure}[b]
\centering \includegraphics[width=1.0\textwidth]{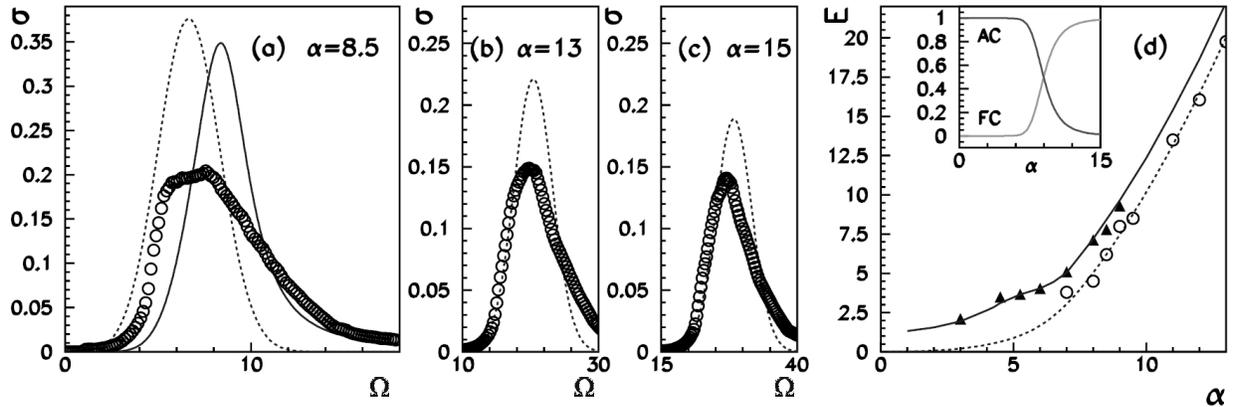}
\caption{{a), b) and c) Comparison of the optical conductivity calculated
with the DQMC method (circles), the extended MFF(solid line) and SCE (dashed
line) for three different values of $\protect\alpha $. d) The energy of the
lower- and higher-frequency features obtained by DQMC (circles and
triangles, respectively) compared (i) with the FC transition energy
calculated from the SCE (dashed line) and (ii) with the energy of the peak
obtained from the extended MFF (solid line). In the inset the weights of
Franck-Condon and adiabatically connected transitions are shown as a
function of $\protect\alpha $. We have used for $\protect\eta $ the value $%
1.3$. ({Reprinted with permission from} Ref.\thinspace\protect\cite{PRL2006}%
. {\copyright 2006 by the American Physical Society}.)}}
\label{figb_graph}
\end{figure}

First, as expected, in the weak-coupling regime, both the extended MFF with
phonon broadening and DSG \cite{DSG72} are in very good agreement with the
DQMC data \cite{Mishchenko2003}, showing significant improvement with
respect to the weak-coupling perturbation approach {\cite{GLF62,DHL1971}}
which provides a good description of the OC spectra only for very small
values of $\alpha$. For {$3\le \alpha \le 6$, DSG predicts the essential
structure of the optical absorption, with a RES-transition gradually
building up for increasing $\alpha$,} {but} underestimates the peak width.
The damping, introduced in the extended MFF approach, becomes crucial {in
this coupling regime}.

Second, comparing the peak and shoulder energies, obtained by DQMC, with the
peak energies, given by MFF, and the FC transition energies from the SCE, it
is concluded \cite{PRL2006} that as $\alpha$ increases from $6$ to $10$ the
spectral weights rapidly switch from the dynamic regime, where the lattice
follows the electron motion, to the adiabatic regime dominated by FC
transitions. In the intermediate electron-phonon coupling regime, $6 <\alpha
<10 $, both adiabatic FC and non-adiabatic dynamical excitations coexist.
For still larger coupling $\alpha \gtrsim 10$, the absorption spectrum
consists of a broad FC-structure, built {of} LO-phonon sidebands.

In summary, the accurate numerical results obtained from DQMC and from the
recent analytical approximations {\cite{PRL2006,DeFilippis2003}} confirm the
essence of the mechanism for the optical absorption of Fr\"{o}hlich
polarons, proposed in Refs. \cite{DSG72,Devreese72} combined with \cite%
{KED69} and do add important new insights.

\subsection{Sum rules for the optical conductivity spectra of {Fr\"ohlich}
polarons}

\label{sum_rules}

In this section {several} sum rules for the optical conductivity spectra of
Fr\"{o}hlich polarons {are applied to test} the DSG approach \cite{DSG72} {%
and} the DQMC results \cite{Mishchenko2003}. The values of the polaron
effective mass for the {DQMC} approach are taken from Ref. \cite%
{Mishchenko2000}. In Tables \ref{tab3} and \ref{tab4}, we show the polaron
ground-state $E_{0}$ and {the zero, $M_0$ and first, $M_1$, frequency moments%
} calculated using the optical conductivity spectra:

\begin{eqnarray}
M_{0} & \equiv\int_{1}^{\Omega_{\max}}\mathrm{Re}\sigma\left( \Omega\right)
d\Omega,  \label{1} \\
M_{1} & \equiv\int_{1}^{\Omega_{\max}}\Omega\mathrm{Re}\sigma\left(
\Omega\right) d\Omega,  \label{2}
\end{eqnarray}
where $\Omega_{\max}$ is the upper value of the frequency available from
Ref. \cite{Mishchenko2003}, and
\begin{equation}
\tilde{M}_{0}\equiv\frac{\pi}{2m^{\ast}}+\int_{1}^{\Omega_{\max}}\mathrm{Re}%
\sigma\left( \Omega\right) d\Omega.  \label{zm}
\end{equation}
Here $m^{\ast}$ is the polaron mass, the optical conductivity is calculated
in units $n_{p}e^{2}/(m\omega_{0}),$ $m^{\ast}$ is measured in units of the
band mass $m$, and the frequency is measured in units of $\omega_{0}$. The
values of $\Omega_{\max}$ are: $\Omega_{\max}=10$ for $\alpha=0.01,$ 1 and
3, $\Omega_{\max}=12$ for $\alpha=4.5,$ 5.25 and 6, $\Omega_{\max}=18$ for $%
\alpha=6.5,$ 7 and 8.

\begin{table}[h]
\caption{Polaron parameters $M_0, M_1, {\tilde M}_0$ obtained from the
diagrammatic Monte Carlo results ({Reprinted with permission from}
\protect\cite{Devreese2005_V}. {\copyright 2006, Societ\`{a} Italiana di
Fisica.})}
\label{tab3}
\newpage
\par
\begin{center}
\begin{tabular}[t]{|l|l|l|l|l|l|}
\hline\hline
$\alpha$ & $M_{0}^{\left( \mathrm{DQMC}\right) }$ & $m^{\ast\left( \mathrm{%
DQMC}\right) }$ & $\tilde{M}_{0}^{\left( \mathrm{DQMC}\right) }$ & $%
M_{1}^{\left( \mathrm{DQMC}\right) }/\alpha$ & $E_{0}^{\left( \mathrm{DQMC}%
\right) }$ \\ \hline
$0.01$ & $0.00249$ & $1.0017$ & $1.5706$ & $0.634$ & $-0.010$ \\
$1$ & $0.24179$ & $1.1865$ & $1.5657$ & $0.65789$ & $-1.013$ \\
$3$ & $0.67743$ & $1.8467$ & $1.5280$ & $0.73123$ & $-3.18$ \\
$4.5$ & $0.97540$ & $2.8742$ & $1.5219$ & $0.862$ & $-4.97$ \\
$5.25$ & $1.0904$ & $3.8148$ & $1.5022$ & $0.90181$ & $-5.68$ \\
$6$ & $1.1994$ & $5.3708$ & $1.4919$ & $0.98248$ & $-6.79$ \\
$6.5$ & $1.30$ & $6.4989$ & $1.5417$ & $1.1356$ & $-7.44$ \\
$7$ & $1.3558$ & $9.7158$ & $1.5175$ & $1.2163$ & $-8.31$ \\
$8$ & $1.4195$ & $19.991$ & $1.4981$ & $1.3774$ & $-9.85$ \\ \hline\hline
\end{tabular}%
\end{center}
\end{table}

\begin{table}[h]
\caption{Polaron parameters $M_0, M_1, {\tilde M}_0$ obtained within the
path-integral approach ({Reprinted with permission from} \protect\cite%
{Devreese2005_V}. {\copyright 2006, Societ\`{a} Italiana di Fisica.})}
\label{tab4}
\newpage
\par
\begin{center}
\begin{tabular}[t]{|l|l|l|l|l|l|}
\hline\hline
$\alpha$ & $M_{0}^{\left( \mathrm{DSG}\right) }$ & $m^{\ast\left( \mathrm{%
Feynman}\right) }$ & $\tilde{M}_{0}^{\left( \mathrm{DSG}\right) }$ & $%
M_{1}^{\left( \mathrm{DSG}\right) }/\alpha$ & $E_{0}^{\left( \mathrm{Feynman}%
\right) }$ \\ \hline
$0.01$ & $0.00248$ & $1.0017$ & $1.5706$ & $0.633$ & $-0.010$ \\
$1$ & $0.24318$ & $1.1957$ & $1.5569$ & $0.65468$ & $-1.0130$ \\
$3$ & $0.69696$ & $1.8912$ & $1.5275$ & $0.71572$ & $-3.1333$ \\
$4.5$ & $1.0162$ & $3.1202$ & $1.5196$ & $0.83184$ & $-4.8394$ \\
$5.25$ & $1.1504$ & $4.3969$ & $1.5077$ & $0.88595$ & $-5.7482$ \\
$6$ & $1.2608$ & $6.8367$ & $1.4906$ & $0.95384$ & $-6.7108$ \\
$6.5$ & $1.3657$ & $9.7449$ & $1.5269$ & $1.1192$ & $-7.3920$ \\
$7$ & $1.4278$ & $14.395$ & $1.5369$ & $1.2170$ & $-8.1127$ \\
$8$ & $1.4741$ & $31.569$ & $1.5239$ & $1.4340$ & $-9.6953$ \\ \hline\hline
\end{tabular}%
\end{center}
\end{table}

The optical conductivity {derived by DSG} \cite{DSG72} exactly satisfies the
sum rule \cite{SR}%
\begin{equation}
\frac{\pi}{2m^{\ast}}+\int_{1}^{\infty}\mathrm{Re}\sigma\left( \Omega\right)
d\Omega=\frac{\pi}{2}.  \label{sr}
\end{equation}
Since the optical conductivity obtained from the {DQMC} results \cite%
{Mishchenko2003} is known only within a limited interval of frequencies $1<
\Omega < \Omega_{\mathrm{max}}$, the integral in Eq. (\ref{zm}) for the
DSG-approach \cite{DSG72} is calculated over the same frequency interval as
for the Monte Carlo results \cite{Mishchenko2003}.

The comparison of the resulting zero frequency moments $\tilde{M}%
_{0}^{\left( \mathrm{DQMC}\right) }$ and $\tilde{M}_{0}^{\left( \mathrm{DSG}%
\right) }$ with each other and with the value $\pi/2=1.5707963...$
corresponding to the right-hand-side of the sum rule (\ref{sr}) shows that
the difference $\left\vert \tilde{M}_{0}^{\left( \mathrm{DQMC}\right) }-%
\tilde {M}_{0}^{\left( \mathrm{DSG}\right) }\right\vert $ on the interval $%
\alpha \leq 8$ is smaller than the absolute value of the contribution of the
``tail''\ of the optical conductivity for $%
\Omega>\Omega_{\max}$ to the integral in the sum rule (\ref{sr}):
\begin{equation}
\int_{\Omega_{\mathrm{max}}}^{\infty}\mathrm{Re}\sigma ^{\left( \mathrm{DSG}%
\right) }\left( \Omega\right) d\Omega \equiv\frac{\pi}{2}-\tilde{M}%
_{0}^{\left( \mathrm{DSG}\right)}.  \label{sr1A}
\end{equation}
Within the accuracy determined by the neglect of the ``tail''\ of the part of the spectrum for $\Omega>\Omega_{%
\max} $, the contribution to the integral in the sum rule (\ref{sr}) for the
optical conductivity obtained from the {DQMC} results \cite{Mishchenko2003}
\textit{agrees well} with that for the optical conductivity found within the
path-integral approach {in} Ref.~\cite{DSG72}. Hence, the conclusion follows
that \textit{the optical conductivity obtained from the {DQMC} results} \cite%
{Mishchenko2003} \textit{satisfies the sum rule} (\ref{sr}) within the
aforementioned accuracy.

\subsection{Optical absorption spectra of continuum-polaron gas}

\label{gas}

For the weak-coupling regime, which is realized in most polar
semiconductors, the ground-state energy of a gas of interacting continuum
polarons has been derived in \cite{LDB1977} by introducing a variational
wave function:
\begin{equation}
\left| \psi_{\text{LDB}}\right\rangle =U\left| \phi\right\rangle \left|
\varphi_{\text{el}}\right\rangle ,  \label{psiLDB}
\end{equation}
where $\left| \varphi_{\text{el}}\right\rangle $ represents the ground-state
many-body wave function for the electron (or hole) system, $\left|
\phi\right\rangle $ is the phonon vacuum and $U$ is a many-body unitary
operator. $U$ defines the LDB-canonical transformation for a fermion gas
interacting with a boson field:
\begin{equation}
U=\exp\left\{\sum_{j=1}^{N}\sum_{\mathbf{q}} \left(f_{\mathbf{q}}d_{\mathbf{q%
}}e^{i\mathbf{q}\cdot\mathbf{r}_{j}} -f_{\mathbf{q}}^{\ast}d_{\mathbf{q}%
}^{+} e^{-i\mathbf{q}\cdot\mathbf{r}_{j}}\right)\right\},  \label{U}
\end{equation}
where $\mathbf{r}_{j}$ represent the position of the $N$ constituent
electrons (or holes). The $f_{\mathbf{q}}$ were determined variationally
\cite{LDB1977}. It may be emphasized that Eq.\,(\ref{U}), although it
appears like a straightforward generalization of the one-particle
transformation in \cite{tomo}, constitutes --- especially in its
implementation --- a nontrivial extension of a one-particle approximation to
a many-body system. An advantage of the LDB-many-polaron canonical
transformations introduced in \cite{LDB1977} for the calculation of the
ground state energy of a polaron gas is that the many-body effects are
contained in the static structure factor of the electron (or hole) system,
which appears in the analytical expression for the energy. Within the
approach the minimum of the total ground-state energy per particle for a
polaron gas lies at lower density than that for the electron gas.

The non-degenerate system of interacting polarons in polar doped insulators
was analysed by Fratini and Qu\'emerais (2000) using a simplified
Feynman-type polaron model. In the low-density limit, the ground state of
the many-polaron system is the Wigner lattice of polarons. With increasing
density depending on the value of $\alpha$, one of the following two
scenarios is possible: (i) the melting of the polaron Wigner lattice {for $%
\alpha < \alpha^*$} and {(ii)} the dissociation of the polarons {for $\alpha
> \alpha^*$} \cite{Fratini2000}.

The LDB-canonical transformation has been fruitfully applied in the theory
of optical absorption spectra of many-polaron systems. In \cite{TDPRB01},
starting from the LDB-many-polaron canonical transformation and the
variational many-polaron wave function introduced in \cite{LDB1977}, the
optical absorption coefficient of a many-polaron gas has been derived. The
real part of the optical conductivity of the many-polaron system is obtained
in an intuitively appealing form, given by Eq. (\ref{opticabsTD}).

This approach to the many-polaron optical absorption allows one to include
the many-body effects {to order $\alpha$} in terms of the dynamical
structure factor $S(\mathbf{k},\Omega -\omega_{\mathrm{L}O})$ of the
electron (or hole) system. The experimental peaks in the mid-infrared
optical absorption spectra of cuprates \cite{Lupi1999}, and manganites \cite%
{Hartinger2004} have been adequately interpreted within this theory. The
many-polaron approach describes the experimental optical conductivity better
than the single-polaron approximations \cite{GLF62,emin1993}. Note that in
\cite{TDPRB01}, like in \cite{Eagles1995}, coexistence of small and {%
Fr\"ohlich} polarons in the same solid is involved.

The optical conductivity of a many-polaron gas {was} investigated in \cite%
{catau2} in a different way by calculating the correction to the dielectric
function of the electron gas, due to the electron-phonon interaction with
variational parameters of a single-polaron Feynman model. A suppression of
the optical absorption of a many-polaron gas as compared to the one-polaron
optical absorption of Refs.\,\cite{DSG72,Devreese72} with increasing density
has been found. Such a suppression {is} expected because of the screening of
the Fr\"{o}hlich interaction with increasing polaron density.

\subsection{Ripplopolarons}

An interesting 2D system consists of electrons on films of liquid He \cite%
{SM73,9}. In this system the electrons couple to the ripplons of the liquid
He, forming ``ripplopolarons''. The effective coupling can be relatively {%
large and} self-trapping can result. The acoustic nature of the ripplon
dispersion at long wavelengths induces the self-trapping. Spherical shells
of charged particles appear in a variety of physical systems, such as
fullerenes, metallic nanoshells, charged droplets and neutron stars. A
particularly interesting physical realization of the spherical electron gas
is found in multielectron bubbles (MEBs) in liquid helium-4. These MEBs are
0.1 $\mu$m -- 100 $\mu$m sized cavities inside liquid helium, that contain
helium vapour at vapour pressure and a nanometer-thick electron layer,
anchored to the surface of the bubble \cite{VolodinJETP26,Albrecht1987}.
They exist as a result of equilibrium between the surface tension of liquid
helium and the Coulomb repulsion of the electrons \cite%
{ShikinJETP27,Salomaa1981}.

Recently proposed experimental schemes to stabilize MEBs
\cite{SilveraBAPS46} have stimulated theoretical investigation of
their properties (see e.~g. \cite{TemperePRL87}). The dynamical
modes of MEB were described by considering the motion of the helium
surface (``ripplons'') and the vibrational modes of the electrons
together. In particular, the case when the ripplopolarons form a
Wigner lattice was analyzed. The interaction energy between the
ripplons and the electrons in the multielectron bubble is derived
from the following considerations: (i) the distance between the
layer electrons and the helium surface is fixed (the electrons find
themselves confined to an effectively 2D surface anchored to the
helium surface) and (ii) the electrons are subjected to a force
field, arising from the electric field of the other electrons. To
study the ripplopolaron Wigner lattice at {nonzero} temperature and
for any value of the electron-ripplon coupling, the variational
path-integral approach \cite{Feynman} has been used. The destruction
of the ripplopolaron Wigner lattice in a MEB occurs through the
dissociation of ripplopolarons.  Below a critical pressure (on the
order of 10$^{4}$ Pa) the ripplopolaron solid will melt into an
electron liquid. This critical pressure is nearly independent of the
number of electrons (except for the smallest bubbles) and is weakly
temperature dependent, up to the helium critical temperature 5.2 K.
This can be understood since the typical lattice potential well in
which the ripplopolaron resides has frequencies of the order of THz
or larger, which correspond to $\sim10$ K.

The new phase that was predicted {in} \cite{TempereEPJ2003}, the
ripplopolaron Wigner lattice, will not be present for electrons on a flat
helium surface. At the values of the pressing field necessary to obtain a
strong enough electron-ripplon coupling, the flat helium surface is no
longer stable against long-wavelength deformations \cite{GorkovJETP18}.
Multi-electron bubbles, with their different ripplon dispersion and the
presence of stabilizing factors such as the energy barrier against
fissioning \cite{TemperePRB67}, allow for much larger electric fields
pressing the electrons against the helium surface. The regime of $N$, $p$, $%
T $ parameters suitable for the creation of a ripplopolaron Wigner lattice
lies within the regime that would be achievable in recently proposed
experiments, aimed at stabilizing multielectron bubbles \cite{SilveraBAPS46}%
. The ripplopolaron Wigner lattice and its melting transition might be
detected by spectroscopic techniques \cite{GrimesPRL42,FisherPRL42} probing
for example the transverse phonon modes of the lattice \cite{DevillePRL53}.

\subsection{Polaron scaling relations}

\label{scaling}

The form of the Fr\"{o}hlich Hamiltonian, Eq.(\ref{frohlichH}), in $n$
dimensions is the same as in 3D, except that now all vectors are $n$%
-dimensional. In this subsection we take $m=\omega_{0}=1$. In 3D the EPI
matrix element is well known, $\left\vert V_{\mathbf{q}}\right\vert ^{2}=2%
\sqrt{2}\pi\alpha/V_{3}q^{2}.$ The interaction coefficient in $n$ dimensions
becomes \cite{PRB33-3926}
\begin{equation}
\left\vert V_{\mathbf{q}}\right\vert ^{2}=\frac{2^{n-3/2}\pi^{\left(
n-1\right) /2}\Gamma\left( \frac{n-1}{2}\right) \alpha}{V_{n}q^{n-1}}
\label{srv}
\end{equation}
with $V_n$ the volume of the $n$-dimensional crystal.

The only difference {between} the model system in $n$ dimensions {and} the
model system in 3D is that now one deals with an $n$-dimensional harmonic
oscillator. Directly following \cite{Feynman}, the variational polaron
energy {was calculated in Ref. \cite{PRB33-3926}}
\begin{eqnarray}
E & =\frac{n\left( v-w\right) }{2}-\frac{n\left( v^{2}-w^{2}\right) }{4v}-%
\frac{2^{-3/2}\Gamma\left( \frac{n-1}{2}\right) \alpha}{\Gamma\left( \frac{n%
}{2}\right) }\int_{0}^{\infty}\frac{e^{-t}}{\sqrt{D_{0}\left( t\right) }}dt
\nonumber \\
& =\frac{n\left( v-w\right) ^{2}}{4v}-\frac{\Gamma\left( \frac{n-1}{2}%
\right) \alpha}{2\sqrt{2}\Gamma\left( \frac{n}{2}\right) }\int _{0}^{\infty}%
\frac{e^{-t}}{\sqrt{D_{0}\left( t\right) }}dt,  \label{ssr5}
\end{eqnarray}
where%
\begin{equation}
D_{0}\left( t\right) =\frac{w^{2}}{2v^{2}}t+\frac{v^{2}-w^{2}}{2v^{3}}\left(
1-e^{-vt}\right) .  \label{sr18}
\end{equation}
In order to {facilitate} a comparison of $E$ for $n$ dimensions with the
Feynman result \cite{Feynman} for 3D,%
\begin{equation}
E_{3\mathrm{D}}\left( \alpha\right) =\frac{3\left( v-w\right) ^{2}}{4v}-%
\frac{1}{\sqrt{2\pi}}\alpha\int_{0}^{\infty}\frac{e^{-t}}{\sqrt {D_{0}\left(
t\right) }}dt,  \label{ssr6}
\end{equation}
it is convenient to rewrite Eq. (\ref{ssr5}) in the form%
\begin{equation}
E_{n\mathrm{D}}\left( \alpha\right) =\frac{n}{3}\left[ \frac{3\left(
v-w\right) ^{2}}{4v}-\frac{1}{\sqrt{2\pi}}\frac{3\sqrt{\pi}\Gamma\left(
\frac{n-1}{2}\right) }{2n\Gamma\left( \frac{n}{2}\right) }\alpha\int
_{0}^{\infty}\frac{e^{-t}}{\sqrt{D_{0}\left( t\right) }}dt\right] .
\label{ssr7}
\end{equation}
The parameters $w$ and $v$ must be determined by minimizing $E$. In the case
of Eq. (\ref{ssr7}) one {should} minimise the expression in the square
brackets. The only difference of this expression from the rhs of Eq. (\ref%
{ssr6}) is that $\alpha$ is multiplied by the factor
\begin{equation}
a_{n}=\frac{3\sqrt{\pi}\Gamma\left( \frac{n-1}{2}\right) }{2n\Gamma\left(
\frac{n}{2}\right) }.  \label{sran}
\end{equation}
This means that the minimizing parameters $w$ and $v$ in $n$D at a given $%
\alpha$ will be exactly the same as those calculated in 3D {with} the Fr\"{o}%
hlich constant {chosen} as $a_{n}\alpha$:
\begin{equation}
v_{n\mathrm{D}}\left( \alpha\right) =v_{3\mathrm{D}}\left(
a_{n}\alpha\right) ,\mathrm{\ }w_{n\mathrm{D}}\left( \alpha\right) =w_{3%
\mathrm{D}}\left( a_{n}\alpha\right) .  \label{srvw}
\end{equation}
Comparing Eq. (\ref{ssr7}) to Eq. (\ref{ssr6}), the {following} scaling
relation \cite{PRB33-3926, prb31-3420, prb36-4442} {is obtained:}
\begin{equation}
E_{n\mathrm{D}}\left( \alpha\right) =\frac{n}{3}E_{3\mathrm{D}}\left(
a_{n}\alpha\right) ,  \label{ssr8}
\end{equation}
where $a_{n}$ is given by Eq. (\ref{sran}). As discussed in Ref. \cite%
{PRB33-3926}, the above scaling relation is not an \textit{exact} relation.
It is valid for the Feynman polaron energy and also for the ground-state
energy to order $\alpha$. The next-order term (i.e., $\alpha^{2}$) {e.~g.}
no longer satisfies Eq. (\ref{ssr8}). The reason is that in the exact
calculation (to order $\alpha^{2}$) the electron {motions} in different
space directions {are} coupled by EPI. No such coupling appears in the
Feynman polaron model; this is the underlying reason for the validity of the
scaling relation for the Feynman approximation.

In Refs. \cite{PD1983,SR,PRB33-3926,prb36-4442}, scaling relations {were}
obtained also for the impedance function, $Z_{n\mathrm{D}}\left(
\alpha;\Omega\right) =Z_{3\mathrm{D}}\left( a_{n}\alpha;\Omega\right)$, the
effective mass and the mobility of a polaron. In the important particular
case of 2D, the scaling relations take the form \cite%
{PRB33-3926,prb31-3420,prb36-4442}:%
\begin{eqnarray}
E_{2\mathrm{D}}\left( \alpha\right) &=&\frac{2}{3}E_{3\mathrm{D}}\left(
\frac{3\pi}{4}\alpha\right) ,  \label{PDScalingRelation} \\
Z_{2\mathrm{D}}\left( \alpha;\Omega\right) &=&Z_{3\mathrm{D}}\left( \frac{%
3\pi }{4}\alpha;\Omega\right) , \\
\frac{m_{2\mathrm{D}}^{\ast}\left( \alpha\right) }{\left( m\right) _{n%
\mathrm{D}}}&=&\frac{m_{3\mathrm{D}}^{\ast}\left( \frac{3\pi}{4}%
\alpha\right) }{\left( m\right) _{3\mathrm{D}}}, \\
\mu_{2\mathrm{D}}\left( \alpha\right) &=&\mu_{3\mathrm{D}}\left( \frac{3\pi}{%
4}\alpha\right) .
\end{eqnarray}

The scaling relations \cite{prb36-4442} can be checked for the path integral
Monte Carlo results \cite{TPC} for the polaron free energy given in 3D and
in 2D for a few values of temperature and for some selected values of $%
\alpha.$ They follow \emph{very closely} the scaling relation of the form
given by Eq.(\ref{PDScalingRelation})\cite{Devreese2005_V}.

\section{Discrete Holstein and Fr\"ohlich polaron}

When the coupling with phonons increases the polaron radius decreases and
becomes of the order of the lattice constant. Then all momenta of the
Brillouin zone contribute to the polaron wave function and the effective
mass approximation cannot be applied. This regime occurs if the
characteristic potential energy $E_{p}$ due to the local lattice deformation
is compared or larger than the half-bandwidth $D$. The strong coupling
regime with the dimensionless coupling constant
\begin{equation}
\lambda \equiv {\frac{E_{p}}{{D}}}\geq 1
\end{equation}%
is called the small or discrete lattice polaron. In general, $E_{p}$ is
expressed as
\begin{equation}
E_{p}={\frac{1}{{2N}}}\sum_{\mathbf{q}}|\gamma (\mathbf{q})|^{2}\omega _{%
\mathbf{q}}
\end{equation}%
for any type of phonons involved in the polaron cloud. For the Fr\"{o}hlich
interaction with optical phonons one obtains $E_{p}\simeq q_{d}e^{2}/\pi
\kappa ,$ where $q_{d}$ is the Debye momentum \cite{alexandrov:1996}. For
example, with parameters appropriate for copper oxides $\varepsilon _{0}\gg
\varepsilon \simeq 5$ and $q_{D}\simeq 0.7\mathring{A}^{-1}$ one obtains $%
E_{p}\simeq 0.6$ eV \cite{eagles1966,alebra2000}. The exact value of $%
\lambda _{c}$ when the continuum (large) polaron transforms into the small
one, depends on the lattice structure, phonon frequency dispersions and the
radius of the electron-phonon interaction, but in most cases the
transformation occurs around $\lambda _{c}\simeq 1$ \cite{alexandrov:2001}.

\subsection{Holstein model}

\label{two-site} For comparison we briefly introduce a small polaron created
by a short-range EPI, which is known as the Holstein polaron. Its main
features are revealed in the simple Holstein model \cite{holb} of two
vibrating molecules and the electron hopping between them. A simplified
version of the model is defined by a two-site Hamiltonian describing the
electron tunnelling between sites $1$ (``left'') and $2$ (``right'')
with the amplitude $t$ and interacting with a vibrational mode of an ion,
placed at some distance in between, Fig.\ref{holstein}:
\begin{equation}
H=t(c_{1}^{\dagger }c_{2}+c_{2}^{\dagger }c_{1})+H_{ph}+H_{e-ph},
\end{equation}%
Here we take the position of an atomic level in the rigid lattice as zero,
and $c_{i}$ annihilates the electron on the left, $i=1$, or on the right, $%
i=2$, site.
\begin{figure}[tbp]
\begin{center}
\includegraphics[angle=-90, width=0.4\textwidth]{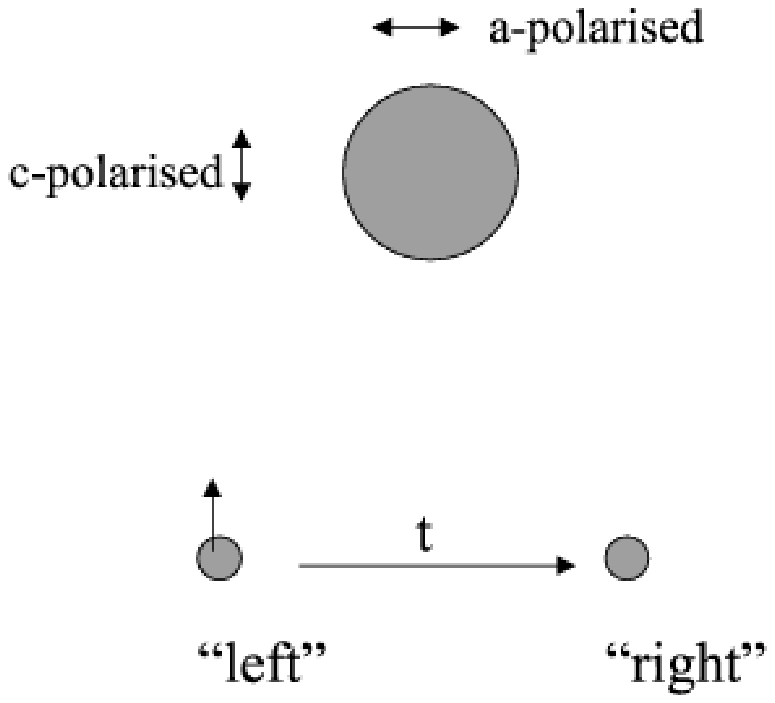} \vskip -0.5mm
\end{center}
\caption{ Electron tunnels between sites $1$ (``left'') and $2$ (``right'') with
the amplitude $t$ and interacts with c-axis or a-axis polarised vibrational
modes of the ion, placed in between.}
\label{holstein}
\end{figure}

The vibration part of the Hamiltonian in this toy model is
\begin{equation}
H_{ph}=-{\frac{1}{{2M}}}{\frac{\partial ^{2}}{{\partial x^{2}}}}+{\frac{%
kx^{2}}{{2}}},
\end{equation}%
where $M$ is the ion mass, $k=M\omega _{0}^{2}$ is the spring constant, and $%
x$ is the ion displacement. The electron-phonon interaction, $H_{e-ph}$,
depends on the polarization of vibrations. If the ion vibrates along the
perpendicular direction to the hopping (in ``c'' -direction, Fig.\ref{holstein}) we have
\begin{equation}
H_{e-ph}=f_{c}x(c_{1}^{\dagger }c_{1}+c_{2}^{\dagger }c_{2}),
\end{equation}%
and
\begin{equation}
H_{e-ph}=f_{a}x(c_{1}^{\dagger }c_{1}-c_{2}^{\dagger }c_{2}),
\end{equation}%
if the ion vibrates along the hopping (``a''\ direction).

The wave-function of the electron and the ion is a linear superposition of
two terms describing the electron on the ``left''\ and on the ``right''\ site, respectively,
\begin{equation}
\psi =[u(x)c_{1}^{\dagger }+v(x)c_{2}^{\dagger }]|0\rangle ,
\end{equation}%
where $|0\rangle $ is the vacuum state describing a rigid lattice without
the extra electron. Substituting $\psi $ into the Schr\"{o}dinger equation, $%
H\psi =E\psi $, we obtain two coupled equations for the amplitudes,
\begin{equation}
(E-f_{a,c}x-H_{ph})u(x)=tv(x),  \label{hol1}
\end{equation}%
\begin{equation}
(E\pm f_{a,c}x-H_{ph})v(x)=tu(x),  \label{hol2}
\end{equation}%
where $+$ and $-$ in the second equation correspond to ``a''\ and ``c''\ polarised
vibrations, respectively. There is the exact solution for the
``c'' -axis polarization, when a change in
the ion position leads to the same shift of the electron energy on the left
and on the right sites,
\begin{equation}
u(x)=u\chi _{n}(x),  \nonumber
\end{equation}%
\begin{equation}
v(x)=v\chi _{n}(x),
\end{equation}%
where $u$ and $v$ are constants and
\begin{equation}
\chi _{n}(x)=\left( \frac{M\omega _{0}}{\pi (2^{n}n!)^{2}}\right)
^{1/4}H_{n}[(x-f_{c}/k)(M\omega _{0})^{1/2}]\exp [-M(x-f_{c}/k)^{2}\omega
_{0}/2],
\end{equation}%
is the harmonic oscillator wave-function. There are two ladders of levels
given by
\begin{equation}
E_{n}^{\pm }=-E_{p}\pm t+\omega _{0}(n+1/2)
\end{equation}%
with $E_{p}=f_{c}^{2}/2k$. Here $H_{n}(\xi )$ are the Hermite polynomials,
and $n=0,1,2,3,...$ . Hence the c-axis single-ion deformation leads to the
polaron level shift but without any renormalisation of the hopping integral $%
t$. In contrast, $a$-polarized vibrations with the opposite shift of the
electron energy on the left and on the right sites, strongly renormalise the
hopping integral. There is no simple general solution of the Holstein model
in this case, but one can find it in two limiting cases, \emph{non-adiabatic}%
, when $t\ll \omega _{0}$ and \emph{adiabatic}, when $t\gg \omega _{0}$.

\subsubsection{Non-adiabatic Holstein polaron}

In the non-adiabatic regime the ion vibrations are fast and the electron
hopping is slow. Hence one can apply a perturbation theory in powers of $t$
to solve
\begin{equation}
\left( \matrix{E-f_ax-H_{ph}&-t\cr -t&E+f_ax-H_{ph}}\right)\left[%
\matrix{u(x)\cr v(x)}\right]=0,
\end{equation}
One takes $t=0$ in zero order, and obtains a two-fold \emph{degenerate}
ground state $[u^{l,r}(x),v^{l,r}(x)]$, corresponding to the polaron
localised on the left ($l$) or on the right ($r$) sites,
\begin{eqnarray}
u^{l}(x)=exp\left[-{\frac{M\omega_0}{{2}}}(x+f_a/k)^{2} \right],\cr %
v^{l}(x)=0
\end{eqnarray}
and
\begin{eqnarray}
u^{r}(x)=0, \cr v^{r}(x)=exp\left[-{\frac{M\omega_0}{{2}}}(x-f_a/k)^{2} %
\right]
\end{eqnarray}
with the energy $E_0=-E_p+\omega_0/2$, where $E_p=f_a^2/2k$. The eigenstates
are found as linear superpositions of two unperturbed states,
\begin{equation}
\left[\matrix{u(x)\cr v(x)}\right]=\alpha\left[\matrix{u^l(x)\cr 0}\right]%
+\beta\left[\matrix{0\cr v^{r}(x)}\right].
\end{equation}
Here the coefficients $\alpha$ and $\beta$ are independent of $x$. The
conventional secular equation for $E$ is obtained, multiplying the first row
by $u^{l}(x)$ and the second row by $v^r(x)$, and integrating over the
vibration coordinate, $x$, each of two equations of the system. The result
is
\begin{equation}
\det\left( \matrix{E-E_0&\tilde{t}\cr \tilde{t}&E-E_0} \right)=0
\end{equation}
with the renormalised hopping integral
\begin{equation}
{\frac{\tilde{t}}{{t}}}= {\frac{\int^{\infty}_{-\infty} dx u^{l}(x) v^{r}(x)%
}{{\int^{\infty}_{-\infty} dx |u^{l}(x)|^{2}}}}.
\end{equation}
The corresponding eigenvalues, $E_{\pm}$ are
\begin{equation}
E_{\pm}=\omega_0/2-E_{p} \pm \tilde{t}.
\end{equation}
The hopping integral splits the degenerate level, as in the rigid lattice,
but an effective `bandwidth' $2\tilde{t}$ is significantly reduced compared
with the bare one,
\begin{equation}
\tilde{t}=t \exp (-2E_p/\omega_0).  \label{adi}
\end{equation}
This polaron band narrowing originates in a small overlap integral of two
displaced oscillator wave functions $u^{l}(x)$ and $v^{r}(x)$.

\subsubsection{Adiabatic Holstein polaron}

\label{adhol} In the adiabatic regime, when $t\gg \omega_0$, the electron
tunnelling is fast compared with the ion motion. Hence one can apply the
Born-Oppenheimer adiabatic approximation \cite{born} taking the wave
function in the form
\begin{equation}
\left[ \matrix {u(x)\cr v(x)}\right]=\chi(x) \left[ \matrix {u_a(x)\cr
v_a(x)}\right].
\end{equation}
Here $u_a(x)$ and $v_a(x)$ are the electron wave functions obeying the
Schr\"odinger equation with the \emph{frozen }ion deformation $x$, i.e.
\begin{equation}
\left( \matrix{E_a(x)-f_ax&-t\cr -t&E_a(x)+f_ax}\right)\left[%
\matrix{u_a(x)\cr v_a(x)}\right]=0.
\end{equation}
The lowest energy level is found as
\begin{equation}
E_a(x)=- \sqrt{(f_ax)^2+t^{2}}.
\end{equation}
$E_a(x)$ together with $kx^2/2$ plays the role of a potential energy term in
the equation for the `vibration' wave function, $\chi(x)$,
\begin{equation}
\left[-{\frac{1}{{2M}}}{\frac{\partial^{2}}{{\partial x^{2}}}}+ {\frac{kx^{2}%
}{{2}}}- \sqrt{(f_ax)^2+t^{2}}\right] \chi(x)=E\chi(x).  \label{chi}
\end{equation}
Terms with the first and second derivatives of the electron wave-functions $%
u_a(x)$ and $v_a(x)$ are small compared with the corresponding derivatives
of $\chi(x)$ in the adiabatic approximation, so they are neglected in Eq.(%
\ref{chi}). As a result we arrive at the familiar double-well potential
problem, where the potential energy $U(x)=kx^{2}/2- \sqrt{(f_ax)^2+t^{2}}$
has two symmetric minima, separated by a barrier. Minima are located
approximately at
\begin{equation}
x_{m}=\pm f_a/k
\end{equation}
in the strong-coupling limit, $E_p\gg t$, and the potential energy near the
bottom of each potential well is about
\begin{equation}
U(x)=-E_{p}+{\frac{k(|x|-f_a/k)^2}{{2}}}.
\end{equation}
If the barrier were impenetrable, there would be the ground state energy
level $E_{0}=-E_p+\omega_0/2$, the same for both wells. The underbarrier
tunnelling results in a splitting of this level $2\tilde{t}$, which
corresponds to a polaron bandwidth in the lattice. It can be estimated using
the quasi-classical approximation as
\begin{equation}
\tilde{t}\propto \exp\left[-2\int_{0}^{x_{m}}p(x)dx\right],  \label{turning}
\end{equation}
where $p(x)=\sqrt{2M[U(x)-E_0]}\approx (Mk)^{1/2}|x-f_a/k|$ is the classical
momentum

Estimating the integral one finds the exponential reduction of the
``bandwidth'',
\begin{equation}
\tilde{t}\propto \exp (-2E_{p}/\omega _{0}),
\end{equation}%
which is the same as in the non-adiabatic regime. Holstein found corrections
to this expression up to terms of the order of $1/\lambda ^{2}$, which
allowed him to improve the exponent and estimate the pre-exponential factor
as
\begin{equation}
\tilde{t}\approx \sqrt{\frac{E_{p}\omega _{0}}{{\pi }}}e^{-\tilde{g}^{2}}.
\label{hol}
\end{equation}%
Here $g^{2}=2E_{p}/\omega _{0}$ and
\begin{equation}
\tilde{g}^{2}=g^{2}\left[ 1-{\frac{1}{{4\lambda ^{2}}}}\ln (4\lambda )-{%
\frac{1}{{8\lambda ^{2}}}}\right] .
\end{equation}%
A more accurate expression for $\tilde{t}$ was obtained in \cite{alekabraya}%
,
\begin{equation}
\tilde{t}\approx \sqrt{\frac{E_{p}\omega _{0}}{{\pi }}}\beta ^{5/2}\lambda
^{1-\beta }[2(1+\beta )]^{-\beta }e^{-\tilde{g}^{2}},  \label{akr}
\end{equation}%
where now $\tilde{g}^{2}=g^{2}\{\beta -[\ln (2\lambda (1+\beta ))]/4\lambda
^{2}\}$, and $\lambda =E_{p}/t$. This expression takes into account the
phonon frequency renormalisation, $\beta \equiv \tilde{\omega _{0}}/\omega
_{0}=\sqrt{1-1/4\lambda ^{2}}$, and the anharmonic corrections of the order
of $1/\lambda ^{2}$ to the turning point $x_{m}$ in Eq.(\ref{turning}). The
term in front of the exponent in Eqs. (\ref{hol},\ref{akr}) differs from $t$
of the non-adiabatic case, Eq.(\ref{adi}). It is thus apparent that the
perturbation approach covers only a part of the entire lattice polaron
region, $\lambda \gtrsim 1$. The upper limit of applicability of the
perturbation theory is given by $t<\sqrt{E_{p}\omega _{0}}$. For the
remainder of the region the adiabatic approximation is more appropriate. A
much lower effective mass of the adiabatic small polaron in the intermediate
coupling region compared with that estimated from the perturbation theory
expression, Eq.(\ref{adi}), is revealed in Eq.(\ref{akr}) \cite{alekabraya}.
The double well potential disappears at $\lambda =\lambda _{c}=0.5$, where
the renormalised phonon frequency $\tilde{\omega _{0}}$ is zero.

The Holstein polaron model can be readily generalized for infinite lattices
(section \ref{LFtransform}). Similar models were used, for instance, in
studies of dissipation \cite{caldeira1981,caldeira1991} and effects of
decoherence in open quantum-mechanical systems.

\subsection{Lang-Firsov canonical transformation}

\label{LFtransform} The kinetic energy is smaller than the interaction
energy as long as $\lambda >1.$ Hence a self-consistent approach to the
discrete (or lattice) polaron problem with EPI of any range is possible with
the ``$1/\lambda $''\ expansion technique
\cite{lan,lan63} on infinite lattices with any type of EPI conserving the
electron site-occupation numbers, and any phonon spectrum. The technique
treats the electron kinetic energy as a perturbation, and can be applied for
multi-polaron systems as well \cite{ale0} (see \ref{mobile}). It is based on
the fact, known for a long time, that there is an analytical exact solution
of the $any-number$ polaron problem in the extreme strong-coupling limit, $%
\lambda \rightarrow \infty $. Following Lang and Firsov one applies the
canonical transformation $e^{S}$ to diagonalise the Hamiltonian, Eq.(\ref%
{hamiltonian}). The diagonalisation is \emph{exact}, if $t(\mathbf{m})=0$
(or $\lambda =\infty $):
\begin{equation}
\tilde{H}=e^{S}He^{-S},
\end{equation}%
where
\begin{equation}
S=-\sum_{\mathbf{q},i}\hat{n}_{i}\left[ u_{i}(\mathbf{q})d_{\mathbf{q}}-H.c.%
\right]   \label{LF}
\end{equation}%
is such that $S^{\dagger }=-S.$ The electron and phonon operators are
transformed as $\tilde{c}_{i}=e^{S}c_{i}e^{-S}$ and $\tilde{d}_{\mathbf{q}%
}=e^{S}d_{\mathbf{q}}e^{-S}$. The result is
\begin{equation}
\tilde{c}_{i}=c_{i}\hat{X}_{i},
\end{equation}%
and
\begin{equation}
\tilde{d}_{\mathbf{q}}=d_{\mathbf{q}}-\sum_{i}\hat{n}_{i}u_{i}^{\ast }(%
\mathbf{q}),
\end{equation}%
where $\hat{X}_{i}=\exp \left[ \sum_{\mathbf{q}}u_{i}(\mathbf{q})d_{\mathbf{q%
}}-H.c.\right] $. The Lang-Firsov transformation shifts the ions to new
equilibrium positions. In a more general sense it changes the boson vacuum.
As a result, the transformed Hamiltonian takes the following form
\begin{equation}
\tilde{H}=\sum_{i,j}\hat{\sigma}_{ij}c_{i}^{\dagger }c_{j}-E_{p}\sum_{i}\hat{%
n}_{i}+\sum_{\mathbf{q}}\omega _{\mathbf{q}}(d_{\mathbf{q}}^{\dagger }d_{%
\mathbf{q}}+1/2)+{\frac{1}{{2}}}\sum_{i\neq j}v_{ij}\hat{n}_{i}\hat{n}_{j},
\label{trans}
\end{equation}%
where
\begin{equation}
\hat{\sigma}_{ij}=t(\mathbf{m-n})\delta _{ss^{\prime }}\hat{X}_{i}^{\dagger }%
\hat{X}_{j}  \label{sigma}
\end{equation}%
is the renormalised hopping integral depending on the phonon operators, and
\begin{equation}
v_{ij}=V_{c}(\mathbf{m-n})-{\frac{1}{{N}}}\sum_{\mathbf{q}}|\gamma (\mathbf{q%
})|^{2}\omega _{\mathbf{q}}\cos [\mathbf{q\cdot (m-n)}]  \label{interaction}
\end{equation}%
is the interaction of polarons, $v_{ij}\equiv v(\mathbf{m-n})$, comprising
their Coulomb repulsion and the interaction via the lattice deformation. In
the extreme infinite-coupling limit, $\lambda \rightarrow \infty ,$ we can
neglect the hopping term of the transformed Hamiltonian. The rest has
analytically determined eigenstates and eigenvalues. The eigenstates $|%
\tilde{N}\rangle =|n_{i},n_{\mathbf{q}}\rangle $ are sorted by the polaron $%
n_{\mathbf{m}s}$ and phonon $n_{\mathbf{q}}$ occupation numbers. The energy
levels are
\begin{equation}
E=-(\mu +E_{p})\sum_{i}n_{i}+{\frac{1}{{2}}}\sum_{i\neq
j}v_{ij}n_{i}n_{j}+\sum_{\mathbf{q}}\omega _{\mathbf{q}}(n_{\mathbf{q}}+1/2),
\label{exact}
\end{equation}%
where $n_{i}=0,1$ and $n_{\mathbf{q}}=0,1,2,3,....\infty $.

\subsubsection{``$1/\protect\lambda $''\
expansion and polaron band}

\label{band} The Hamiltonian $\tilde{H}$ in zero order with respect to the
hopping describes localised polarons and independent phonons, which are
vibrations of ions relative to new equilibrium positions depending on the
polaron occupation numbers. The middle of the electron band is shifted down
by the polaron level-shift $E_{p}$ due to the potential well created by
lattice deformation. Importantly the phonon frequencies remain unchanged in
this limit at any polaron density, $n$. At finite $\lambda $ and $n$ there
is a softening of phonons $\delta \omega _{0}$ of the order of $\omega
_{0}n/\lambda ^{2}$ \cite{alecap,alecapgob,gobcap,ale2} (the initial paper
on the phonon renormalisation \cite{alecap} predicting $\delta \omega
_{0}\propto 1/\lambda $ was subsequently corrected \cite%
{alecapgob,gobcap,ale2}). Interestingly the optical phonon can be mixed with
a low-frequency polaronic plasmon forming a new excitation,
``plasphon'', which was proposed in \cite%
{ale2} as an explanation of the anomalous phonon mode splitting observed in
cuprates \cite{splitting}.

Now let us discuss the $1/\lambda $ expansion. First we restrict the
discussion to a single-polaron problem with no polaron-polaron interaction.
The finite hopping term leads to the polaron tunnelling because of
degeneracy of the zero order Hamiltonian with respect to the site position
of the polaron. To see how the tunnelling occurs we apply the perturbation
theory using $1/\lambda $ as a small parameter. The proper Bloch set of $N$%
-fold degenerate zero order eigenstates with the lowest energy ($-E_{p}$) of
the unperturbed Hamiltonian is
\begin{equation}
|\mathbf{k},0\rangle ={\frac{1}{\sqrt{N}}}\sum_{\mathbf{m}}c_{\mathbf{m}%
s}^{\dagger }\exp (i\mathbf{k\cdot m})|0\rangle ,
\end{equation}%
where $|0\rangle $ is the vacuum, and $N$ is the number of sites. By
applying the textbook perturbation theory one readily calculates the
perturbed energy levels. Up to the second order in the hopping integral they
are given by
\begin{equation}
E_{\mathbf{k}}=-E_{p}+\epsilon _{\mathbf{k}}-\sum_{\mathbf{k^{\prime }},n_{%
\mathbf{q}}}{\frac{|\langle \mathbf{k},0|\sum_{i,j}\hat{\sigma}%
_{ij}c_{i}^{\dagger }c_{j}|\mathbf{k^{\prime }},n_{\mathbf{q}}\rangle |^{2}}{%
{\sum_{\mathbf{q}}\omega _{\mathbf{q}}n_{\mathbf{q}}}},}  \label{pol}
\end{equation}%
where $|\mathbf{k^{\prime }},n_{\mathbf{q}}\rangle $ are the exited states
of the unperturbed Hamiltonian with one electron and at least one real
phonon. The second term in this equation, which is linear with respect to
the bare hopping $t(\mathbf{m})$, describes the polaron-band dispersion \cite%
{aledyn},
\begin{equation}
\epsilon _{\mathbf{k}}=\sum_{\mathbf{m}}t(\mathbf{m})\,e^{-g^{2}(\mathbf{m}%
)}\exp (-i\mathbf{k\cdot m}),  \label{width}
\end{equation}%
where
\begin{equation}
g^{2}(\mathbf{m})={\frac{1}{{2N}}}\sum_{\mathbf{q}}|\gamma (\mathbf{q}%
)|^{2}[1-\cos (\mathbf{q\cdot m})]
\end{equation}%
is the \textit{band-narrowing factor} at zero temperature.

The third term, quadratic in $t(\mathbf{m})\,$, yields a negative $\mathbf{k}
$-\emph{independent} correction to the polaron level-shift of the order of $%
1/\lambda ^{2}$, and a small correction to the polaron band dispersion, Eq.(%
\ref{width}) \cite{ale2,gog,kudfir}. The correction to the level
shift is due to polaronic hops onto a neighbouring site with no
deformation around it. As any second order correction this
transition shifts the energy down by an amount of about
$-t^{2}(\mathbf{m})/E_{p}$. It has little to do with the polaron
effective mass and the polaron tunneling mobility because the
lattice deformation does not follow the electron. The polaron hops
back and forth many times (about $e^{g^{2}}$) ``waiting''\ for a
sufficient lattice deformation to appear around neighbouring site
$\mathbf{n}$. Only after the deformation around the neighbouring
site is created does the polaron tunnel onto the next site together
with the deformation.

\subsubsection{Temperature effect on the polaron band}

\label{collapse}

Let us now analyse the temperature dependence of the polaron bandwidth,
which is determined by the average of the multiphonon operator, Eq.(\ref%
{sigma}),
\begin{equation}
\langle\langle \hat{X}_{i}^{\dagger }\hat{X}_{j}\rangle\rangle \equiv \prod_{%
\mathbf{q} }\langle \langle \exp [u_{i}^{\ast }(\mathbf{q} )d_{\mathbf{q}
}^{\dagger }-H.c.]\exp [u_{j}(\mathbf{q} )d_{\mathbf{q}\nu }-H.c.]\rangle
\rangle.  \label{identity}
\end{equation}
Here the double angular brackets correspond to quantum as well as
statistical averages of any operator $\hat{A}$ with the Gibbs distribution,
\begin{equation}
\langle \langle \hat{A}\rangle \rangle =\sum_{\nu }e^{({\Omega -E_{\nu })/T}%
}\langle \nu |\hat{A}|\nu \rangle \equiv Tr\{e^{(\Omega -\tilde{H})/T}\hat{A}%
\},
\end{equation}
where $\Omega $ is the thermodynamic potential and $|\nu \rangle $ are the
eigenstates of $\tilde{H}$ with the eigenvalues $E_{\nu }$. An operator
identity $\exp(\hat{A}+\hat{B})=\exp(\hat{A}\exp(\hat{B})exp(-[\hat{A},\hat{B%
}]/2)$ is instrumental. It is applied when the commutator $[\hat{A},\hat{B}]$
is a number. The identity allows us to write
\begin{eqnarray}
e^{[u_{i}^{\ast }(\mathbf{q} )d_{\mathbf{q} }^{\dagger }-H.c.]}e^{[u_{j}(%
\mathbf{q} )d_{\mathbf{q}}-H.c.]} &=&e^{\alpha ^{\ast }d_{\mathbf{q}%
}^{\dagger }}e^{-\alpha d_{\mathbf{q}}}e^{-|\alpha |^{2}/2}\times \\
&&e^{[u_{i}(\mathbf{q} )u_{j}^{\ast }(\mathbf{q} )-u_{i}^{\ast }(\mathbf{q}%
)u_{j}(\mathbf{q} )]/2}.  \nonumber
\end{eqnarray}
Quantum and statistical averages are calculated by expanding the exponents
in the trace as
\begin{equation}
\langle\langle e^{\alpha ^{\ast }d^{\dagger }}e^{-\alpha d}\rangle\rangle
=(1-p)\sum_{N=0}^{\infty }\sum_{n=0}^{N}p^{N}(-1)^{n}\frac{\left| \alpha
\right| ^{2n}}{(n!)^{2}}N(N-1)\times ...\times (N-n+1),  \label{part}
\end{equation}
where we dropped the phonon and site quantum numbers for transparency. Here $%
p=\exp (-\omega _{\mathbf{q}\nu }/T),$ so that a single-mode phonon
partition function is $Z_{ph}=1/(1-p)$. Eq.(\ref{part}) can be written in
the form \cite{lan}
\begin{equation}
\langle\langle e^{\alpha ^{\ast }d^{\dagger }}e^{-\alpha d}\rangle\rangle
=(1-p)\sum_{n=0}^{N}(-1)^{n}\frac{\left| \alpha \right| ^{2n}}{(n!)^{2}}p^{n}%
\frac{d^{n}}{dp^{n}}\sum_{N=0}^{\infty }p^{N}.
\end{equation}
Taking the sum over $N$, $\sum_{N=0}^{\infty }p^{N}=1/(1-p)$, and
differentiating it $n$ times yield $n!$ in the numerator, after which the
series over $n$ turns out equal to
\begin{equation}
\langle\langle e^{\alpha ^{\ast }d^{\dagger }}e^{-\alpha d}\rangle\rangle
=e^{-\left| \alpha \right| ^{2}n_{\omega }},
\end{equation}
where $n_{\omega }=[\exp (\omega _{\mathbf{q}}/T)-1]^{-1}$ is the
Bose-Einstein phonon distribution function. Collecting all multipliers one
finally obtains
\begin{equation}
\langle\langle \hat{\sigma}_{ij}\rangle\rangle=T(\mathbf{m-n})\delta
_{ss^{\prime }}\exp \left( -{\frac{1}{{2N}}}\sum_{\mathbf{q }}|\gamma (%
\mathbf{q} )|^{2}[1-\cos (\mathbf{q\cdot m})]\coth \frac{{\omega _{\mathbf{q
}}}}{2T}\right),  \label{width2}
\end{equation}
with the zero-temperature limit given by Eq.(\ref{width}).

The small-polaron band is exponentially narrow. Hence, one can raise a
concern about its existence in real solids \cite{ran2}. At zero temperature
the perturbation term of the transformed Hamiltonian conserves the momentum
because all off-diagonal matrix elements vanish,
\begin{equation}
\langle \mathbf{k},0|\sum_{i,j}\hat{\sigma}_{i,j}c_{i}^{\dagger }c_{j}|%
\mathbf{k^{\prime }},0\rangle =0
\end{equation}
if $\mathbf{k}\neq \mathbf{k^{\prime }}$. The emission of a single
high-frequency phonon is impossible for any $\mathbf{k}$ because of the
energy conservation. The \textit{polaron} half-bandwidth is exponentially
reduced,
\begin{equation}
w\approx De^{-g^{2}},
\end{equation}
and it is usually less than the optical phonon energy $\omega _{0}$ ($g^{2}$
is about $D\lambda /\omega _{0}).$ Hence, there is no damping of the polaron
band at $T=0$ caused by optical phonons, no matter how strong the
interaction is. The phonons ``dress'' the electron and coherently follow its
motion. However, at finite temperatures the simultaneous emission and
absorption of phonons is possible. Moreover the polaron bandwidth shrinks
with increasing temperature because the phonon-averaged hopping integrals
depend on temperature , Eq.(\ref{width2}). For high temperatures, $T\gg
\omega _{0}/2,$ the band narrows exponentially as $w\approx De^{-T/T_{0}}$,
where
\[
T_{0}^{-1}={\frac{1}{{N}}}\sum_{\mathbf{q}}|\gamma (\mathbf{q} )|^{2}{\omega
_{\mathbf{q }}^{-1}}[1-\cos (\mathbf{q\cdot m})].
\]
On the other hand, the two-phonon scattering of polarons becomes more
important with increasing temperature. One can estimate this scattering rate
by applying the Fermi-Dirac golden rule,
\begin{equation}
{\frac{1}{{\tau }}}=2\pi \left\langle \sum_{\mathbf{q},\mathbf{q^{\prime }}%
}\left| M_{\mathbf{qq}^{\prime }}\right| ^{2}\delta (\epsilon _{\mathbf{k}%
}-\epsilon _{\mathbf{k}+\mathbf{q}-\mathbf{q^{\prime }}})\right\rangle,
\end{equation}
where the corresponding matrix element is
\[
M_{\mathbf{qq}^{\prime }}=\sum_{i,j}\left\langle \mathbf{k}+\mathbf{q}-%
\mathbf{q^{\prime }},n_{\mathbf{q}}-1,n_{\mathbf{q^{\prime }}}+1\right| \hat{%
\sigma}_{i,j}c_{i}^{\dagger }c_{j}\left| \mathbf{k},n_{\mathbf{q}},n_{%
\mathbf{q^{\prime }}}\right\rangle \mathbf{.}
\]
For simplicity we consider the momentum independent $\gamma (\mathbf{q}%
)=\gamma _{0}$ and $\omega _{\mathbf{q}}=\omega _{0}.$ Expanding $\hat{\sigma%
}_{ij}$-operators in powers of the phonon creation and annihilation
operators one estimates the matrix element of the two-phonon scattering as $%
M_{\mathbf{qq}^{\prime }}\approx N^{-1}w\gamma _{0}^{2}\sqrt{n_{\mathbf{q}%
}(n_{\mathbf{q^{\prime }}}+1)}.$ Using this estimate and the polaron density
of states (DOS), $\rho_{p}(\xi )\equiv N^{-1}\sum_{\mathbf{k}}\delta (\xi
-\epsilon _{\mathbf{k}})\approx 1/2w$ one obtains \cite{alexandrov:1995}
\begin{equation}
{\frac{1}{{\tau }}}\approx w\gamma _{0}^{4}n_{\omega }(1+n_{\omega }),
\label{damping}
\end{equation}
where $n_{\omega }=[\exp (\omega _{0}/T)-1]^{-1}$ is the phonon distribution
function.

The polaron band is well defined, if $1/\tau <w$, which is satisfied for a
temperature range $T\leq T_{\min }\approx \omega _{0}/\ln \gamma _{0}^{4}$
about half of the characteristic phonon frequency for relevant values of $%
\gamma _{0}^{2}$. At higher temperatures the incoherent thermal activated
hopping dominates in the polaron dynamics \cite{yam,sew,holb,lan}, and the
polaron states are no longer the Bloch states. When the optical phonon
frequencies are exceptionally high (i.e. about $1000$K as in
high-temperature superconductors \cite{splitting}) lattice polarons are in
the Bloch states in the relevant range of temperatures, where the Boltzmann
kinetic theory with renormalised energy spectrum is applied.

\subsection{Discrete Fr\"ohlich polaron at strong coupling}

\label{range} The narrowing of the band and the polaron effective mass
strongly depend on the range of EPI \cite{alexandrov:1996,eagles:1969}. Let
us compare the small Holstein polaron (SHP) formed by the zero-range EPI and
a small polaron formed by the long-range (Fr\"{o}hlich) interaction, which
we refer to as the small or discrete Fr\"{o}hlich polaron (SFP). We use the
real-space representation of $H_{e-ph}$ \cite{alexandrov:1999},
\begin{equation}
H_{e-ph}=\sum_{\mathbf{n},i}f(\mathbf{m-n})\xi _{\mathbf{n}}\hat{n}_{i}.
\end{equation}
with the normal coordinate at site $\mathbf{n}$
\begin{equation}
\xi _{\mathbf{n}}=\sum_{\mathbf{q}}(2NM\omega _{\mathbf{q}})^{-1/2}e^{i%
\mathbf{q\cdot n}}d_{\mathbf{q}}+H.c.
\end{equation}
and the force between the electron at site $\mathbf{m}$ and the normal
coordinate $\xi _{\mathbf{n}},$%
\begin{equation}
f(\mathbf{m})=N^{-1}\sum_{\mathbf{q}}\gamma (\mathbf{q})(M\omega _{\mathbf{q}%
}^{3})^{1/2}e^{i\mathbf{q\cdot m}}.
\end{equation}

In general, there is no simple relation between the polaron level-shift $%
E_{p}$ and the exponent $g^{2}$ of the mass enhancement. This relation
depends on the form of EPI. Indeed for EPI with a single dispersionless
phonon mode, $\omega _{\mathbf{q}}=\omega_0$, one obtains
\begin{equation}
E_{p}=\frac{1}{2M\omega_0^{2}}\sum_{\mathbf{m}}f^{2}(\mathbf{m}),
\end{equation}
and
\begin{equation}
g^{2}=\frac{1}{2M\omega_0^{3}}\sum_{\mathbf{m}}\left[ f^{2}(\mathbf{m})-f(%
\mathbf{m})f(\mathbf{m+a})\right] ,
\end{equation}
where $\mathbf{a}$ is the primitive lattice vector. In the nearest-neighbour
approximation the effective mass renormalisation is given by $m^{\ast
}/m=e^{g^{2}}$ where $1/m^{\ast }=\partial ^{2}\epsilon _{\mathbf{k}%
}/\partial k^{2}$ at $k\rightarrow 0$ is the inverse polaron mass.
\begin{figure}[tbp]
\begin{center}
\includegraphics[angle=-90,width=0.55\textwidth]{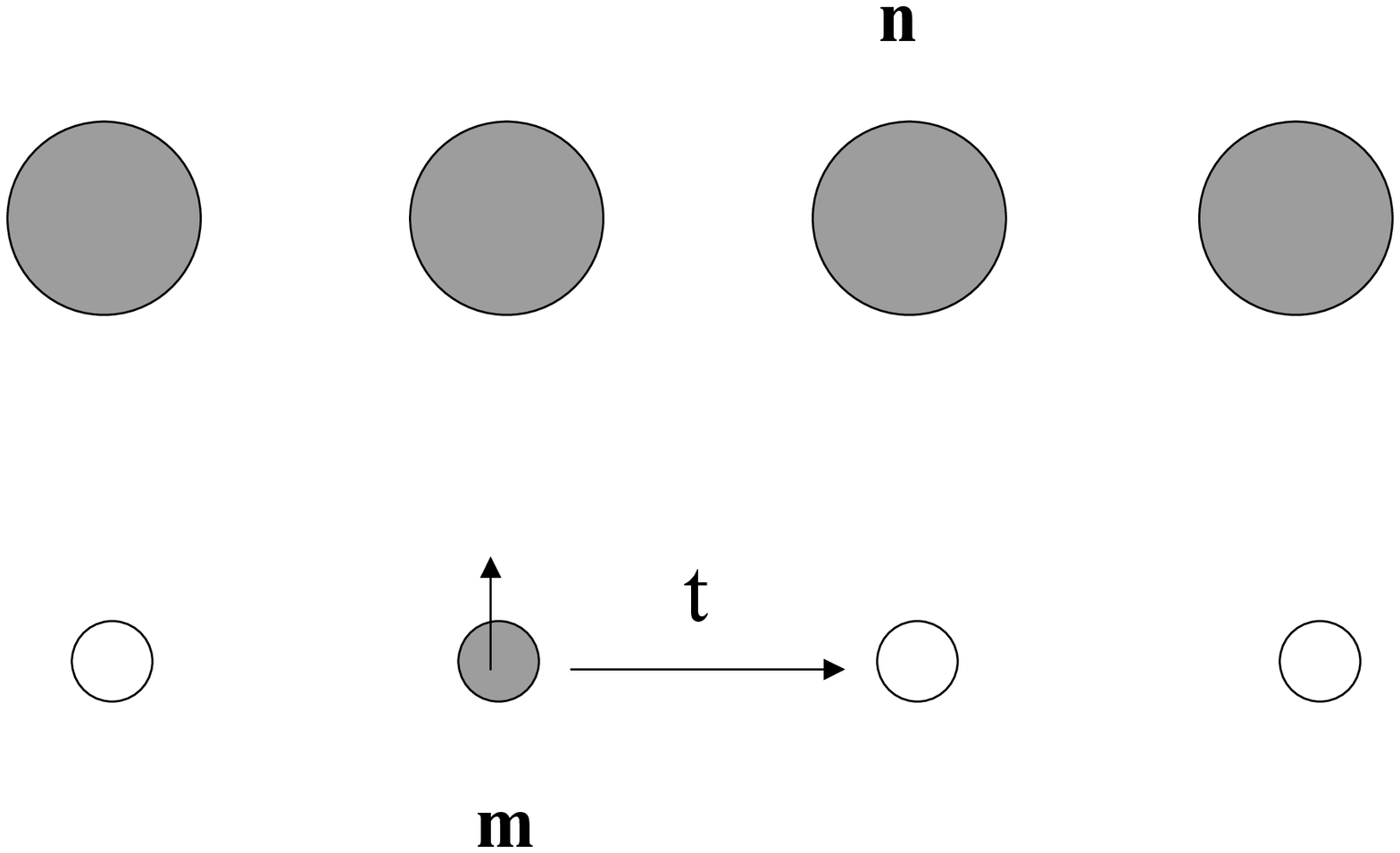} \vskip -0.5mm
\end{center}
\caption{A one-dimensional model of the lattice polaron on chain $\mathbf{m}$
interacting with displacements of all ions of another chain $\mathbf{n}$ ($t$
is the polaron hopping integral along chain $\mathbf{m}$).}
\label{chain}
\end{figure}
If the interaction is short-ranged, $f(\mathbf{m})=\kappa \delta _{\mathbf{m}%
,0}$ (the Holstein model), then $g^{2}=E_{p}/\omega_0 $. Here $\kappa $ is a
constant. In general, one has $g^{2}=\gamma E_{p}/\omega_0 $ with the
numerical coefficient
\begin{equation}
\gamma =\frac{1-\sum_{\mathbf{m}}f(\mathbf{m})f(\mathbf{m+a})}{\sum_{\mathbf{%
n}}f^{2}(\mathbf{n})},
\end{equation}
which might be less than $1$. To estimate $\gamma $ let us consider a
one-dimensional chain model with the long-range Coulomb interaction between
the electron on one chain ($\mathbf{m}$) and ion vibrations of another chain
($\mathbf{n}),$ polarised in the direction perpendicular to the chains \cite%
{alexandrov:1999}, Fig.\ref{chain}. The corresponding force is given by
\begin{equation}
f(\mathbf{m-n})=\frac{\kappa }{(|\mathbf{m}-\mathbf{n}|^{2}+1)^{3/2}}.
\label{force}
\end{equation}
Here the distance along the chains $|\mathbf{m}-\mathbf{n}|$ is measured in
units of the lattice constant $a$, the inter-chain distance is also $a$, and
we take $a=1$. For this long-range interaction we obtain $E_{p}=1.27\kappa
^{2}/(2M\omega_0 ^{2})$, $g^{2}=0.49\kappa ^{2}/(2M\omega_0 ^{3})$ and $%
g^{2}=0.39E_{p}/\omega_0$ . Thus the effective mass renormalisation in the
non-adiabatic regime is much smaller than in the Holstein model, roughly as $%
m_{SFP}^{\ast }\propto (m_{SHP}^{\ast })^{1/2}$ in units of $m$. An
analytical solution of a two-site single electron system interacting with
many vibrating ions of a lattice via a long-range Fr\"ohlich EPI found that
SFP is also several orders of magnitude lighter than SHP in the adiabatic
regime \cite{aleyav}.

Another interesting point is that the size of SFP and the length, over which
the distortion spreads, are \emph{different}. In the strong-coupling limit the
polaron is almost localised on one site $\mathbf{m}$. Hence, the size of its
wave function is the atomic size. On the other hand, the ion displacements,
proportional to the displacement force $f(\mathbf{m-n})$, spread over a
large distance. Their amplitude at a site $\mathbf{n}$ falls with the
distance as $|\mathbf{m-n}|^{-3}$ in the one-dimensional model, Fig.\ref%
{chain}. The polaron cloud (i.e. lattice distortion) is more extended than
the polaron itself. Such polaron tunnels with a larger probability than the
Holstein polaron due to a smaller \emph{relative} lattice distortion around two
neighbouring sites. For a short-range EPI the \emph{entire} lattice
deformation disappears at one site and then forms at its neighbour, when the
polaron tunnels from site to site. Therefore $\gamma =1$ and the polaron is
very heavy already at $\lambda \approx 1$. On the contrary, if the
interaction is long-ranged, only a fraction of the total deformation changes
every time the polaron tunnels from one site to its neighbour, and $\gamma $
is smaller than $1.$

\subsection{Effect of dispersive phonons}

A lighter mass of SFP compared with the nondispersive SHP is a generic
feature of any dispersive electron-phonon interaction. For example, a
short-range interaction with dispersive acoustic phonons ($\gamma (\mathbf{q}%
)\propto 1/q^{1/2},\omega _{\mathbf{q}}\propto q$) also leads to a lighter
polaron in the strong-coupling regime compared with SHP \cite%
{alexandrov:2003}. Even within the Holstein model with the local
(intramolecular) EPI the dispersion of phonon frequencies is a vital
ingredient since nondispersive phonons might lead to a divergent site jump
probability of polarons \cite{yam}. Importantly the comprehensive studies of
the molecular Holstein Hamiltonian, in which the dispersive features of the
phonon spectrum are taken into account, found much lower values of the
polaron mass compared with the non-dispersive model \cite%
{zoli:1998,zoli:2000,zoli:2004}. In those studies the $1/\lambda$
perturbation theory based on the standard Lang-Firsov (LF) and the
variational (modified) MLF transformation of the molecular Holstein
Hamiltonian with dispersive phonons has been applied for 1D, 2D and 3D
lattices including the second-order corrections as in Eq.(\ref{pol}),
\begin{equation}
H =-t\sum_{\langle i,j\rangle} c_{i}^{\dagger }c_{j}+g\omega_0\sum_{i}\hat{n}%
_{i}( d_{i} + d^{\dagger}_{i})+\sum_{\mathbf{q}}\omega _{\mathbf{q}}(d_{%
\mathbf{q} }^{\dagger }d_{\mathbf{q}}+1/2),  \label{holham}
\end{equation}
where $d_{\mathbf{q}}$ is the Fourier transform of $d_i$. The phonon
dispersion has been modeled using the intermolecular first neighbours force
constant, $M\omega_{1}^2$, which yields, for example, in 1D case $\omega^2
_{q}=\omega_0^2/2+\omega_1^2 +(\omega_0^2/4+\omega_0^2\omega_1^2 \cos q
+\omega_1^2)^{1/2}$. MLF improves the convergence of the $1/\lambda$
perturbation series by introducing a suitable variational parameter $%
\lambda_{\mathbf{q}}$ in the LF transformation, Eq.(\ref{LF}), as \cite%
{chatterjee:2000,chatterjee:2003}
\begin{equation}
S\Rightarrow S_{MLF}=\sum_{\mathbf{q,k}}\lambda_{\mathbf{q}} c^{\dagger}_{%
\mathbf{k+q}}c_{\mathbf{k}}(d_{\mathbf{q} }-d^{\dagger}_{-\mathbf{q}} ).
\label{MLF}
\end{equation}

As a result the polaron mass converges to much lower values when the phonon
dispersion is introduced, in particular in the adiabatic regime. There is a
continuous mass enhancement whose abruptness is significantly smoothed for
the largest values of the phonon dispersion, similar to the SFP discussed
above. While a phase transition is ruled out in the single-electron Holstein
Hamiltonian, where the ground state energy is analytic in the EPI strength
\cite{lowen0,lowen}, a crossover from more itinerant to more self-trapped
behaviour may be identified as a rather sudden event in the adiabatic regime
\cite{emin:1976,eagles:1969,kalosakas:1998,zheng:2003}.

\subsection{All-coupling discrete polaron}

\label{all}

\subsubsection{Holstein model at any coupling}

During past twenty years significant efforts were directed towards the
extension of the weak and strong-coupling perturbation lattice polaron
theories to the intermediate region of the relevant parameters, $\lambda
\sim 1$, and $\omega_0/t \sim 1$. It was argued \cite{gog,fir,ale2} that the
expansion parameter is actually $1/2z\lambda^{2}$, so the analytical
strong-coupling expansion in powers of $1/\lambda$ might have a wider region
of applicability than one can expect using simple physical arguments (i.e. $%
\lambda >1$). However, it has not been clear how fast the expansion
converges.

Kudinov and Firsov (1997) developed the analytical approach to the two-site
Holstein model by the use of the expansion technique, which provides the
electronic and vibronic terms as well as the wave functions and all
correlation functions in any order of powers of $t$. They have found the
exponential reduction factor in $all$ orders of the $1/\lambda$ perturbation
expansion. On the other hand, the corrections to the atomic level were found
as small as $1/\lambda^{2}$ rather than exponential in agreement with the
conventional second-order result, Eq.(\ref{width}). Chatterjee and Das
(2000) studied the same problem for any coupling within the perturbative
expansion combined with MLF, Eq.(\ref{MLF}), and MLF with a \emph{squeezing}
canonical transformation \cite{zheng1988}, $\exp(\tilde{S})$, where $\tilde{S%
}=\alpha (d_id_i-d_i^{\dagger}d_i^{\dagger})$. Using two variational
parameters introduced by MLF and squeezing transformations allows for very
good convergence of the $1/\lambda$ perturbation series even in the
near-adiabatic regime, $\omega_0/t\gtrsim 0.5$, where the conventional $%
1/\lambda$ expansion shows poor convergence. These studies also showed that
the region of the parameters of the Holstein model, where neither week nor
strong-coupling perturbation analytical methods are applicable, is rather
narrow. A semi-analytical approach to the solution of two coupled
differential equations, Eq.(\ref{hol1},\ref{hol2}), of the Holstein model in
the whole parameter space has been proposed in \cite{wang} based on the
coherent-state expansion of $u(x),v(x)$. These authors obtained the
recursive relations among the expansion coefficients, allowing for highly
accurate numerical solutions, which agree well with those by MLF method in
the weak- and strong-coupling regimes. The deviation from the MLF solution
in the intermediate-coupling regime implies that MLF misses some higher
order correlation terms. A continued fraction analytical solution of the
two-site Holstein model was derived by \cite{capone2002} as for a related
model in quantum optics \cite{swain}. In practice it also requires some
truncation of the infinite phonon Hilbert space. Finally all Green's
functions for the two-site Holstein-Hubbard model (HHM) were derived in
terms of continued fractions \cite{berciu}.

Numerical results obtained by different methods actually show that the
ground state energy (about $-E_{p}$) is not very sensitive to the
parameters, while the effective mass and the bandwidth strongly depend on
the polaron size, EPI range, and the adiabatic ratio, $\omega_0/t$. Several
methods exist for numerical simulations of lattice polarons. They include
exact diagonalisation (ED) \cite%
{Kongeter:1990,ran0,Marsiglio:1993,alekabraya,Fehske:1995,Wellein:1996,Marsiglio:1995,Stephan:1996,Capone:1997,barisic,fehskebook}%
, the global-local (GL) \cite{romero} and other advanced variational methods
\cite{Trugman:2001,Ku:2002}, quantum Monte-Carlo (QMC) algorithms \cite%
{Hirsch:1982,Hirsch:1983,Fradkin:1983,De Raedt:1982,De Raedt:1983,De
Raedt:1984,De Raedt:1985,berger,hohen,hohenadler,linden2007}, density matrix
renormalization group (DMRG) \cite%
{Jeckelmann:1998,Jeckelmann:1999,Zhang:1999}, continuous-time QMC \cite%
{kornilovitch1998,kornilovitch1999,alexandrov:1999,spencer2005,jim}, and
diagrammatic QMC \cite{Prokofev,Mishchenko2000,Macridin}. The methods vary
in accuracy and versatility, and, combined together, can provide all the
polaron properties of interest in the entire space of model parameters. On
the other hand, ED suffers from the necessary truncation of the phonon
Hilbert space, especially at strong couplings and low phonon frequencies
(even then, the total Hilbert space is huge, reducing the number of sites
and leading to poor momentum resolution), DMRG cannot easily handle
long-range interactions, diagrammatic QMC and ED are inconvenient in
calculating the density of states, and path-integral CTQMC slows down at
small frequencies \cite{fehskebook,MN2006,kornbook,linden2007}. In numerical
analysis of polaron models, a complex approach is needed where each method
is employed to calculate what it does best.

Until recently most numerical studies were performed on the Holstein model
(i.e with zero-range EPI). Reliable results for the intermediate region were
obtained using ED of vibrating clusters \cite%
{alekabraya,Fehske:1995,Wellein:1996,Marsiglio:1995,Stephan:1996,Capone:1997,fehskebook,ran0,ran,ran2}%
. Taking as a measure of the polaron kinetic energy the correlation function
$t_{eff}=\langle -t(c_{1}^{\dagger }c_{2}+c_{2}^{\dagger }c_{1})\rangle $
(here $c_{1,2}$ are annihilation operators on the `left' and `right' sites
of the Holstein model) one might doubt of the Lang-Firsov approach \cite%
{ran,ran2}, since this correlation function is much larger than the small
polaron bandwidth. However, applying the $1/\lambda $ expansion up to the
second order in $t$ one obtains the numerical $t_{eff}$ very close to the
perturbation $t_{LF}$ in the strong-coupling regime, $\lambda >1$, \cite%
{aledyn,fkka}
\begin{equation}
t_{eff}\simeq t_{LF}\equiv -t\exp \left( -{\frac{\lambda t}{{\omega }}}%
\right) -{\frac{t}{{\lambda }}},  \label{kin}
\end{equation}%
with $\lambda \equiv 2E_{p}/t$. Here only the first exponential term
describes the true coherent tunnelling, while the second term describes the
correction to the middle of the polaron band owing to the virtual
``back-forth''\ transitions to the
neighbouring site, as discussed above (\ref{band}). The main contribution to
$t_{eff}$ comes from the second-order term lowering the middle of the band
\cite{ale2,aledyn,gog,kudfir}, rather than from the polaron-transport
related first term (see also \cite{ran3,ran4}). Comparing the analytical
expression, Eq.(\ref{kin}) with the numerically calculated $t_{eff}$ one
confirms that the Holstein-Lang-Firsov approach is asymptotically exact both
in the non-adiabatic and adiabatic regimes \cite{aledyn}, if the
second-order correction is taken into account.

The numerical diagonalisation of the two-site-one-electron Holstein model
\cite{alekabraya} shows that the first-order term of the $1/\lambda$
perturbation theory describes well the polaron bandwidth in the \emph{%
non-adiabatic} regime $for$ $all$ $values$ of the coupling constant. There
is no agreement in the \emph{adiabatic } region, where the first order
perturbation expression, Eq.(\ref{adi}), $overestimates$ the polaron mass by
a few orders of magnitude. A poor convergence of the perturbation expansion
is due to appearance of the familiar double-well potential \cite{holb} in
the adiabatic limit. The tunnelling probability is extremely sensitive to
the shape of this potential. The splitting of levels for the two-site
cluster is well described by the Holstein quasi-classical formula
generalised for the intermediate coupling in Eq.(\ref{akr}). While the small
Holstein polaron is only a few times heavier than the bare (unrenormalised)
electron in a wide range of the coupling for a moderate adiabatic ratio $%
\omega_0/t \gtrsim 1$, it becomes very heavy in the adiabatic regime and for
the strong coupling \cite{alekabraya}.

\subsubsection{Holstein polaron in infinite lattices}

\label{short-range} A number of other independent numerical results proved
that ``by the use of the Holstein approximation and the
canonical Lang-Firsov approach with appropriate corrections, one obtains an
excellent estimate of the coherent bandwidth in both adiabatic and
non-adiabatic (strong-coupling) regimes''\ \cite{feh2}.
\begin{figure}[h]
\centering
\includegraphics[width=0.5\textwidth, angle=-90]{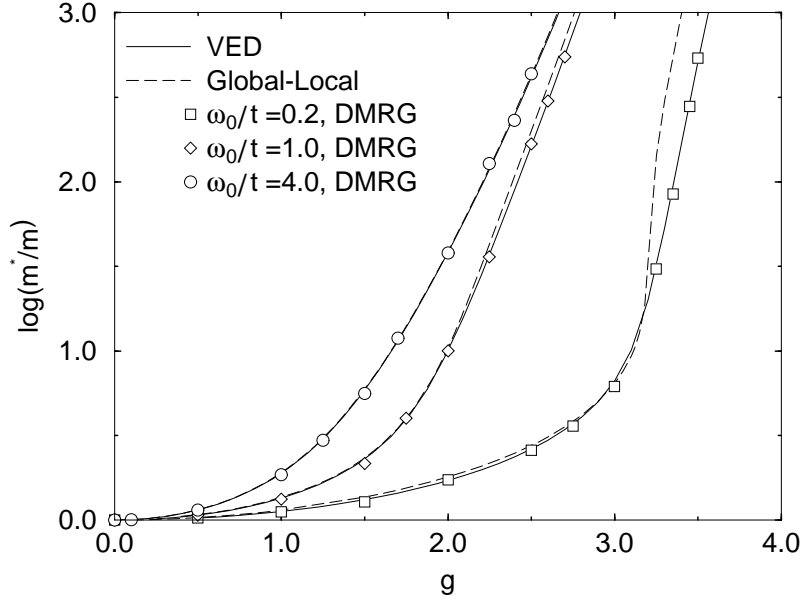}
\caption{Logarithm of the polaron effective mass in 1D Holstein model as a
function of $g$ (after \protect\cite{fehskebook}). VED results (full lines)
were obtained operating repeatedly $L=20~$ times with the off-diagonal
pieces of the Holstein Hamiltonian. For comparison GL (global-local) results
(dashed lines) are included~\protect\cite{romero}. Open symbols, indicating
the value of $\protect\omega _{0}/t$, are DMRG results~\protect\cite%
{Jeckelmann:1998}. }
\label{fig:mass0}
\end{figure}
In particular the elaborate variational ED (VED) \cite{BTB99,fehskebook}
provides an exact numerical solution of the Holstein crystal model, Eq.(\ref%
{holham}), in any dimension. In contrast to finite-lattice ED, it yields the
ground-state energy which is a variational bound for the exact energy in the
thermodynamic limit. Figure~\ref{fig:mass0} shows the effective mass
computed by VED~\cite{BTB99} in comparison with global-local (GL) and DMRG
methods. $m^{\ast }$ is obtained from $m/m^{\ast }=(1/2t)\partial
^{2}E_{k}/\partial k^{2}$, $k\rightarrow 0$, where $m_{0}=1/(2t)$ is the
rigid band mass (here the lattice constant is $a=1$). In the intermediate
coupling regime where VED gives an energy accuracy of 12 decimal places, one
can calculate the effective mass accurately (6-8 decimal places) \cite%
{fehskebook}.

In Fig.~\ref{fig:mass0} the parameters span different physical regimes
including weak and strong coupling, and low and high phonon frequency. There
is good agreement between VED and GL away from strong coupling and excellent
agreement in all regimes with DMRG results. DMRG calculations are not based
on finite-$k$ calculations due to a lack of periodic boundary conditions, so
they extrapolate the effective mass from the ground state data using chains
of different sizes, which leads to larger error bars and demands more
computational effort. There is no phase transition in the ground state of
the model, but the polaron becomes extremely heavy in the strong coupling
regime. The crossover to a regime of large polaron mass is more rapid in
adiabatic regime, i.e. at small $\omega_0/t$.

Unifying systematization of the crossovers between the different
polaron behaviors in one dimension was obtained in terms of quantum
to classical, weak coupling to strong coupling, adiabatic to
nonadiabatic, itinerant to self-trapped polarons and large to small
polarons \cite{barisic2}. It was argued that the relationship
between various aspects of polaron states can be specified by five
regimes: the weak-coupling regime, the regime of large adiabatic
polarons, the regime of small adiabatic polarons, the regime of
small nonadiabatic (Lang-Firsov) polarons, and the transitory regime
of small  polarons for which the adiabatic and nonadiabatic
contributions are inextricably mixed in the polaron dispersion
properties.  Holstein polarons in three-dimensional anisotropic
lattices have been recently studied using  a variational ED
technique, which provides highly accurate results for the polaron
mass and polaron radius \cite{alvermann}.  Varying the anisotropy
\cite{alvermann} demonstrated how a polaron evolves from a
one-dimensional to a three-dimensional quasiparticle and  showed
that even the local Holstein interaction leads to an enhancement of
anisotropy in charge-carrier motion.

For analyzing all-coupling polarons in more complex lattices the
continuous-time path-integral quantum Monte-Carlo algorithm (CTQMC) is
ideally suited. The algorithm is formulated in real space \cite%
{kornilovitch1998,kornbook} and based on the analytical integration over
phonon degrees of freedom introduced by Feynman (1955) and on an earlier
numerical implementation in discrete time by De Raedt and Lagendijk
(1982,1983,1984,1985). CTQMC introduced two critical improvements. Firstly,
formulation in continuous imaginary time eliminated errors caused by the
Trotter slicing and made the method numerically exact for any strength of
EPI. Secondly, introduction of twisted boundary conditions in imaginary time
\cite{KornilovitchPike,kornilovitch1998} enabled calculation of polaron
effective masses, spectra and the densities of states (DOS) in any
dimensions in infinite lattices.

The polaron action, obtained by analytical integration over phonon degrees
of freedom, is a functional of the polaron path in imaginary time $\mathbf{r}%
(\tau)$. It is given by the following double integral

\begin{eqnarray}
A[\mathbf{r}(\tau )] &=&\frac{z\lambda \bar{\omega}}{2\Phi _{0}(0,0)}%
\int_{0}^{\bar{\beta}}\int_{0}^{\bar{\beta}}d\tau d\tau ^{\prime -\bar{\omega%
}\bar{\beta}/2}\left( e^{\bar{\omega}(\bar{\beta}/2-|\tau -\tau ^{\prime
}|)}+e^{-\bar{\omega}(\bar{\beta}/2-|\tau -\tau ^{\prime }|)}\right) \Phi
_{0}[\mathbf{r}(\tau ),\mathbf{r}(\tau ^{\prime })]  \nonumber \\
&+&\frac{z\lambda \bar{\omega}}{\Phi _{0}(0,0)}\int_{0}^{\bar{\beta}%
}\int_{0}^{\bar{\beta}}d\tau d\tau ^{\prime -\bar{\omega}\tau }e^{-\bar{%
\omega}(\bar{\beta}-\tau ^{\prime })}\left( \Phi _{\Delta \mathbf{r}}[%
\mathbf{r}(\tau ),\mathbf{r}(\tau ^{\prime })]-\Phi _{0}[\mathbf{r}(\tau ),%
\mathbf{r}(\tau ^{\prime })]\right) :,  \label{eq:seven}
\end{eqnarray}%
\begin{equation}
\Phi _{\Delta \mathbf{r}}[\mathbf{r}(\tau ),\mathbf{r}(\tau ^{\prime
})]=\sum_{\mathbf{m}}\bar{f}_{\mathbf{m}}[\mathbf{r}(\tau )]\bar{f}_{\mathbf{%
m}+\Delta \mathbf{r}}[\mathbf{r}(\tau ^{\prime })]:,  \label{eq:eight}
\end{equation}%
where the vector $\Delta \mathbf{r}=\mathbf{r}(\bar{\beta})-\mathbf{r}(0)$
is the difference between the end points of the polaron path, $\bar{\beta}%
=t/T$, $\bar{\omega}=\omega _{0}/t$, and $\bar{f}_{\mathbf{m}}(\mathbf{n}%
)=f_{\mathbf{m}}(\mathbf{n})/\kappa $ (see also Eq.(\ref{force})). From this
starting point, the polaron is simulated using the Metropolis Monte-Carlo
method. The electron path is continuous in time with hopping events (or
kinks) introduced or removed from the path with each Monte-Carlo step. From
this ensemble, various physical properties may be computed, in particular,
the ground state energy, the number of phonons in the polaron
``cloud''\ and the polaron band energy
spectrum,
\begin{equation}
\epsilon _{\mathbf{k}}-\epsilon _{0}=-\lim_{\bar{\beta}\rightarrow \infty }%
\frac{1}{\bar{\beta}}\ln \langle \cos (\mathbf{k}\cdot \Delta \mathbf{r}%
)\rangle :,  \label{eq:eleven}
\end{equation}%
where $\mathbf{k}$ is the quasi momentum. By expanding this expression in
small $\mathbf{k}$, the $i$-th component of the inverse effective mass is
obtained as
\begin{equation}
\frac{1}{m_{i}^{\ast }}=\lim_{\bar{\beta}\rightarrow \infty }\frac{1}{\bar{%
\beta}}\langle (\Delta \mathbf{r}_{i})^{2}\rangle :.  \label{eq:twelve}
\end{equation}%
Thus the inverse effective mass is the diffusion coefficient of the polaron
path in the limit of the infinitely long ``diffusion
time''\ $\bar{\beta}$. CTQMC has enabled accurate analysis
of models with long-range electron-phonon interactions \cite%
{alexandrov:1999,spencer2005,jim} and a model with anisotropic electron
hopping \cite{kornilovitch1999}.

All numerical results confirm gross polaronic features well understood
analytically by \cite{holb} and others both in the non-adiabatic and
adiabatic regimes. A great power of numerical methods is the ability to
calculate an entire polaron spectrum, and the polaron DOS $\rho(E) = N^{-1}
\sum_{\mathbf{k}} \delta(E - E_{\mathbf{k}} + E_0)$, in the whole parameter
space. The coherent part of the spectrum, $\epsilon_{\mathbf{k}}$, possesses
an interesting property of flattening at large lattice momenta in the
adiabatic limit, $t \gg \omega_0$ \cite%
{Wellein:1996,Wellein:1997,romero,romero1999,Stephan:1996,fehskebook}. In
the weak-coupling limit, the flattening can be readily understood as
hybridization between the bare electron spectrum and a phonon mode \cite%
{levinson}. The resulting polaron dispersion is cosine-like at small $%
\mathbf{k}$ and flat at large $\mathbf{k}$. As a result the polaron DOS
should be peaked close to the \emph{top} of the polaron band. Exact VED \cite%
{fehskebook} and CTQMC \cite{kornbook} calculations have confirmed that this
is indeed the case for the short-range EPI. At small $\omega_0$, DOS
develops a massive peak at the top of the band. With increasing $\omega_0$,
the polaron spectrum approaches the cosine-like shape in full accordance
with the Lang-Firsov non-adiabatic formula. The respective DOS gradually
assumes the familiar shape of the tight-binding band with renormalised
hopping integrals. However, at small-to-moderate $\omega_0/t$ in two and
three dimensions, the bottom half of the polaron band contains a tiny
minority of the total number of states, so that the system's responses will
be dominated by the states in the peak.

In the long-range model (\ref{force}) DOS is much closer to the
tight-binding shape than the Holstein DOS at the same parameters \cite{jim}.
The polaron spectrum and DOS is another manifestation of the extremity of
the Holstein model. Long-range EPI removes those peculiarities and makes the
shape of polaron bands closer to the $1/\lambda$ expansion results.

\subsubsection{Discrete Fr\"ohlich polaron at any coupling}

\label{lightSFP} As discussed above (\ref{range}) the lattice polaron mass
strongly depends on the radius of EPI. So does the range of the
applicability of the analytical $1/\lambda $ expansion theory. The theory
appears almost exact in a wide region of the Fr\"{o}hlich EPI, Eq.(\ref%
{force}), for which the exact polaron mass was calculated with CTQMC
algorithm in \cite{alexandrov:1999}.

At large $\lambda $ ($>1.5$) SFP was found to be much lighter than SHP in
agreement with the analytical results (\ref{range}), while the large Fr\"{o}%
hlich polaron (i.e. at $\lambda <1$) was $heavier$ than the large Holstein
polaron with the same binding energy \cite{alexandrov:1999}. The mass ratio $%
m_{FP}^{\ast }/m_{HP}^{\ast }$ is a non-monotonic function of $\lambda $.
The effective mass of the small Fr\"{o}hlich polaron, $m_{FP}^{\ast
}(\lambda )$ is well fitted by a single exponent, which is $e^{0.73\lambda }$
for $\omega_0=t$ and $e^{1.4\lambda }$ for $\omega_0 =0.5 t$. The exponents
are remarkably close to those obtained with the Lang-Firsov transformation, $%
e^{0.78\lambda }$ and $e^{1.56\lambda }$, respectively. Hence, in the case
of the Fr\"{o}hlich interaction the transformation is perfectly accurate
even in the moderate adiabatic regime, $\omega_0 /t\leq 1 $ for $any$
coupling strength. It is not the case for the Holstein polaron. If the
interaction is short-ranged, the same analytical technique is applied only
in the non-adiabatic regime $\omega_0 /t>1$.

An important question about polaron properties also involves the effects of
screening on the electron-phonon interaction. Unscreened EPI makes polarons
very mobile \cite{alexandrov:1996}, which leads to strong effects even on
the qualitative physical properties of the polaron gas. Modeling the
screening effects, the interaction force between electrons and phonons was
introduced in \cite{spencer2005} in the form of a screened discrete
Fr\"ohlich interaction,
\begin{equation}
f_{\mathbf{m}}(\mathbf{n})=\frac{\kappa} {[(\mathbf{m}-\mathbf{n})^2+1]^{3/2}%
}\exp\left(-\frac{|\mathbf{m}-\mathbf{n}|} {R_{\mathrm{sc}}}\right).
\label{screened}
\end{equation}
It describes EPI of holes with c-axis polarised lattice distortions, which
has been suggested as the relevant electron-phonon interaction in the
cuprates \cite{alexandrov:1996}.

The CTQMC polaron mass for the one-dimensional lattice with the screened Fr%
\"{o}hlich EPI, Eq.(\ref{screened}) is shown in Fig.\ref{LR:lnmass} at four
different values of the screening length, $R_{\mathrm{sc}}\to 0$ (the
short-range Holstein interaction), $R_{\mathrm{sc}}=1$, $R_{\mathrm{sc}}=3$,
and $R_{\mathrm{sc}}\to\infty$ (the non-screened Fr\"{o}hlich interaction).

An important observation is evident from the plot of $\ln (m^{\ast }/m_{0})$
against $\lambda $ shown in figure (\ref{LR:lnmass}). At intermediate and
large couplings (that is, in the transition and small polaron regions),
altering the value of $R_{\mathrm{sc}}$ has a \textit{dramatic} effect on
the effective mass. For example, the non-screened Fr\"{o}hlich polaron is
over $10^{3}$ times ``lighter''\ than the
Holstein polaron at $\lambda =4$, and over $10^{4}$ times lighter at $%
\lambda =5$. It is apparent from the above results for the screened Fr\"{o}%
hlich model that as $R_{\mathrm{sc}}$ increases (from Holstein to Fr\"{o}%
hlich), the QMC results move, in general, closer to the ``$%
1/\lambda $''\ expansion over the \textit{entire} range of $%
\lambda $. That is, the ``$1/\lambda $''\
expansion becomes generally more applicable as the range of EPI increases
(as well as with increasing $\bar{\omega}$).

The polaron features for the long-range EPI were also investigated by
extending a variational approach previously proposed for the study of
systems with local (Holstein) coupling \cite{perroni2004}. The ground state
spectral weight, the average kinetic energy, the mean number of phonons, and
the electron-lattice correlation function were calculated for a wide range
of model parameters focusing on the adiabatic regime. A strong mixing of
electronic and phononic degrees of freedom even for small values of EPI was
found in the adiabatic case due to the long-range interaction.
\begin{figure}[tbp]
\includegraphics[angle=-90,width=0.80\textwidth]{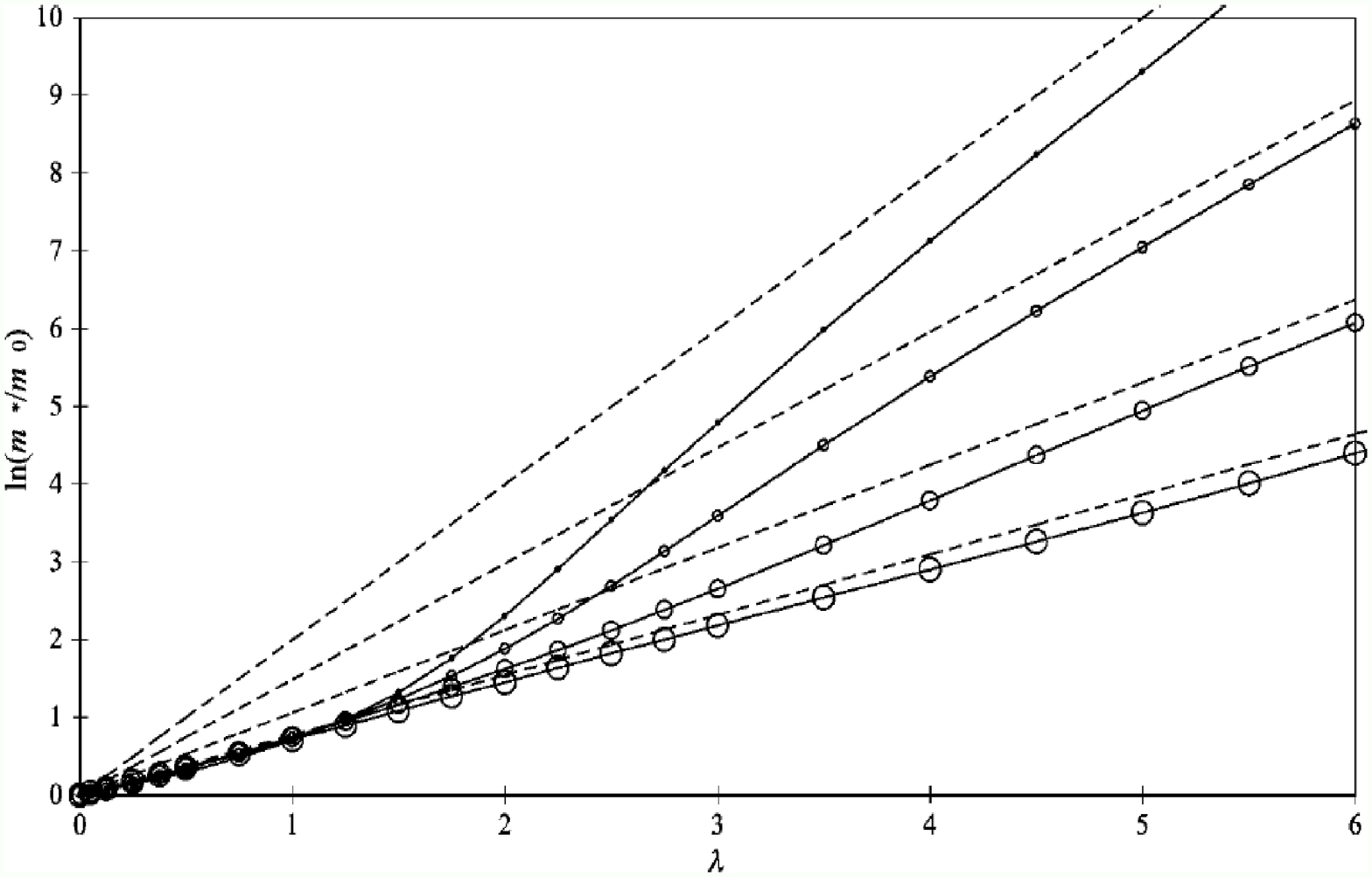} \vskip -0.5mm
\caption{The logarithm of the effective mass for screening lengths $R_{%
\mathrm{sc}}=0,1,3,\infty$ (increasing size of circles) versus $\protect%
\lambda$ at $\bar{\protect\omega}=1$ \protect\cite{spencer2005}. At
intermediate and strong coupling, decreasing the value of $R_{\mathrm{sc}}$
dramatically increases the effective mass. The curves tend to the
strong-coupling analytical result (dashed lines).}
\label{LR:lnmass}
\end{figure}

\subsection{Isotope effect on the polaron mass and the polaron band-structure%
}

\label{isotope} There is a qualitative difference between ordinary metals
and polaronic (semi)conductors. The renormalised effective mass of electrons
is independent of the ion mass $M$ in ordinary metals (where the Migdal
adiabatic approximation is believed to be valid, see below), because $%
\lambda $ does not depend on the isotope mass. However, when electrons form
polarons dressed by lattice distortions, their effective mass $m^{*}$
depends on $M$ through $m^{*}= m \exp (\gamma E_p/\omega_0)$, in the
strong-coupling limit. Here the phonon frequency depends on the ion mass, so
that there is a large polaronic isotope effect (PIE) on the carrier mass
with the carrier mass isotope exponent $\alpha_{m^*} = (1/2)\ln (m^*/m)$ as
predicted in \cite{aleiso}, in contrast to the zero isotope effect in
ordinary metals (see \cite{alexandrov:2003} for more details). The effect
was found experimentally in cuprates \cite{Zhao:1995,Zhao:1997,Khasanov:2004}
and manganites \cite{pet}. More recent high resolution angle resolved
photoemission spectroscopy (ARPES) also revealed a complicated isotope
effect on the whole band-structure in cuprate superconductors depending on
the electron energy and momentum \cite{Gweon:2004}.

PIE in the intermediate region of parameters was calculated using the
dynamic mean-field approximation (DMFT) \cite{pietronero2005,millis2002},
and CTQMC algorithm \cite{korniso,spencer2005,jim}. Importantly $%
\alpha_{m^*} $ and the effective mass averaged over dimensions are related
to the critical temperature isotope exponent, $\alpha=-d \ln T_c/d\ln M$, of
a (bi)polaronic superconductor as
\begin{equation}
\alpha = \alpha_{m^*} \left( 1 - \frac{m/m^*}{\lambda - \mu_c} \right) \: ,
\label{alpha}
\end{equation}
where $\mu_c$ is the Coulomb pseudo-potential \cite{aleiso,alexandrov:2003}.

In the adiabatic regime the isotope effect on the polaron mass does not
fully represent the isotope effect on the vast majority of polaron states,
in particular for the short-range EPI, so that additional insight can be
gained from the isotope effect on the entire polaron spectrum \cite{korniso}%
. The isotope effect on polaron spectrum and DOS in $d=2$ is illustrated in
Fig~\ref{fig:IsotopeSpectrum}. The ratio of the two phonon frequencies, $%
\omega = 0.80 \, t$ and $0.75 \, t$, has been chosen to roughly correspond
to the substitution of $^{16}$O for $^{18}$O in complex oxides. One can see
that the polaron band shrinks significantly, by 20-30\%, for both polaron
types. The middle panels show the isotope exponents on spectrum points
calculated as
\begin{equation}
\alpha_{\mathbf{k}} = \frac{1}{2} \frac{\langle \omega \rangle}{\langle
\epsilon_{\mathbf{k}} \rangle} \frac{\Delta \epsilon_{\mathbf{k}}}{\Delta
\omega} \: ,  \label{eq:fortyoneone}
\end{equation}
where the angular brackets denote the mean value of either the two
frequencies or of the two energy values. An interesting observation is that $%
\alpha_{\mathbf{k}}$ of the Fr\"ohlich polaron is roughly independent of $%
\mathbf{k}$ ($\pm 10\%$). In the Holstein case $\alpha_{\mathbf{k}}$ dips in
the vicinity of the $\mathrm{\Gamma}$ point.
\begin{figure}[tbp]
\centering
\includegraphics[width=11.7cm]{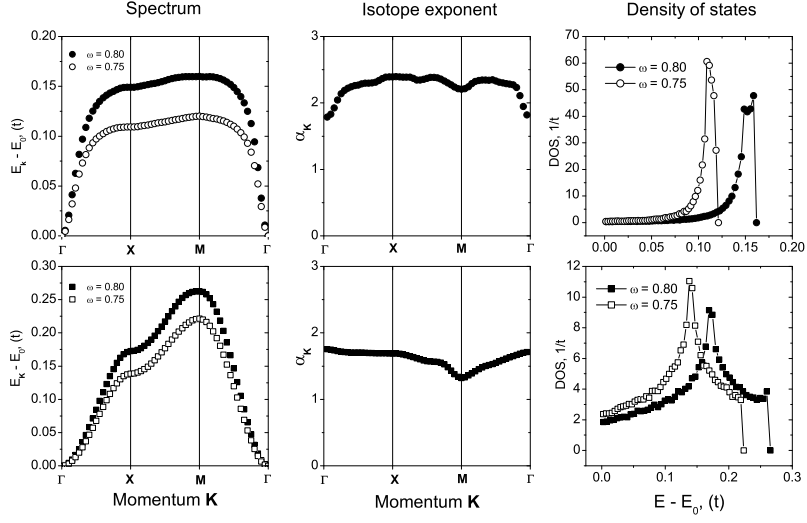}
\caption{ Isotope effect on polaron spectrum and density of states. \emph{%
Top row}\/: the $2d$ Holstein polaron at $\protect\lambda = 1.2$. Left:
polaron spectrum at $\protect\omega = 0.80$ and $0.75$ (in units of $t$).
Middle: the isotope exponent for each $\mathbf{k}$-point. Right: the density
of states for the two frequencies. \emph{Bottom row}\/: the same for the
discreate $2d$ Fr\"ohlich polaron at $\protect\lambda = 2.4$ (after
\protect\cite{korniso}). }
\label{fig:IsotopeSpectrum}
\end{figure}

Unconventional isotope effects as observed in high-temperature
superconducting cuprates \cite{Khasanov:2004}, were also explained by
polaron formation stemming from the coupling to the particular quadrupolar $%
Q(2)$-type phonon mode \cite{bussmanniso}. Polaronic isotope effects in the
spectral function of strongly correlated systems were numerically studied by
\cite{fratini2005} in the framework of the Holstein-Hubbard model and by
\cite{Mishchenko2006} using the extended $t-J$ model including EPI.

\subsection{Jahn-Teller polaron}

\label{sec:fourfive} The density-displacement EPIs discussed above
are not only possible types of the electron-lattice coupling.
Examples of other types are the Su-Schrieffer-Heeger (SSH) EPI
\cite{SSH} ( for recent path-integral results for SSH polarons see
\cite{zolibook}), and the Jahn-Teller (JT) EPI \cite{Jahn:1937}. In
the former, the lattice deformation is coupled to electron kinetic
energy, while JT involves a multidimensional electron basis and a
multidimensional representation of the deformation group. The JT
interaction is active in some molecules and crystals of high point
symmetry, and it has served as a guide in the search for
high-temperature superconductivity \cite{Bednorz1988}. Later on
JT-type EPIs were widely discussed in connection with cuprate and
other high-T$_c$
superconductors (see, for example, \cite%
{muller:2000,alekabful,gunnarsson,Mihailovic:2001,Mertelj}) and
colossal-magnetoresistance manganites (see, for example, Refs. \cite%
{millis:1995,roder,yunoki,bussmann,edwards2002}).

The simplest model of the JT interaction is the $E \otimes \, e$ interaction
\cite{Kanamori1960} that describes a short-range coupling between
twice-degenerate $e_g$ electronic levels $(c_1, c_2)$ and a local
double-degenerate vibron mode $(\zeta,\eta)$. The Hamiltonian reads ( here
we follow \cite{kornbook})
\begin{eqnarray}
H_{\mathrm{JT}} = & - & t \sum_{\langle \mathbf{n n^{\prime }} \rangle }
\left( c^{\dagger}_{\mathbf{n^{\prime }}1} c_{\mathbf{n}1} + c^{\dagger}_{%
\mathbf{n^{\prime }}2} c_{\mathbf{n}2} \right)  \nonumber \\
& - & \kappa \sum_{\mathbf{n}} \left[ \left( c^{\dagger}_{\mathbf{n}1} c_{%
\mathbf{n}2} + c^{\dagger}_{\mathbf{n}2} c_{\mathbf{n}1} \right) \eta_{%
\mathbf{n}} + \left( c^{\dagger}_{\mathbf{n}1} c_{\mathbf{n}1} -
c^{\dagger}_{\mathbf{n}2} c_{\mathbf{n}2} \right) \zeta_{\mathbf{n}} \right]
\nonumber \\
& + & \sum_{\mathbf{n}} \left[ - \frac{1}{2M} \left( \frac{\partial^2}{%
\partial \zeta^2_{\mathbf{n}}} + \frac{\partial^2}{\partial \eta^2_{\mathbf{n%
}}} \right) + \frac{M \omega^2}{2} \left( \zeta^2_{\mathbf{n}} + \eta^2_{%
\mathbf{n}} \right) \right] \: .  \label{eq:fortyone}
\end{eqnarray}
The symmetry of the interaction ensures the same coupling parameter $\kappa$
for the two phonons. Because the ionic coordinates of different cells are
not coupled, the model describes a collection of separate clusters that are
linked only by electron hopping. To relate the Hamiltonian to more realistic
situations, phonon dispersion should be added \cite%
{Mihailovic:2001,Allen1999}.

An important property of the $E \otimes \, e$ interaction is the absence of
an exact analytical solution in the atomic limit $t = 0$. Here, in contrast
with the Holstein and Fr\"ohlich EPI, the atomic limit is described by two
coupled \emph{partial} differential equations for the electron doublet $%
\psi_{1,2}(\zeta,\eta)$. At large couplings, however, the elastic energy
assumes the Mexican hat shape and the phonon dynamics separates into radial
oscillatory motion and azimuthal rotary motion. This results in an
additional pre-exponential factor $\propto \kappa$ in the ion overlap
integral, leading to the effective mass $m^{\ast}_{\mathrm{JT}}/m = (2/\pi
g)^{1/2} \exp{g^2}$, where $g^2 = \kappa^2/2M\omega^3$ \cite{Takada2000}.

A path integral approach to Hamiltonian (\ref{eq:fortyone}) was developed by
Kornilovitch (2000). Its details are found in \cite{kornbook}. Because there
are two electron orbitals, the electron path must be assigned an additional
orbital index (or \textit{colour}) $a = 1,2$. Colour 1 (or 2) of a given
path segment means that it resides in the first (second) atomic orbital of
the electron doublet. There is a difference between the two phonons. Phonon $%
\zeta$ is coupled to electron density, like in the Holstein case. The
difference from the Holstein is that the direction of the force changes to
the opposite when the electron changes orbitals. In contrast, phonon $\eta$
is coupled to orbital changes themselves: the more often the electron
changes orbitals, the more ``active'' is $\eta$. Discrete orbital changes
are analogous to electron hops between discrete lattice sites, and as such
are associated with ``kinetic orbital energy''. Phonon coupling to orbital
changes is analogous to phonon coupling to electron hopping in the SSH model
\cite{SSH}.

The JT polaron properties can be calculated with no approximations using QMC
algorithm. The effective mass, spectrum, and density of states are obtained
as for the conventional lattice polarons. Most properties behave similarly
to those of the $d=3$ Holstein polaron at the same phonon frequency \cite%
{Kornilovitch2000}. For example, the kinetic energy sharply decreases by
absolute value between $\lambda = 1.2$ and $1.4$. The JT polaron mass is
slightly larger at the small to intermediate coupling, but several times
smaller at the strong coupling. This non-monotonic behaviour of the ratio of
the JT and Holstein masses was later confirmed by accurate variational
calculations \cite{ElShawish2003}, although in that work the JT polaron (and
bipolaron) was investigated in one spatial dimension. The relative lightness
of the JT polaron is consistent with Takada's result mentioned above \cite%
{Takada2000}. Finally, the density of JT polaron states features the same
peak at the top of the band, caused by the spectrum flattening at large
polaron momenta, see Fig. (\ref{fig:IsotopeSpectrum}).

In summary, the local character of the JT interaction and the independence
of vibrating clusters result in the same extremity of polaron properties as
in the $3d$ Holstein model. One could expect that either a long-range JT
interaction or phonon dispersion will soften the sharp polaron features and
make JT polarons more mobile \cite{kornbook}.

\section{Response of discrete polarons}

\label{response}

\subsection{Hopping mobility}

Studies of lattice polarons allowed for a theoretical understanding of low
mobility problem \cite{Ioffe} of many ``poor''\ conductors, where an estimate of the mean-free path
yields values much lower than the lattice constant. Transport properties of
lattice polarons depend strongly on temperature. For $T$ lower than the
characteristic phonon energy polaron kinetics is the Boltzmann kinetics of
heavy particles tunnelling in the narrow band \cite{holb}. However at higher
temperatures the polaron band collapses (\ref{collapse}), and the transport
is diffusive via thermally activated jumps of polarons from site to site
\cite{yam,holb,lan63,lan68,appel,fir,klinger1979,mott1979}.

The $1/\lambda$ perturbation expansion is particularly instrumental in
calculating hopping mobilities and optical absorption coefficients. Applying
the canonical transformation (\ref{LFtransform}) and singling out the
diagonal coherent tunnelling in the polaron band one can write the
Hamiltonian as
\begin{equation}
\tilde{H}=H_{p}+H_{ph}+H_{int},
\end{equation}
where $H_{p}=\sum_{\mathbf{k}}\xi_{\mathbf{k}}c_{\mathbf{k}}^{\dagger }c_{%
\mathbf{k}}$ is the ``free'' $polaron$ contribution (here we drop the spin),
$H_{ph}=\sum_{\mathbf{q}}\omega _{\mathbf{q}}(d_{\mathbf{q}}^{\dagger }d_{%
\mathbf{q}}+1/2)$ is the phonon part, and $\xi_{\mathbf{k}}=Z^{\prime }E_{%
\mathbf{k}}-\mu $ is the renormalised (polaron) band dispersion. The
chemical potential $\mu $ includes the polaron level shift $-E_{p}$, and it
could also include all higher orders in $1/\lambda $ corrections to the
polaron spectrum, which are independent of $\mathbf{k}$, Eq.(\ref{width}).
The band-narrowing factor $Z^{\prime}$ is defined as
\begin{equation}
Z^{\prime }={\frac{\sum_{\mathbf{m}}t(\mathbf{m})e^{-g^{2}(\mathbf{m})}\exp
(-i\mathbf{k\cdot m})}{{\sum_{\mathbf{m}}t(\mathbf{m})\exp (-i\mathbf{k\cdot
m})}}},  \label{z}
\end{equation}
which is $Z^{\prime }=\exp (-\gamma E_{p}/\omega ))$ where $\gamma\leq 1$
depends on the range of EPI and phonon frequency dispersions (\ref{range}).
The interaction term $H_{int}$ comprises the polaron-polaron interaction,
Eq.(\ref{interaction}), and the residual \emph{polaron-phonon} interaction
\begin{equation}
H_{p-ph}\equiv \sum_{i\neq ,j}[\hat{\sigma}_{ij}-\left\langle \hat{\sigma}%
_{ij}\right\rangle]c_{i}^{\dagger }c_{j},
\end{equation}
where $\left\langle \hat{\sigma}_{ij}\right\rangle $ means averaging with
respect to the bare phonon distribution. In the framework of the
single-polaron problem one neglects the polaron-polaron interaction and
treats $H_{p-ph}$ as a perturbation.

The motion of the small polaron at high temperatures is a random walk
consisting of steps from site to site \cite{holb}. Holstein calculated the
probability $W$ for the hop of a small polaron to a neighbouring site. He
suggested that the random walk was a Markovian process. For such processes
the diffusion coefficient is given by $D\simeq a^{2}W$, where $W$ is the
hopping probability. The only term in the polaronic Hamiltonian, which
changes the phonon occupation numbers, is the polaron-phonon interaction $%
H_{p-ph}$. The nearest neighbour hopping probability in the second order is
\begin{equation}
W=2\pi \langle\sum_{j} |\langle j|H_{p-ph}|i\rangle|^{2} \delta \left(\sum_{%
\mathbf{q}}\omega_{\mathbf{q}}(n^{j}_{\mathbf{q}}-n^{i}_{\mathbf{q}%
})\right)\rangle,  \label{W}
\end{equation}
where $|i\rangle$ and $|j\rangle$ are the eigenstates of $H_{0}$
corresponding to the polaron on site $i$ with $n^{i}_{\mathbf{q}}$ phonons
in each phonon mode and the polaron on the neighbouring site $j$ with $%
n^{j}_{\mathbf{q}}$ phonons, respectively. Replacing the $\delta$ function
in Eq.(\ref{W}) by the integral yields
\begin{equation}
W=t^2e^{-2g^{2}}\int_{-\infty}^{\infty}d\tau \left[\exp \left({\frac{1}{{N}}}%
\sum_{\mathbf{q}}|\gamma(\mathbf{q})|^{2}[1-\cos(\mathbf{q}\cdot \mathbf{a})]%
{\frac{\cos(\omega_{\mathbf{q}}\tau)}{{\sinh{\frac{\omega_{\mathbf{q}}}{{2T}}%
}}}}\right)-1\right]e^{-i0^{+}|\tau|}.
\end{equation}
The integration over $\tau$ is performed using a saddle-point approximation
by allowing a finite phonon frequency dispersion $\delta \omega \ll \omega_{%
\mathbf{q}}$. Expanding $\cos(\omega_{\mathbf{q}}\tau)$ in powers of $\tau$,
and using the Einstein relation $\mu=eD/T$ one obtains the hopping mobility,
\begin{equation}
\mu_{h}=ea^{2}{\frac{\pi^{1/2}t^2}{{2T(E_{a}T)^{1/2}}}}e^{-E_{a}/T}.
\end{equation}
where
\begin{equation}
E_{a}={\frac{T}{{N}}}\sum_{\mathbf{q}}|\gamma(\mathbf{q})|^{2} [1-\cos(%
\mathbf{q}\cdot \mathbf{a})]\tanh(\omega_{\mathbf{q}}/4T)
\end{equation}
is the activation energy, which is $half$ of the polaron level shift $E_{p}$
for the Holstein EPI. This expression is applicable at $T\gg \omega_0/2$ and
$t^2/\omega_0 (E_aT) \ll 1$. Importantly, increasing the EPI range
diminishes the value of $E_{a}$ further. The hopping mobility $\mu_{h}\equiv
\sigma_{h}/ne\sim exp(-E_{a}/T)$ can be below $ea^{2}\simeq 1 cm^{2}/Vs$,
which is the lowest limit for the Boltzmann theory to be applied. Within the
Boltzmann theory such a low mobility corresponds to the mean free path $l<a$%
, which is not a reasonable result.

Holstein suggested that at low temperatures $T< \omega_{0}/2\ln\gamma_0$,
there is the ordinary Boltzmann transport in momentum space, but in the
narrow polaron band. According to \cite{holb}, the transition from the band
regime to the hopping regime occurs when the uncertainty in the polaron
energy becomes comparable to the width of the polaron band. If phonons
dominate in the scattering, the polaron mobility, $\mu=\mu_t+\mu_h$
decreases when the temperature increases from zero to some $T_{\min }$,
where it is at minimum, because its Boltzmann part, $\mu_t$, falls down due
to an increasing number of phonons, while the hopping part, $\mu_h$, remains
small. However $\mu$ increases above $T_{\min },$ due to the thermal
activated hopping. There is a wide temperature range around $T_{\min }$
where the thermal activated hopping still makes a small contribution to the
conductivity, but the uncertainty in the polaron band is already significant
\cite{fir,fir2}. The polaron transport theory requires a special
diagrammatic technique \cite{lan63,fir,fir2,lan68} and a
conditional-probability function description \cite{kudfir65,kudfir66} in
this region.

The transverse conductivity $\sigma _{xy}$ and the Hall coefficient $%
R_{H}=\sigma _{xy}/H\sigma _{xx}^{2}$ of lattice polarons can be calculated
with the Peierls substitution \cite{peierls},
\begin{equation}
t(\mathbf{m}-\mathbf{n})\rightarrow t(\mathbf{m}-\mathbf{n})e^{-ie\mathbf{A}(%
\mathbf{m})\cdot (\mathbf{m}-\mathbf{n})},
\end{equation}%
which is a fair approximation if the magnetic field, B, is weak compared
with the atomic field $eBa^{2}<<1$. Here $\mathbf{A}(\mathbf{r})$ is the
vector potential, which can be also time dependent. Within the Boltzmann
theory the sign of the Hall coefficient $R_{H}\simeq \pm 1/en$ depends on
the type of carriers (holes or electrons) and the Hall mobility $\mu
_{H}\equiv R_{H}\sigma _{xx}$ is the same as the drift mobility $\mu _{t}$
up to a numerical factor of the order of 1. The calculations of the hopping
Hall current similar to those of the hopping conductivity \cite%
{friedmanho,emin,bryksinfirsov,bryksinbook} show that the Hall mobility
depends on the symmetry of the crystal lattice and has nothing in common
with the hopping mobility, neither with respect to the temperature
dependence and even nor with respect to the sign. In particular for
hexagonal lattices three-site hops yield \cite{friedmanho}
\begin{equation}
\mu _{H}=ea^{2}{\frac{\pi ^{1/2}t}{{(12E_{a}T)^{1/2}}}}e^{-E_{a}/3T}
\end{equation}%
with the \emph{same sign} for electrons and holes. The activation energy of
the Hall mobility is three times less than that of the hopping mobility. In
cubic crystals, the hopping Hall effect is governed by four-site hops. The
four-site calculations in \cite{emin,bryksinfirsov} gave the Hall mobility
with the ``normal''\ sign depending on the
type of carriers.

The Hall conductivity and resistivity of strongly localized electrons at low
temperatures and small magnetic fields strongly depend on frequency, the
size of a sample \cite{entin1995}, and on a magnetic order like, for example
in ferromagnetic (Ga,Mn)As \cite{burkov}. In the presence of the spin-orbit
interaction each hopping path acquires a spin-dependent phase factor of the
same form as that in a perpendicular (to the 2D system) magnetic field,
which leads to spin accumulation and spin-Hall effects \cite%
{bryksinbook,entin2005}.

\subsection{Optical conductivity}

\label{optcond} One of the fingerprints of lattice polarons is the frequency
($\omega$) and temperature dependence of their mid-infrared (MIR)
conductivity $\sigma(\omega)$. In the low frequency and low temperature
region, where the tunneling band transport operates \cite{hola,holb}, the
conductivity acquires the conventional Drude form,
\begin{equation}
\sigma(\omega)={\frac{ne\mu_{t}}{{1+(\omega \tau)^{2}}}},
\end{equation}
where the transport relaxation time $\tau$ may be frequency dependent
because of the narrow band. For high (mid-infrared) frequencies, well above
the polaron bandwidth ($\omega\gg w$) but below the interband gap, the Drude
law is not applied. In this frequency region one can apply the generalised
Einstein relation $\sigma(\omega)=eD(\omega)/\omega$, where $%
D(\omega)=a^{2}W(\omega)$, and $W(\omega)$ is the hopping probability of the
absorption of the energy quantum $\omega$ \cite{holb}. The number of nearest
neighbour transitions per second with the absorption of a photon of the
energy $\omega$ is given by the Fermi-Dirac golden rule,
\begin{equation}
W^{-}=2\pi \langle\sum_{j}|\langle i|H_{p-ph}|j\rangle|^{2} \delta
\left(\sum_{\mathbf{q}}\omega_{\mathbf{q}}(n^{j}_{\mathbf{q}}-n^{i}_{\mathbf{%
q}})+\omega\right)\rangle,
\end{equation}
and with the emission,
\begin{equation}
W^{+}=2\pi \langle\sum_{j}|\langle i|H_{p-ph}|j\rangle|^{2} \delta
\left(\sum_{\mathbf{q}}\omega_{\mathbf{q}}(n^{j}_{\mathbf{q}}-n^{i}_{\mathbf{%
q}})-\omega\right)\rangle.
\end{equation}
As a result one obtains
\begin{eqnarray}
W(\omega)&=&W^{-}-W^{+}=2t^2e^{-2g^{2}}\sinh (\omega/2T)\cr &\times&
\int_{-\infty}^{\infty}d\tau e^{-i\omega \tau}\left[\exp \left({\frac{1}{{N}}%
}\sum_{\mathbf{q}}|\gamma(\mathbf{q})|^{2}[1-\cos(\mathbf{q}\cdot \mathbf{a}%
)]{\frac{\cos(\omega \tau)}{{\sinh\omega_{\mathbf{q}}/2T}}}\right)-1\right].
\end{eqnarray}
As in the case of the hopping mobility discussed above the integral over $%
\tau$ is calculated using the saddle-point approximation \cite{boettger:1985}%
,
\begin{equation}
\sigma(\omega)=\sigma_{h} {\frac{T \sinh(\omega/2T)}{{\omega}}}%
e^{\omega^{2}/4\delta},
\end{equation}
where
\begin{equation}
\delta={\frac{1}{{2N}}}\sum_{\mathbf{q}}|\gamma(\mathbf{q})|^{2}[1-\cos(%
\mathbf{q}\cdot \mathbf{a})]{\frac{\omega_{\mathbf{q}}^{2}}{{\sinh(\omega_{%
\mathbf{q}}/2T)}}}.  \nonumber
\end{equation}
For high temperatures, $T>>\omega_0/2$, this expression simplifies as
\begin{equation}
\sigma (\omega)=ne^{2}a^{2}{\frac{\pi^{1/2}t^2[1-e^{-\omega/T}]}{{2 \omega
(E_{a}T)^{1/2}}}} \exp\left[-{\frac{(\omega-4E_{a})^{2}}{{16E_{a}T}}}\right].
\label{mir}
\end{equation}
Here $\sigma_h=ne \mu_h$ and $n$ is the atomic density of polarons. The
frequency dependence of the MIR conductivity has a form of an asymmetric
Gaussian peak centered at $\omega=4E_{a}$ with the half-width $4\sqrt{E_{a}T}
$ \cite{eag,klinger1963,reik1963}. According to the Franck-Condon principle,
the position of the ions is not changed during an optical transition.
Therefore the frequency dependence of the MIR conductivity can be understood
in terms of transitions between the adiabatic levels of the two-site
Holstein model (\ref{adhol}). The polaron, say, in the left potential well
absorbs a photon through the vertical transition to the right well, where
the deformation is lacking, without any change in the molecular
configuration. The photon energy required to excite the polaron from the
bottom of the well is $\omega \simeq 2E_{p}$, which corresponds to the
maximum of $\sigma(\omega)$ in Eq.(\ref{mir}). The main contribution to the
absorption comes from the states near the bottom with the energy of the
order of $T$. The corresponding photon energies are found in the interval $%
2E_{p}\pm \sqrt{8E_{p}T}$ in agreement with Eq.(\ref{mir}). For low
temperatures $T<\omega_0/2$ the half-width of the MIR maximum is about $\sim
\sqrt{E_{p}\omega_0}$ rather than $\sim\sqrt{E_{p}T}$ \cite{emin1993}. The
optical absorption of small polarons is distinguished from that of large
polarons (\ref{comparison}) by the shape and the temperature dependence.
Their comparison \cite{emin1973,emin1993,calvani2001} shows a more
asymmetric and less temperature dependent mid-infrared (MIR) absorption of
large polarons compared with that of small polarons. The high-frequency
behavior of the optical absorption of small polarons is described by a
Gaussian decay \cite{reik1972}, Eq.(\ref{mir}), while for large polarons it
is much slower power law $\omega^{-5/2}$ \cite{Devreese72}.

Many materials are characterised by intermediate values of EPI, $\lambda\sim
1$, which requires an extension of the theory of optical absorption to the
crossover region from continuous to lattice polarons. The intermediate
coupling and frequency regime has been inaccessible for an analytical or
semi-analytical analysis, with an exception of infinite spatial dimensions,
where DMFT yields reliable results ~\cite{sumi,ciuchi,zeyher,fratini}. DMFT
treats the local dynamics exactly, but it cannot account for the spatial
correlations being important in real finite-dimensional systems.
Nevertheless DMFT allows one to address the intermediate coupling and
adiabaticity regimes in the absorption not covered by the standard small
polaron theory, where qualitatively new features arise. In particular, the
optical absorption exhibits a reentrant behavior, switching from
weak-coupling-like to polaronic-like upon increasing the temperature, and
sharp peaks with a nonmonotonic temperature dependence emerge at
characteristic phonon frequencies \cite{fratini2006}.

Earlier ED studies of the lattice polaron absorption were limited to small 2
to 10-site 1D and 2D clusters in the Holstein model ~\cite%
{alekabrayb,wellein98,Capone:1997,feh2}. The optical absorption occurs as
energy is transferred between the electromagnetic field and the phonons via
the charge carriers. The vibration energy must be capable of being
dissipated. Hence, using ED one has to introduce some continuous density of
phonon states, or a phonon lifetime, which makes MIR absorption to be finite
\cite{alekabrayb}. As a result one obtains a fair agreement between ED
absorption spectra and the analytical results, Eq.(\ref{mir}), in the strong
coupling limit as far as a smooth part of $\omega$ dependence is concerned.
The MIR conductivity occurs much more asymmetric in the intermediate
coupling region than in the strong coupling regime, and it shows an
additional oscillating superstructure corresponding to a different spectral
weight of the states with a different number of virtual phonons in the
polaron cloud.

More recent ED, VED and a kernel polynomial method (KPM) (for a review of
KPM see \cite{weisse2006}) allowed for numerical calculations of lattice
polaron properties in the Holstein model in the whole parameter range on
fairly large systems ~\cite{ElShawish2003,schubert2005,barisic}. The main
signature of optical conductivity for intermediate-to-strong EP coupling in
the adiabatic regime \cite{fehskebook} is that the spectrum is strongly
asymmetric, which is also characteristic for rather large polarons (\ref%
{comparison}), as observed in cuprate superconductors
\cite{Mih:1990}, perovskite tungsten bronzes \cite{ruscher} and many
other doped insulators. Importantly, the weaker decay at the
high-energy side meets
the experimental findings for many polaronic materials like $\mathrm{TiO_2}$~%
\cite{kudinov69} even better than standard small-polaron theory. At larger
EPI a more pronounced and symmetric maximum appears in the low-temperature
optical response. When the phonon frequency becomes comparable to the
electron transfer amplitude different absorption bands appear, which can be
classified according to the number of phonons involved in the optical
transition \cite{fehskebook}.

\subsection{Spectral function of discrete strong-coupling polarons}

\label{spectral} The polaron problem has the exact solution, Eq.(\ref{exact}%
), in the extreme infinite-coupling limit, $\lambda =\infty ,$ for any EPI
conserving the on-site occupation numbers of electrons. For a finite but
strong coupling $1/\lambda $ perturbation expansion is applied. Importantly,
the analytical perturbation theory becomes practically exact in a wider
range of the adiabatic parameter and of the coupling constant for the
long-range Fr\"{o}hlich interaction (\ref{lightSFP}).

Keeping this in mind, let us calculate the polaron spectral function in the
first order in $1/\lambda $ \cite{alexandrovranninger92,alexandrovchanun2000}%
. We can neglect $H_{p-ph}$ in the first-order if $1/\lambda \ll 1$. To
understand spectral properties of a single polaron we also neglect the
polaron-polaron interaction. Then the energy levels are
\begin{equation}
E_{\tilde{m}}=\sum_{_{\mathbf{k}}}\xi _{_{\mathbf{k}}}n_{\mathbf{k}}+\sum_{%
\mathbf{q}}\omega _{\mathbf{q}}[n_{\mathbf{q}}+1/2],
\end{equation}
where $\xi_{\mathbf{k}}$ is the small polaron band dispersion with respect
to the chemical potential, and the transformed eigenstates $\left| \tilde{m}%
\right\rangle $ are sorted by the polaron Bloch-state occupation numbers, $%
n_{\mathbf{k}}=0,1$, and the phonon occupation numbers, $n_{\mathbf{q}%
}=0,1,2,...,\infty .$

The spectral function of any system described by quantum numbers $m,n$ with
the eigenvalues $E_{n},E_{m}$ is defined as
\begin{equation}
A(\mathbf{k},\omega )\equiv \pi (1+e^{-\omega /T})e^{\Omega
/T}\sum_{n,m}e^{-E_{n}/T}\left\vert \left\langle n\right\vert c_{\mathbf{k}%
}\left\vert m\right\rangle \right\vert ^{2}\delta (\omega _{nm}+\omega ).
\end{equation}%
It is real and positive, $A(\mathbf{k},\omega )>0$, and obeys the sum rule
\begin{equation}
\frac{1}{\pi }\int_{-\infty }^{\infty }d\omega A(\mathbf{k},\omega )=1.
\label{sumrule2}
\end{equation}%
The matrix elements of the electron operators can be written as
\begin{equation}
\left\langle n\right\vert c_{\mathbf{k}}\left\vert m\right\rangle =\frac{1}{%
\sqrt{N}}\sum_{\mathbf{m}}e^{-i\mathbf{k\cdot m}}\left\langle \tilde{n}%
\right\vert c_{i}\hat{X}_{i}\left\vert \tilde{m}\right\rangle
\end{equation}%
using the Wannier representation and the Lang-Firsov transformation. Now,
applying the Fourier transform of the $\delta $-function, the spectral
function is expressed as
\begin{eqnarray}
A(\mathbf{k},\omega ) &=&\frac{1}{2}\int_{-\infty }^{\infty }dte^{i\omega t}%
\frac{1}{N}\sum_{\mathbf{m,n}}e^{i\mathbf{k\cdot (n-m)}}\times  \\
&&\left\{ \langle \langle c_{i}(t)\hat{X}_{i}(t)c_{j}^{\dagger }\hat{X}%
_{j}^{\dagger }\rangle \rangle +\langle \langle c_{j}^{\dagger }\hat{X}%
_{j}^{\dagger }c_{i}(t)\hat{X}_{i}(t)\rangle \rangle \right\} .  \nonumber
\end{eqnarray}%
Here the quantum and statistical averages are performed for independent
polarons and phonons, therefore $\langle \langle c_{i}(t)\hat{X}_{i}(t)\hat{X%
}_{j}^{\dagger }c_{i}^{\dagger }\rangle \rangle =\langle \langle
c_{i}(t)c_{j}^{\dagger }\rangle \rangle \langle \langle \hat{X}_{i}(t)\hat{X}%
_{j}^{\dagger }\rangle \rangle .$ The Heisenberg free-polaron operator
evolves with time as $c_{\mathbf{k}}(t)=c_{\mathbf{k}}\exp (-i\xi _{\mathbf{k%
}}t),$ so that
\begin{eqnarray}
\langle \langle c_{i}(t)c_{i}^{\dagger }\rangle \rangle  &=&\frac{1}{N}\sum_{%
\mathbf{k}^{\prime }\mathbf{,k}^{\prime \prime }}e^{i(\mathbf{k}^{\prime }%
\mathbf{\cdot m-k}^{\prime \prime }\mathbf{\cdot n)}}\langle \langle c_{%
\mathbf{k}^{\prime }}(t)c_{\mathbf{k}^{\prime \prime }}^{\dagger }\rangle
\rangle = \\
&&\frac{1}{N}\sum_{\mathbf{k}^{\prime }}[1-\bar{n}(\mathbf{k}^{\prime })]e^{i%
\mathbf{k}^{\prime }\mathbf{\cdot (m-n)-}i\xi _{\mathbf{k}^{\prime }}t},
\nonumber \\
\langle \langle c_{i}^{\dagger }c_{i}(t)\rangle \rangle  &=&\frac{1}{N}\sum_{%
\mathbf{k}^{\prime }}\bar{n}(\mathbf{k}^{\prime })e^{i\mathbf{k}^{\prime }%
\mathbf{\cdot (m-n)-}i\xi _{\mathbf{k}^{\prime }}t}
\end{eqnarray}%
where $\bar{n}(\mathbf{k)}=[1+\exp \xi _{\mathbf{k}}/T]^{-1}$ is the
Fermi-Dirac distribution function of polarons. The Heisenberg free-phonon
operator evolves in a similar way, $d_{\mathbf{q}}(t)=d_{\mathbf{q}}\exp
(-i\omega _{\mathbf{q}}t),$ so that
\begin{equation}
\langle \langle \hat{X}_{i}(t)\hat{X}_{j}^{\dagger }\rangle \rangle
=\prod_{_{\mathbf{q}}}\langle \langle \exp [u_{i}(\mathbf{q,}t)d_{\mathbf{q}%
}-H.c.]\exp [-u_{j}(\mathbf{q})d_{\mathbf{q}}-H.c.]\rangle \rangle ,
\end{equation}%
where $u_{i,j}(\mathbf{q,}t)=u_{i,j}(\mathbf{q})\exp (-i\omega _{\mathbf{q}%
}t).$ This average is calculated using the operator identity, as in Eq.(\ref%
{identity}),
\begin{equation}
\langle \langle \hat{X}_{i}(t)\hat{X}_{j}^{\dagger }\rangle \rangle =\exp
\left\{ -\frac{1}{2N}\sum_{\mathbf{q}}|\gamma (\mathbf{q})|^{2}f_{\mathbf{q}%
}(\mathbf{m-n,}t)\right\} ,  \label{cor}
\end{equation}%
where
\begin{equation}
f_{\mathbf{q}}(\mathbf{m,}t)=[1-\cos (\mathbf{q\cdot m)}\cos (\omega _{%
\mathbf{q}}t)]\coth \frac{\omega _{\mathbf{q}}}{2T}+i\cos (\mathbf{q\cdot m)}%
\sin (\omega _{\mathbf{q}}t).
\end{equation}%
Here we used the symmetry of $\gamma (-\mathbf{q})=\gamma (\mathbf{q})$, so
that terms containing $\sin (\mathbf{q\cdot m)}$ have disappeared.

To proceed with the analytical results we consider low temperatures, $T\ll
\omega _{\mathbf{q}},$ when $\coth (\omega _{\mathbf{q}}/2T)\approx 1.$
Expanding the exponent in Eq.(\ref{cor}) and performing summation over $%
\mathbf{m}, \mathbf{n}$, $\mathbf{k}^{\prime }$ and integration over time we
arrive at \cite{alexandrovchanun2000}
\begin{equation}
A(\mathbf{k},\omega )=\sum_{l=0}^{\infty }\left[ A_{l}^{(-)}(\mathbf{k}%
,\omega )+A_{l}^{(+)}(\mathbf{k},\omega )\right] ,  \label{spectr2}
\end{equation}
where
\begin{eqnarray}
A_{l}^{(-)}(\mathbf{k},\omega ) &=&\pi Z\sum_{\mathbf{q}_{1},...\mathbf{q}%
_{l}}{\frac{\prod_{r=1}^{l}|\gamma (\mathbf{q}_{r})|^{2}}{{(2N)^{l}l!}}%
\times } \\
&&\left[ 1-\bar{n}\left( {\mathbf{k-}\sum_{r=1}^{l}\mathbf{q}_{r}}\right) %
\right] {\delta }\left( {\omega -}\sum_{r=1}^{l}\omega _{\mathbf{q}_{r}}{%
-\xi }_{{\mathbf{k-}\sum_{r=1}^{l}\mathbf{q}_{r}}}\right) ,  \nonumber
\end{eqnarray}

\begin{eqnarray}
A_{l}^{(+)}(\mathbf{k},\omega ) &=&\pi Z\sum_{\mathbf{q}_{1},...\mathbf{q}%
_{l}}{\frac{\prod_{r=1}^{l}|\gamma (\mathbf{q}_{r})|^{2}}{{(2N)^{l}l!}}%
\times } \\
&&\bar{n}\left( {\mathbf{k+}\sum_{r=1}^{l}\mathbf{q}_{r}}\right) {\delta }%
\left( {\omega +}\sum_{r=1}^{l}\omega _{\mathbf{q}_{r}}{-\xi }_{{\mathbf{k+}%
\sum_{r=1}^{l}\mathbf{q}_{r}}}\right) ,  \nonumber
\end{eqnarray}
and $Z=\exp ( -(2N)^{-1}\sum_{\mathbf{q}}|\gamma (\mathbf{q})|^{2}).$

Clearly Eq.(\ref{spectr2}) is in the form of the perturbative multi-phonon
expansion. Each contribution $A_{l}^{(\pm )}(\mathbf{k},\omega )$ to the
spectral function describes the transition from the initial state $\mathbf{k}
$ of the polaron band to the final state $\mathbf{k}\mathbf{\pm }$ ${%
\sum_{r=1}^{l}\mathbf{q}_{r}}$ with the emission (or absorption) of $l$
phonons.

The $1/\lambda $ expansion result, Eq.(\ref{spectr2}), is different from the
conventional spectral function of metallic electrons coupled to phonons in
the Migdal-Eliashberg theory \cite{mig,eli}. There is no imaginary part of
the self-energy since the exponentially small (at low temperatures)
polaronic damping, Eq.(\ref{damping}), is neglected. Instead EPI leads to
the coherent dressing of electrons by phonons, and phonon ``side-bands''. The spectral function of the polaronic
carriers comprises two different parts. The first ($l=0$) $\mathbf{k}$%
-dependent \textit{coherent} term arises from the polaron band tunnelling,
\begin{equation}
A_{coh}(\mathbf{k},\omega )=\left[ A_{0}^{(-)}(\mathbf{k},\omega
)+A_{0}^{(+)}(\mathbf{k},\omega )\right] =\pi {Z\delta (\omega -\xi }_{%
\mathbf{k}}{)}.
\end{equation}%
The spectral weight of the coherent part is suppressed as $Z\ll 1.$ However
in the case of the Fr\"{o}hlich interaction the effective mass is less
enhanced, since the band narrowing factor $Z^{\prime }$, Eq.(\ref{z}), in $%
\xi _{\mathbf{k}}=Z^{\prime }E_{\mathbf{k}}-\mu $ is large compared with $Z$
\cite{alexandrovchanun2000}.

The second \textit{incoherent} part $A_{incoh}(\mathbf{k},\omega )$
comprises all the terms with $l\geq 1.$ It describes excitations accompanied
by emission and absorption of phonons. We notice that its spectral density
spreads over a wide energy range of about twice the polaron level shift $%
E_{p}$, which might be larger than the unrenormalised bandwidth $2D$ in the
rigid lattice without phonons. On the contrary, the coherent part shows a
dispersion only in the energy window of the order of the polaron bandwidth, $%
w=Z^{\prime }D$. It is interesting that there is some $\mathbf{k}$
dependence of the $incoherent$ background as well, if the EPI matrix element
and/or phonon frequencies depend on $\mathbf{q}$. Only in the Holstein model
with the short-range dispersionless e-ph interaction $\gamma (\mathbf{q)=}%
\gamma _{0}$ and $\omega _{\mathbf{q}}=\omega _{0}$ the incoherent part is
momentum independent \cite{mahan:1990},
\begin{eqnarray}
A_{incoh}(\mathbf{k},\omega ) &=&\pi \frac{{Z}}{N}\sum_{l=1}^{\infty }{\frac{%
\gamma _{0}^{2l}}{{2}^{l}{l!}}\times } \\
&&\sum_{\mathbf{k}^{\prime }}\left\{ \left[ 1-\bar{n}\left( \mathbf{k}%
^{\prime }\right) \right] {\delta }\left( {\omega -}l\omega _{0}{-\xi }_{%
\mathbf{k}^{\prime }}\right) +\bar{n}\left( \mathbf{k}^{\prime }\right) {%
\delta }\left( {\omega +}l\omega _{0}{-\xi }_{\mathbf{k}^{\prime }}\right)
\right\} .  \nonumber
\end{eqnarray}

As soon as we know the spectral function, different Green's functions (GF)
are readily obtained using their analytical properties. For example, the
temperature GF is given by the integral
\begin{equation}
\mathcal{G}(\mathbf{k},\omega _{k})=\frac{1}{\pi }\int_{-\infty }^{\infty
}d\omega ^{\prime }\frac{A(\mathbf{k},\omega ^{\prime })}{i\omega
_{k}-\omega ^{\prime }}.
\end{equation}
where $\omega _{k}=\pi T(2k+1),k=0,\pm 1,\pm 2,...$. Calculating the
integral we find in the Holstein model \cite{alexandrovranninger92}
\begin{equation}
\mathcal{G}(\mathbf{k},\omega _{n})=\frac{Z}{i{\omega }_{n}{-\xi }_{\mathbf{k%
}}}+\frac{{Z}}{N}\sum_{l=1}^{\infty }{\frac{\gamma _{0}^{2l}}{{2}^{l}{l!}}}%
\sum_{\mathbf{k}^{\prime }}\left\{ \frac{1-\bar{n}\left( \mathbf{k}^{\prime
}\right) }{i{\omega }_{n}{-}l\omega _{0}{-\xi }_{\mathbf{k}^{\prime }}}+%
\frac{\bar{n}\left( \mathbf{k}^{\prime }\right) }{i{\omega }_{n}{+}l\omega
_{0}{-\xi }_{\mathbf{k}^{\prime }}}\right\} .  \label{GFHol}
\end{equation}
Here the first term describes the coherent tunnelling in the narrow polaron
band while the second $\mathbf{k}$-independent sum describes the phonon-side
bands.

\subsection{Spectral function of discrete all-coupling polarons}

The spectral function, Eq.(\ref{GFHol}), satisfies the major sum rule, Eq.(%
\ref{sumrule2}). However the higher-momentum integrals, $\int_{-\infty
}^{\infty }d\omega \omega ^{p}A(\mathbf{k},\omega )$ with $p>0,$ calculated
using Eq.(\ref{GFHol}), differ from the exact values \cite{kornsum} by an
amount proportional to $1/\lambda .$ The difference is due to a partial
``undressing''\ of high-energy excitations
in the side-bands, which is beyond the first order $1/\lambda $ expansion. A
rather accurate Green's function of the Holstein polaron has been recently
obtained by summing all the diagrams, but with each diagram averaged over
its free propagators' momenta \cite{berciu2006a,berciu2006b}. The resulting
Green's function satisfies exactly the first six spectral weight sum rules.
All higher sum rules are satisfied with great accuracy, becoming
asymptotically exact for coupling both much larger and much smaller than the
free particle bandwidth.

The spectral properties of the Holstein model in a wider parameter range
have been studied numerically using ED (see \cite%
{ran0,alekabrayb,wellein98,Capone:1997,feh2,ElShawish2003,schubert2005,fehskebook,loos}%
, and references therein). At the weak EPI, the electronic spectrum of 1D
Holstein model is nearly unaffected for energies below the phonon emission
threshold \cite{fehskebook}. Hence, the renormalised dispersion $\epsilon_k$
nearly coincides with the tight-binding cosine band (shifted $\propto E_p$)
up to some $k_X$, where the phonon energy intersects the bare electron band
with the familiar flattening effect (\ref{isotope}). Reaching the
intermediate EPI (polaron crossover) regime a coherent band separates from
the rest of the spectrum [$k_X\to\pi$]. At the same time its spectral weight
becomes smaller and will be transferred to the incoherent part, where
several sub-bands emerge. In the strong-coupling case the coherent
quasi-particle absorption band becomes extremely narrow and its bandwidth
approaches the strong-coupling result. The incoherent part of the spectrum
carries most of the spectral weight consisting of a sequence of sub-bands
separated in energy by $\omega_0$, in agreement with the analytical results.

Effects of the finite-range EPI on the spectral properties of lattice
polarons have been studied numerically by \cite{fehske2002} using exact
Lanczos diagonalisation method in the framework of the 1D model of \cite%
{alexandrov:1999}. The polaron band-structure has been calculated in
agreement with the analytical and CTQMC results \ref{lightSFP}. The optical
absorption of lattice polarons with a finite-range (Fr\"ohlich-type) EPI has
been found similar to the continuous-polaron absorption (\ref{comparison})
for all EPI strengths. The polaron features due to EPIs with different
coupling ranges were also investigated in the framework of the variational
approach \cite{cataudella2005}. The ground-state energy, the spectral
weight, the average kinetic energy, the mean number of phonons, and the
electron-lattice correlation function were calculated for the system with
coupling to local and nearest-neighbor lattice displacements and compared
with the long-range case. As in \cite{alexandrovchanun2000} a substantially
different mass renormalisation compared with the coherent weight reduction, $%
Z \ll Z^{\prime }$, was found for the finite-range EPI.

\section{Bipolaron}

\subsection{Polaron-polaron interaction}

Polarons interact with each other, cf. Eq.(\ref{interaction}) for small
polarons. The range of the deformation surrounding the Fr\"{o}hlich polarons
is quite large, and their deformation fields are overlapped at finite
density. Taking into account both the long-range attraction of polarons
owing to the lattice deformations $and$ their direct Coulomb repulsion, the
residual \textit{long-range} interaction turns out rather weak and repulsive
in ionic crystals \cite{alexandrov:1995}. In the long-wave limit ($q\ll \pi
/a$) the Fr\"{o}hlich EPI dominates in the attractive part, but polarons
repel each other at large distances, $|\mathbf{m-n}|\gg a$,
\begin{equation}
v(\mathbf{m-n)}={\frac{e^{2}}{{\varepsilon _{0}|\mathbf{m-n}|}}>0}.
\end{equation}%
The Fr\"{o}hlich EPI nearly nullifies the bare Coulomb repulsion, if $%
\varepsilon _{0}\gg 1$, but cannot overscreen it at large distances.

Considering the polaron-phonon interaction in the multi-polaron system we
have to take into account the dynamic properties of the polaron response
function \cite{aledyn}. One may erroneously believe that the long-range Fr%
\"{o}hlich EPI becomes a short-range (Holstein) one due to screening of
ionic potentials by heavy polaronic carriers. In fact, as distinct from
large polarons, small polarons cannot screen high-frequency optical
vibrations because their renormalised plasma frequency is comparable with or
even less than the phonon frequency in the strong-coupling regime \cite{ale2}%
. The small-polaron plasma frequency is rather low due to the large static
dielectric constant, $\varepsilon _{0}\gg 1,$ in ionic lattices and the
enhanced polaron mass $m^{\ast }\gg m$. If ${\omega }_{0}>\omega _{p}$, the
singular behaviour of the Fr\"{o}hlich EPI, $\gamma (\mathbf{q})\sim 1/q$,
is unaffected by screening. Polarons are too slow to screen high-frequency
crystal field oscillations. As a result, EPI with high-frequency optical
phonons in ionic solids remains unscreened at any density of polarons.

Another important issue is a possibility of the Wigner crystallization of
polarons. Because the net long-range interaction is relatively weak, a
relevant dimensionless parameter $r_{s}=m^{\ast }ae^{2}/\varepsilon
_{0}(4\pi n/3)^{1/3}$ is not very large in ionic semiconductors. The Wigner
crystallization appears around $r_{s}\simeq 100$ or larger, which
corresponds to the atomic density of polarons $n\leq 10^{-6}$ with $%
\varepsilon _{0}=30$ and $m^{\ast }=5m$. This estimate tells us that
sufficiently mobile small polarons can be in a liquid state \cite{aledyn} at
substantial doping levels, however they can be crystallised at low doping
(see \ref{gas}).

When the short-range deformation and molecular-type (i.e. Holstein) EPIs are
added to the Fr\"{o}hlich interaction, two polarons attract each other at a
short distance of about the lattice constant. Then, owing to a narrow band,
two lattice polarons easily form a local tightly bound state, i.e. a $small$
bipolaron \cite%
{beni1974,anderson1975,mott1975,aleran1981,aleran1981b,aub1993}. One can
estimate the coupling constant $\lambda $ and the adiabatic ratio $\omega
_{0}/t,$ at which the small bipolaron ``instability''\ occurs \cite{aledyn}. The characteristic
attractive potential is $|v|=D/(\lambda -\mu _{c})$, where $\mu _{c}$ is the
dimensionless Coulomb repulsion, and $\lambda $ includes the interaction
with all phonon modes. The radius of the potential is about $a$. In three
dimensions a bound state of two attractive particles appears, if $|v|\geq
\pi ^{2}/8m^{\ast }a^{2}$. Substituting the polaron mass, $m^{\ast
}=[2a^{2}t]^{-1}\exp (\gamma \lambda D/\omega _{0})$, we find
\begin{equation}
\frac{t}{{\omega }_{0}}\leq (\gamma z\lambda )^{-1}\ln \left[ {\frac{\pi ^{2}%
}{{4z(\lambda -\mu }_{c}{)}}}\right] .
\end{equation}%
As a result, small bipolarons form at $\lambda \geq \mu _{c}+\pi ^{2}/4z$,
which is almost independent of the adiabatic ratio.

The formation of small bipolarons is closely related to a negative
$U$-center model \cite{mott1975}. Starting from initial works by
Eagles \cite{eag2} and Legget \cite{leg} this model received
particular attention in connection with high-temperature
superconductivity \cite{micr}. However, in using the negative
Hubbard $U$ model, one has to realize that this model, which
predicts a smooth crossover from Cooper pairs to real-space pairs
\cite{noz}, cannot account for the polaron-bipolaron crossover with
the increasing EPI. The essential effect of the polaron
band-narrowing, which is responsible for high critical temperatures
in the model of polaronic superconductors \cite{ale0}, is missing in
the negative Hubbard $U$ model.

\subsection{ Holstein bipolaron}

\label{mobile}

The attractive energy of two small polarons is generally much larger than
the polaron bandwidth, which allows for a consistent treatment of small
bipolarons \cite{ale0,aleran1981,aleran1981b}. Under this condition the
hopping term in the transformed Hamiltonian $\tilde{H}$ is a small
perturbation of the ground state of immobile bipolarons and free phonons,
\begin{equation}
\tilde{H}=H_{0}+H_{pert},
\end{equation}
where
\begin{equation}
H_{0}={\frac{1}{{2}}}\sum_{i,j}v_{ij}c_{i}^{\dagger }c_{j}^{\dagger
}c_{j}c_{i}+\sum_{\mathbf{q,\nu }}\omega _{\mathbf{q\nu }}[d_{\mathbf{q\nu }%
}^{\dagger }d_{\mathbf{q\nu }}+1/2]
\end{equation}
and
\begin{equation}
H_{pert}=\sum_{i,j}\hat{\sigma}_{ij}c_{i}^{\dagger }c_{j}  \label{pertub}
\end{equation}

Let us first discuss the dynamics of \textit{onsite }\ bipolarons, which are
the ground state of the system with the Holstein non-dispersive EPI \cite%
{beni1974,anderson1975,mott1975,aleran1981,aub1993,aub1995}. The
onsite bipolaron is formed if $2E_{p}>U$, where $U$ is the onsite
Coulomb correlation energy (the Hubbard $U$). The intersite
polaron-polaron interaction, Eq.(\ref{interaction}), is just the
Coulomb repulsion since the phonon mediated attraction between two
polarons on different sites is zero in the Holstein model. Two or
more onsite bipolarons as well as three or more polarons cannot
occupy the same site because of the Pauli exclusion principle.
Hence, bipolarons repel single polarons and each other. Their
binding energy, $\Delta =2E_{p}-U,$ is larger than the polaron
half-bandwidth, $\Delta \gg w,$ so that there are no unbound
polarons in the ground state. $H_{pert}$, Eq.(\ref{pertub}),
destroys bipolarons in the first order. Hence it has no diagonal
matrix elements. Then the bipolaron dynamics is described by the use
of another canonical transformation $\exp (S_{2})$
\cite{aleran1981}, which eliminates the first order of $H_{pert}$,
\begin{equation}
(S_{2})_{fp}=\sum_{i,j}{\frac{\langle f|\hat{\sigma}_{ij}c_{i}^{\dagger
}c_{j}|p\rangle }{{E_{f}-E_{p}}}}.  \label{s2}
\end{equation}
Here $E_{f,p}$ and $|f\rangle ,|p\rangle $ are the energy levels and the
eigenstates of $H_{0}$. Neglecting the terms of the order higher than $%
(w/\Delta )^{2}$ one obtains
\begin{equation}
(H_{b})_{ff^{\prime }}\equiv \left( e^{S_{2}}\tilde{H}e^{-S_{2}}\right)
_{ff^{\prime }},
\end{equation}
\begin{eqnarray*}
(H_{b})_{ff^{\prime }} &\approx &(H_{0})_{ff^{\prime }}-{\frac{1}{{2}}}%
\sum_{\nu }\sum_{i\neq i^{\prime },j\neq j^{\prime }}\langle f|\hat{\sigma}%
_{ii^{\prime }}c_{i}^{\dagger }c_{i^{\prime }}|p\rangle \langle p|\hat{\sigma%
}_{jj^{\prime }}c_{j}^{\dagger }c_{j^{\prime }}|f^{\prime }\rangle \times \\
&&\left( {\frac{1}{{E_{p}-E_{f^{\prime }}}}}+{\frac{1}{{E_{p}-E_{f}}}}%
\right) .  \label{bipham}
\end{eqnarray*}
The \textit{\ bipolaronic} Hamiltonian $H_{b}$ is defined in the subspace $%
|f\rangle ,|f^{\prime }\rangle $ with no single (unbound) polarons. On the
other hand, the intermediate \textit{bra} $\langle p|$ and \textit{ket} $%
|p\rangle $ refer to configurations involving two unpaired polarons and any
number of phonons. Hence we have
\begin{equation}
E_{p}-E_{f}=\Delta +\sum_{\mathbf{q,\nu }}\omega _{\mathbf{q\nu }}\left( n_{%
\mathbf{q\nu }}^{p}-n_{\mathbf{q\nu }}^{f}\right) ,
\end{equation}
where $n_{\mathbf{q\nu }}^{f,p}$ are the phonon occupation numbers $%
(0,1,2,3...\infty )$.

The lowest eigenstates of $H_{b}$ are in the subspace, which has only doubly
occupied $c_{\mathbf{m}s}^{\dagger }c_{\mathbf{m}s^{\prime }}^{\dagger
}|0\rangle $ or empty $|0\rangle $ sites. On-site bipolaron tunnelling is a
two-step transition. It takes place via a single polaron tunneling to a
neighbouring site. The subsequent tunnelling of its ``partner''\ to the same site restores the initial energy
state of the system. There are no $real$ phonons emitted or absorbed because
the (bi)polaron band is narrow. Hence we can average $H_{b}$ with respect to
phonons,
\begin{eqnarray}
H_{b} &=&H_{0}-i\sum_{\mathbf{m\neq m}^{\prime },s}\sum_{\mathbf{n\neq n}%
^{\prime },s^{\prime }}t(\mathbf{m-m}^{\prime })t(\mathbf{n-n}^{\prime
})\times  \\
&&c_{\mathbf{m}s}^{\dagger }c_{\mathbf{m}^{\prime }s}c_{\mathbf{n}s^{\prime
}}^{\dagger }c_{\mathbf{n}^{\prime }s^{\prime }}\int_{0}^{\infty
}dte^{-i\Delta t}\Phi _{\mathbf{mm}^{\prime }}^{\mathbf{nn}^{\prime }}(t).
\nonumber
\end{eqnarray}%
Here $\Phi _{\mathbf{mm}^{\prime }}^{\mathbf{nn}^{\prime }}(t)$ is a
multiphonon correlator,
\begin{equation}
\Phi _{\mathbf{mm}^{\prime }}^{\mathbf{nn}^{\prime }}(t)\equiv \langle
\langle \hat{X}_{i}^{\dagger }(t)\hat{X}_{i^{\prime }}(t)\hat{X}%
_{j}^{\dagger }\hat{X}_{j^{\prime }}\rangle \rangle ,
\end{equation}%
which can be readily calculated as \cite{ale2}
\begin{eqnarray}
\Phi _{\mathbf{mm}^{\prime }}^{\mathbf{nn}^{\prime }}(t) &=&e^{-g^{2}(%
\mathbf{m-m}^{\prime })}e^{-g^{2}(\mathbf{n-n}^{\prime })}\times  \\
&&\exp \left\{ \frac{1}{2N}\sum_{\mathbf{q,\nu }}|\gamma (\mathbf{q,}\nu
)|^{2}F_{\mathbf{q}}(\mathbf{m,m}^{\prime },\mathbf{n,n}^{\prime })\frac{%
\cosh \left[ \omega _{\mathbf{q\nu }}\left( \frac{1}{2T}-it\right) \right] }{%
\sinh \left[ \frac{\omega _{\mathbf{q\nu }}}{2T}\right] }\right\} ,
\nonumber
\end{eqnarray}%
where $F_{\mathbf{q}}(\mathbf{m,m}^{\prime },\mathbf{n,n}^{\prime })=\cos [%
\mathbf{q\cdot (n}^{\prime }-\mathbf{m})]+\cos [\mathbf{q\cdot (n}-\mathbf{m}%
^{\prime })]-\cos [\mathbf{q\cdot (n}^{\prime }-\mathbf{m}^{\prime })]-\cos [%
\mathbf{q\cdot (n}-\mathbf{m})].$

Taking into account that there are only bipolarons in the subspace, where $%
H_{b}$ operates, one can rewrite the Hamiltonian in terms of the creation $%
b_{\mathbf{m}}^{\dagger }=c_{\mathbf{m}\uparrow }^{\dagger }c_{\mathbf{m}%
\downarrow }^{\dagger }$ and annihilation $b_{\mathbf{m}}=c_{\mathbf{m}%
\downarrow }c_{\mathbf{m}\uparrow }$ operators of singlet pairs \cite%
{aleran1981}:
\begin{eqnarray}
H_{b} &=&-\sum_{\mathbf{m}}\left[ \Delta +{\frac{1}{{2}}}\sum_{\mathbf{%
m^{\prime }}}v^{(2)}(\mathbf{m-m}^{\prime })\right] n_{\mathbf{m}}+ \\
&&\sum_{\mathbf{m\neq m^{\prime }}}\left[ t_{b}(\mathbf{m-m}^{\prime })b_{%
\mathbf{m}}^{\dagger }b_{\mathbf{m^{\prime }}}+{\frac{1}{{2}}}\bar{v}(%
\mathbf{m-m}^{\prime })n_{\mathbf{m}}n_{\mathbf{m^{\prime }}}\right] .
\nonumber  \label{biham2}
\end{eqnarray}%
Here $n_{\mathbf{m}}=b_{\mathbf{m}}^{\dagger }b_{\mathbf{m}}$ is the
bipolaron site-occupation operator, $\bar{v}(\mathbf{m-m}^{\prime })=4v(%
\mathbf{m-m}^{\prime })+v^{(2)}(\mathbf{m-m}^{\prime })$ is the
bipolaron-bipolaron interaction including the direct polaron-polaron
interaction $v(\mathbf{m-m}^{\prime })$ and a repulsive correlation energy,
\begin{equation}
v^{(2)}(\mathbf{m-m}^{\prime })=2i\int_{0}^{\infty }d\tau e^{-i\Delta \tau
}\Phi _{\mathbf{mm^{\prime }}}^{\mathbf{m^{\prime }m}}(\tau ).
\end{equation}%
This additional repulsion appears since a virtual hop of one of two polarons
of the pair is forbidden, if the neighbouring site is occupied by another
pair. The bipolaron transfer integral, $t_{b}$ is of the second order in $t(%
\mathbf{m})$
\begin{equation}
t_{b}(\mathbf{m-m}^{\prime })=-2it^{2}(\mathbf{m-m}^{\prime
})\int_{0}^{\infty }d\tau e^{-i\Delta \tau }\Phi _{\mathbf{mm^{\prime }}}^{%
\mathbf{mm^{\prime }}}(\tau ).  \label{tb}
\end{equation}%
The multiphonon correlator is simplified for dispersionless phonons at $T\ll
\omega _{0}$ as
\begin{eqnarray*}
\Phi _{\mathbf{mm^{\prime }}}^{\mathbf{mm}^{\prime }}(t) &=&e^{-2g^{2}(%
\mathbf{m-m^{\prime })}}\exp \left[ -2g^{2}(\mathbf{m-m^{\prime })}%
e^{-i\omega _{0}t}\right] , \\
\Phi _{\mathbf{mm^{\prime }}}^{\mathbf{m}^{\prime }\mathbf{m}}(t)
&=&e^{-2g^{2}(\mathbf{m-m^{\prime })}}\exp \left[ 2g^{2}(\mathbf{m-m^{\prime
})}e^{-i\omega _{0}t}\right] ,
\end{eqnarray*}%
which yields \cite{alekab0}
\begin{equation}
t(\mathbf{m)}=-{\frac{2t^{2}(\mathbf{m})}{{\Delta }}}e^{-2g^{2}(\mathbf{m)}%
}\sum_{l=0}^{\infty }\frac{[-2g^{2}(\mathbf{m)}]^{l}}{l{!(1+l\omega
_{0}/\Delta )}}  \label{bmass}
\end{equation}%
and
\begin{equation}
v^{(2)}(\mathbf{m})={\frac{2t^{2}(\mathbf{m})}{{\Delta }}}e^{-2g^{2}(\mathbf{%
m)}}\sum_{l=0}^{\infty }\frac{[2g^{2}(\mathbf{m)}]^{l}}{l{!(1+l\omega
_{0}/\Delta )}}.  \label{rep}
\end{equation}%
When $\Delta \ll \omega _{0},$ we can keep the first term only with $l=0$ in
the bipolaron hopping integral, Eq.(\ref{tb}). In this case the bipolaron
half-bandwidth $zt(\mathbf{a)}$ is of the order of $2w^{2}/(z\Delta )$.
However, if the bipolaron binding energy is large, $\Delta \gg \omega _{0},$
the bipolaron bandwidth dramatically decreases proportionally to $e^{-4g^{2}}
$ in the limit $\Delta \rightarrow \infty $. This limit is not realistic
since $\Delta =2E_{p}-V_{c}<2g^{2}\omega _{0}$. In a more realistic regime, $%
\omega _{0}<\Delta <2g^{2}\omega _{0}$, Eq.(\ref{tb}) yields \cite{alekab0}
\begin{equation}
t_{b}(\mathbf{m)}\approx {\frac{2\sqrt{2\pi }t^{2}(\mathbf{m})}{\sqrt{\omega
_{0}\Delta }}}\exp \left[ -2g^{2}-{\frac{\Delta }{{\omega }_{0}}}\left(
1+\ln \frac{{2g^{2}(\mathbf{m)}\omega _{0}}}{\Delta }\right) \right] .
\end{equation}%
On the contrary, the bipolaron-bipolaron repulsion, Eq.(\ref{rep}), has no
small exponent in the limit $\Delta \rightarrow \infty $, $v^{(2)}\propto
D^{2}/\Delta .$ Together with the direct Coulomb repulsion the second order $%
v^{(2)}$ ensures stability of bipolarons against clustering.

Interestingly the high-temperature behavior of the bipolaron bandwidth is
just opposite to that of the small polaron bandwidth. While the polaron band
collapses with increasing temperature (\ref{collapse}), the bipolaron band
becomes wider \cite{bryksin1988},
\begin{equation}
t_b(\mathbf{m)}\propto \frac{1}{\sqrt{T}}\exp \left[ -{\frac{E_{p}+\Delta }{{%
2T}}}\right]
\end{equation}
for $T>\omega _{0}$.

The hopping conductivity of strong-coupling on-site bipolarons in the
Holstein-Hubbard model (HHM) was found small compared with the hopping
conductivity of thermally excited single polarons \cite{bryksin1988}.
However, as the frequency of the electric field increases, the dominant role
in the optical absorption is gradually transferred to bipolarons at low
temperatures. Like in the single-polaron case the bipolaron optical
absorption can be estimated using the Franck-Condon principle which states
that optical transitions take place without any change in the nuclear
configuration. The corresponding analysis \cite{bryksin1984} shows that the
absorption coefficient of light by the on-site bipolaron has three Gaussian
peaks located at frequencies $\Omega =4E_{a},8E_{a}-U$ and $16E_{a}$. The
lowest peak corresponds to the absorption by single thermally excited
polarons. The highest peak is due to the shakeoff of phonons without
dissociation of the bipolaron while the main central peak is the absorption
involving dissociation. A generalisation of the Franck-Condon principle for
the optical absorption of intersite bipolarons with a finite-range EPI was
given by \cite{alebraabs}.

The optical absorption and single-particle spectral functions of the
bipolaron in 1D HHM in the whole range of parameters were calculated using
ED on a 2-site cluster with two electrons \cite{alekabrayb}, and more
recently in \cite{hohenadler2005} using the cluster perturbation theory
(CPT). The latter allowed one to calculate the spectrum at continuous wave
vectors and to find pronounced deviations (e.g. noncosine dispersions) of
the bipolaron band structure from a simple tight-binding band due to an
important contribution from the next-nearest-neighbour hoppings.

Treating phonons classically in the extreme adiabatic limit \cite%
{aub1993,aub1995,aub,proville,aubbook} found along with the onsite bipolaron
($S0$) also an anisotropic pair of polarons lying on two neighboring sites,
i.e. the $intersite$ bipolaron ($S1$). Such bipolarons were originally
hypothesized in \cite{alexandrov:1991} to explain the anomalous nuclear
magnetic relaxation (NMR) in cuprate superconductors. The intersite
bipolaron could take a form of a ``quadrisinglet''\ ($QS$) in 2D HHM, where the electron
density at the central site is 1 and ``1/4''\ on the four nearest neigbouring sites. In a certain
region of $U$, where $QS$ is the ground state, the double-well potential
barrier, which usually pins polarons and bipolarons to the lattice,
depresses to almost zero, so that adiabatic lattice bipolarons can be rather
mobile (see \cite{aubbook} for more details).

\cite{sil} investigated the stability of the bipolaronic phase in HHM using
a modified Lang-Firsov variational transformation with on-site and
nearest-neighbour lattice distortions. There is a critical on-site Hubbard
repulsion $U_c$ below which the bipolaronic phases are stable for a fixed
electron-phonon coupling. In the absence of on-site repulsion, bipolaronic
phases are stable over the entire range of electron-phonon coupling for one
dimension, whereas there is a critical electron-phonon coupling for
formation of a stable bipolaron in two and three dimensions. Mobile $S1$
bipolarons were found in 1D HHM using variational methods also in the non-
and near-adiabatic regimes with dynamical quantum phonons \cite%
{lamagna,trugman}. The intersite bipolaron with a relatively small effective
mass is stable in a wide region of the parameters of HHM due to both
exchange and nonadiabaticity effects \cite{lamagna}. Near the strong
coupling limit the mobile $S1$ bipolaron has an effective mass of the order
of a single Holstein polaron mass, so that one should not rule out the
possibility of a superconducting state of $S1$ bipolarons with $s$ or $d$%
-wave symmetry in HHM \cite{trugman}. The recent diagrammatic Monte Carlo
study \cite{Macridin} found $S1$ bipolarons for large $U$ at intermediate
and large EPI and established the phase diagram of 2D HHM, comprising large
and small unbound polarons, $S0$ and $S1$ domains. Ref. \cite{Macridin}
emphasised that the transition to the bound state and the properties of the
bipolaron in HHM are very different from bound states in the attractive
(negative $U$) Hubbard model without EPI \cite{micnas1981}.

The two-dimensional many-body HHM was examined within a fluctuation-based
effective cumulant approach by \cite{hakioglu} confirming that the numerical
results on systems with finite degrees of freedom (\ref{all}) can be
qualitatively extended to the systems with large degrees of freedom. When
the electron-electron repulsion $U$ is dominant, the transition is to a Mott
insulator; when EPI dominates, the transition is to a bipolaronic state. In
the former case, the transition was found to be second order in contrast to
the transition to the bipolaronic state, which is first order for larger
values of $U$ \cite{kollerb}.

For a very strong electron-phonon coupling polarons become self-trapped on a
single lattice site and bipolarons are on-site singlets. A finite bipolaron
mass appears only in the second order of polaron hopping, Eq.(\ref{tb}), so
that on-site bipolarons might be very heavy in the Holstein model with the
short-range EPI. As a result the model led some authors to the conclusion
that the formation of itinerant small polarons and bipolarons in real
materials is unlikely \cite{ran,ran2}, and high-temperature bipolaronic
superconductivity is impossible \cite{and2,chakraverty}.

\subsection{Continuum Fr\"ohlich bipolaron}

\label{largebip}

While polarons repel each other at large distances, two $large$ polarons can
be also bound into a $large$ bipolaron by an exchange interaction even
without any additional EPI but the Fr\"{o}hlich one \cite%
{vinet1957,vinet1961,kochetov1977,suprun1982,adam1989,mukh1982,hiramoto1985,verbist1990,verbist1991,bassani1991,kashirina}
(see also the reviews \cite{SF1994,Devreese98}). Large bipolarons in
the continuum limit are referred to as \textit{Fr\"{o}hlich
bipolarons}. Besides the Fr\"{o}hlich coupling constant, $\alpha $,
the Fr\"{o}hlich bipolaron energy depends also on the dimensionless
parameter $U$, a measure for the
strength of the Coulomb repulsion between the two electrons \cite%
{verbist1990,verbist1991}, $U=(e^{2}/\varepsilon \omega _{0})\sqrt{m\omega
_{0}}$. In the discussion of bipolarons the ratio $\eta =\varepsilon
/\varepsilon _{0}$ of the high-frequency (electronic) and static dielectric
constants is important ($0\leq \eta \leq 1$). The following relation exists
between {$U$ and $\alpha $}: $U=\frac{\sqrt{2}\alpha }{1-\eta }.$ Only
values of $U$ satisfying the inequality $U\geq \sqrt{2}\alpha $ have a
physical meaning. It was shown that bipolaron formation is favoured by
larger values of $\alpha $ and by smaller values of $\eta $.

Intuitive arguments suggesting that the Fr\"ohlich bipolaron is stabilised
in going from 3D to 2D had been given, but the first quantitative analysis
based on the Feynman path integral was presented in Refs. \cite%
{verbist1990,verbist1991}. The conventional condition for bipolaron
stability is $E_{\mathrm{bip}} \leq 2E_{\mathrm{pol}}$, where $E_{\mathrm{pol%
}}$ and $E_{\mathrm{bip}}$ denote the ground-state energies of the polaron
and bipolaron at rest, respectively. From this condition it follows that the
Fr\"ohlich bipolaron with zero spin is stable (given the effective Coulomb
repulsion between electrons) if the electron-phonon coupling constant is
larger than a certain critical value: $\alpha \geq \alpha_c$.

\begin{figure}[t]
\begin{center}
\includegraphics[width=0.5\textwidth]{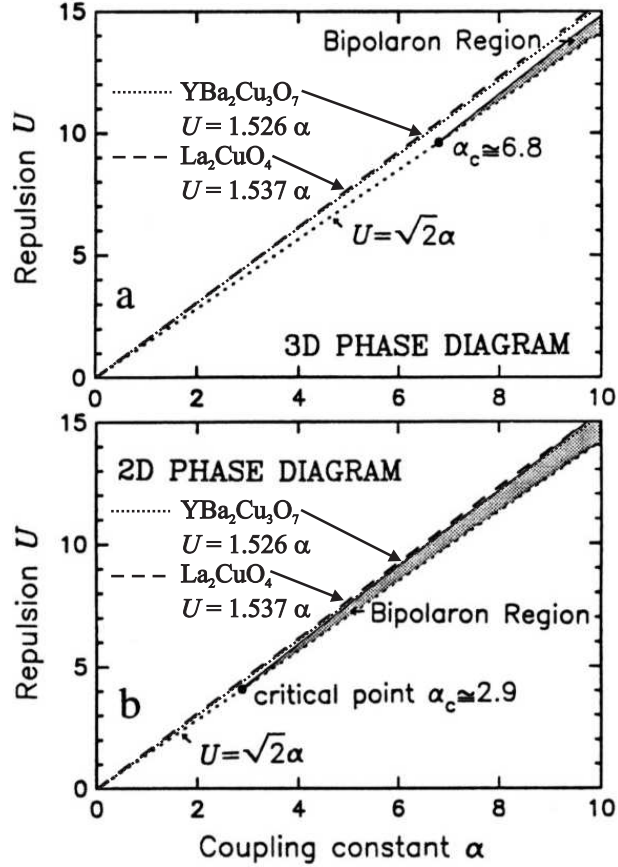}
\end{center}
\caption{The stability region for bipolaron formation in 3D (a) and
in 2D (b). The dotted line $U=\protect\sqrt{2}\alpha$ separates the
physical region $(U\geq\protect\sqrt{2} \alpha)$ from the
non-physical $(U\leq\protect\sqrt{2}\alpha)$. The shaded area is the
stability region in physical space. The dashed {(dotted)}
{``characteristic line''} {$U=1.537\alpha$ ($U=1.526\alpha$)} is
determined by $U=\protect\sqrt{2}\alpha/(1-\varepsilon/\varepsilon_{0})$
where we took the experimental values $\varepsilon=4$ and
$\varepsilon_0=50$ for La$_2$CuO$_4$ {($\varepsilon=4.7$ \cite{Genzel1989}
and $\varepsilon_{0}=64.7$ calculated using the experimental data of
\cite{Genzel1989,Kress1988} for YBa$_2$Cu$_3$O$_7$)}. The critical
points $\alpha_c=6.8$ for 3D and $\alpha_c=2.9$ for 2D are
represented as full dots. (Reprinted with permission after Ref.
\protect\cite{verbist1990}. \copyright 1990 by Elsevier.) }
\label{bipolaron}
\end{figure}

A ``phase-diagram'' for the two continuum polarons --- bipolaron system is
shown in Fig. \ref{bipolaron} for 3D and for 2D. The Fr\"ohlich coupling
constant as high as 6.8 is needed to allow for bipolaron formation in 3D.
The confinement of the bipolaron in two dimensions facilitates bipolaron
formation at smaller $\alpha$. This can be shown using the PD-type scaling
relation between the free energies $F$ in two dimensions $F_{\mathrm{2}%
D}(\alpha, U, \beta)$ and in three dimensions $F_{\mathrm{3}D}(\alpha, U,
\beta)$ \cite{verbist1990,verbist1991},
\begin{equation}
F_{2D}(\alpha, U, \beta) = \frac{2}{3} F_{3D}(\frac{3\pi}{4}\alpha, \frac{%
3\pi}{4}U, \beta).  \label{ScalingBip}
\end{equation}
According to Eq. (\ref{ScalingBip}), the critical value of the coupling
constant for bipolaron formation $\alpha_c$ turns out to scale with a factor
$3\pi/4\approx 2.36$ or $\alpha^{(2D)}_c = \alpha^{(3D)}_c/2.36$. From Fig. %
\ref{bipolaron}b it is seen that bipolarons in 2D can be stable for $%
\alpha\ge 2.9$, a domain of coupling constants which is definitely realised
in several solids. The ``characteristic line'' $U=1.526\alpha$ for the
material parameters of YBa$_2$Cu$_3$O$_7$ enters the region of bipolaron
stability in 2D at a value of $\alpha$ which is appreciably smaller than in
the case of La$_2$CuO$_4$. This fact suggests YBa$_2$Cu$_3$O$_7$ as a good
candidate for the occurrence of stable Fr\"ohlich bipolarons.

An analytical strong-coupling asymptotic expansion in inverse powers of the
electron-phonon coupling constant for the large bipolaron energy at $T=0$
was derived in \cite{Smondyrev:1995}
\begin{equation}
E_{3D}(\alpha, u) = -\frac{2\alpha^2}{3\pi} A(u) - B(u) + O(\alpha^{-2}),
\end{equation}
where the coefficients are closed analytical functions of the ratio $%
u=U/\alpha$:
\begin{equation}
A(u) = 4 -2\sqrt{2}u\left(1+\frac{u^2}{128}\right)^{3/2} + \frac{5}{8}u^2-
\frac{u^4}{512}
\end{equation}
and for $B(u)$ see the above-cited paper. The scaling relation (\ref%
{ScalingBip}) allows one to find the bipolaron energy in two dimensions as
well.

The stability of bipolarons has also been examined with the use of operator
techniques with a variational approach \cite{bassani1991}. The bipolaron is
bound if the electron-phonon coupling constant $\alpha $ is larger than $%
\sim 6$ in three dimensions and larger than $\sim 2$ in two dimensions,
provided the ratio $\eta =\varepsilon /\varepsilon _{0}$ is smaller than a
critical value $\eta _{c}$ which depends on $\alpha $. The critical value $%
\eta _{c}$ is larger in the two-dimensional case than in the
three-dimensional one. The bipolaron radius is shown to be of the order of a
few polaron radii. The results of \cite{bassani1991} and \cite%
{verbist1990,verbist1991} tend to qualitatively confirm each other.
Furthermore, bipolaron states obtained in \cite{bassani1991} under the
assumption that the total linear momentum is conserved, have intrinsically
high mobility.

In the framework of the renewed interest in bipolaron theory after the
discovery of high-T$_c$ superconductivity, an analysis of the optical
absorption by large \cite{emin1993,devreese1995} and small \cite{alebraabs}
bipolarons was given.

\subsection{Discrete strong-coupling Fr\"ohlich bipolaron}

\label{super}

The Holstein model is an extreme polaron model, and typically yields the
highest possible values of the (bi)polaron mass in the strong coupling
regime (except the case when the lattice vibrations are polarised along the
hopping direction \cite{Trugman:2001}). Many doped ionic lattices are
characterized by poor screening of high-frequency optical phonons and they
are more appropriately described by the finite-range Fr\"{o}hlich EPI \cite%
{alexandrov:1996}. The unscreened Fr\"{o}hlich EPI provides relatively light
lattice polarons (\ref{range}) and also ``superlight''\ small bipolarons, which are several orders of
magnitude lighter than bipolarons in HHM \cite%
{alexandrov:1996,alekor2,jimcrab}.

To illustrate the point let us consider a generic ``Coulomb-Fr\"{o}hlich''\ model (CFM) on a lattice, which
explicitly includes the finite-range Coulomb repulsion and the strong \emph{%
long-range} EPI \cite{alexandrov:1996,alekor2}. The implicitly present
(infinite) Hubbard $U$ prohibits double occupancy and removes the need to
distinguish the fermionic spin, if we are interested in the charge rather
than spin excitations. Introducing spinless fermion annihilation operators $%
c_{\mathbf{n}}$ and phonon annihilation operators $d_{\mathbf{m}}$, the
Hamiltonian of CFM is written in the real space representation as \cite%
{alekor2}
\begin{eqnarray}
H &=&\sum_{\mathbf{n\neq n^{\prime }}}T(\mathbf{n-n^{\prime }})c_{\mathbf{n}%
}^{\dagger }c_{\mathbf{n^{\prime }}}+{\frac{1}{{2}}}\sum_{\mathbf{n\neq
n^{\prime }}}V_{c}(\mathbf{n-n^{\prime }})c_{\mathbf{n}}^{\dagger }c_{%
\mathbf{n}}c_{\mathbf{n^{\prime }}}^{\dagger }c_{\mathbf{n^{\prime }}}+ \\
&&\omega _{0}\sum_{\mathbf{n\neq m}}g(\mathbf{m-n})(\mathbf{e}\cdot \mathbf{e%
}_{\mathbf{m-n}})c_{\mathbf{n}}^{\dagger }c_{\mathbf{n}}(d_{\mathbf{m}%
}^{\dagger }+d_{\mathbf{m}})+\omega _{0}\sum_{\mathbf{m}}\left( d_{\mathbf{m}%
}^{\dagger }d_{\mathbf{m}}+\frac{1}{2}\right) ,  \nonumber
\end{eqnarray}%
where $T(\mathbf{n})$ is the bare hopping integral in a rigid lattice.

If we are interested in the non- or near-adiabatic limit and the strong EPI,
the kinetic energy is a perturbation. Then the model can be grossly
simplified using the Lang-Firsov canonical transformation. In particular
lattice structures like a staggered triangular ladder in Fig.\ref{ladder}
the intersite lattice bipolarons tunnel already in the first order in $t(%
\mathbf{n})$. That allows us to average the transformed Hamiltonian over
phonons to obtain its polaronic part as $H_{p}=H_{0}+H_{pert}$, where
\[
H_{0}=-E_{p}\sum_{\mathbf{n}}c_{\mathbf{n}}^{\dagger }c_{\mathbf{n}}+{\frac{1%
}{{2}}}\sum_{\mathbf{n\neq n^{\prime }}}v(\mathbf{n-n^{\prime }})c_{\mathbf{n%
}}^{\dagger }c_{\mathbf{n}}c_{\mathbf{n^{\prime }}}^{\dagger }c_{\mathbf{%
n^{\prime }}},
\]
and
\[
H_{pert}=\sum_{\mathbf{n\neq n^{\prime }}}t(\mathbf{n-n^{\prime }})c_{%
\mathbf{n}}^{\dagger }c_{\mathbf{n^{\prime }}}.
\]
is a perturbation. $E_{p}$ is the familiar polaron level shift,
\begin{equation}
E_{p}=\omega \sum_{\mathbf{m} }g^{2}(\mathbf{m-n})(\mathbf{e}\cdot \mathbf{e}%
_{\mathbf{m-n}})^{2},  \label{polshift}
\end{equation}
which is independent of $\mathbf{n}$. The polaron-polaron interaction is
\begin{equation}
v(\mathbf{n-n^{\prime }})=V_{c}(\mathbf{n-n^{\prime }})-V_{ph}(\mathbf{%
n-n^{\prime }}),
\end{equation}
where
\begin{equation}
V_{ph}(\mathbf{n-n^{\prime }}) =2\omega _{0}\sum_{\mathbf{m}}g(\mathbf{m-n}%
)g(\mathbf{m-n^{\prime }}) (\mathbf{e}\cdot \mathbf{e}_{\mathbf{m-n}})(%
\mathbf{e}\cdot \mathbf{e}_{\mathbf{m-n^{\prime }}}).  \label{v}
\end{equation}
The transformed hopping integral is $t(\mathbf{n-n^{\prime }})=T(\mathbf{%
n-n^{\prime }})\exp [-g^{2}(\mathbf{n-n^{\prime }})]$ with
\begin{eqnarray}
g^{2}(\mathbf{n-n^{\prime }}) &=&\sum_{\mathbf{m}}g(\mathbf{m-n})(\mathbf{e}%
_\cdot \mathbf{e}_{\mathbf{m-n}})\times \\
&&\left[ g(\mathbf{m-n})(\mathbf{e}\cdot \mathbf{e}_{\mathbf{m-n}})-g(%
\mathbf{m-n^{\prime }})(\mathbf{e}\cdot \mathbf{e}_{\mathbf{m-n^{\prime }}})%
\right]  \nonumber
\end{eqnarray}
at $T \ll \omega _{0}$. The mass renormalization exponent can be expressed
via $E_{p}$ and $V_{ph}$ as
\begin{equation}
g^{2}(\mathbf{n-n^{\prime }})=\frac{1}{\omega _{0}}\left[ E_{p}-\frac{1}{2}%
V_{ph}(\mathbf{n-n^{\prime }})\right] .  \label{g2}
\end{equation}
When $V_{ph}$ exceeds $V_{c}$ the full interaction becomes negative and
polarons form pairs. The real space representation allows us to elaborate
more physics behind the lattice sums in $V_{ph}$ \cite{alekor2}. When a
carrier (electron or hole) acts on an ion with a force $\mathbf{f}$, it
displaces the ion by some vector $\mathbf{x}=\mathbf{f}/k$. Here $k$ is the
ion's force constant. The total energy of the carrier-ion pair is $-\mathbf{f%
}^{2}/(2k)$. This is precisely the summand in Eq.(\ref{polshift}) expressed
via dimensionless coupling constants. Now consider two carriers interacting
with the \emph{same} ion. The ion displacement is $\mathbf{x}=(\mathbf{f}%
_{1}+\mathbf{f}_{2})/k$ and the energy is $-\mathbf{f}_{1}^{2}/(2k)-\mathbf{f%
}_{2}^{2}/(2k)-(\mathbf{f}_{1}\cdot \mathbf{f}_{2})/k$. Here the last term
should be interpreted as an ion-mediated interaction between the two
carriers. It depends on the scalar product of $\mathbf{f}_{1}$ and $\mathbf{f%
}_{2}$ and consequently on the relative positions of the carriers with
respect to the ion. If the ion is an isotropic harmonic oscillator, as we
assume here, then the following simple rule applies. If the angle $\phi $
between $\mathbf{f}_{1}$ and $\mathbf{f}_{2}$ is less than $\pi /2$ the
polaron-polaron interaction will be attractive, otherwise it will be
repulsive. In general, some ions will generate attraction, and some ions -
repulsion between polarons.
\begin{figure}[tbp]
\begin{center}
\includegraphics[angle=-0,width=0.5\textwidth]{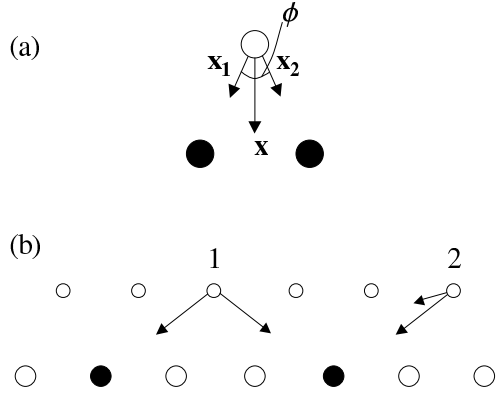}
\end{center}
\caption{Mechanism of the polaron-polaron interaction. (a) Together, two
polarons (solid circles) deform the lattice more effectively than
separately. An effective attraction occurs when the angle $\protect\phi$
between $\mathbf{x}_{1}$ and $\mathbf{x}_{2}$ is less than $\protect\pi /2$.
(b) A mixed situation: ion 1 results in repulsion between two polarons while
ion 2 results in attraction. (After \protect\cite{alekor2})}
\label{attraction}
\end{figure}

The overall sign and magnitude of the interaction is given by the lattice
sum in Eq.(\ref{v}). One should note that according to Eq.(\ref{g2}) an
attractive EPI reduces the polaron mass (and consequently the bipolaron
mass), while repulsive EPI enhances the mass. Thus, the long-range EPI
serves a double purpose. Firstly, it generates an additional inter-polaron
attraction because the distant ions have a small angle $\phi $. This
additional attraction helps to overcome the direct Coulomb repulsion between
polarons. And secondly, the Fr\"{o}hlich EPI makes lattice bipolarons
lighter.

The many-particle ground state of $H_{0}$ depends on the sign of the
polaron-polaron interaction, the carrier density, and the lattice structure.
Following \cite{alekor2}, we consider the staggered ladder, Fig.\ref{ladder}%
, assuming that all sites are isotropic two-dimensional harmonic
oscillators. For simplicity, we also adopt the nearest-neighbour
approximation for both interactions, $g(\mathbf{l})\equiv g$, $V_{c}(\mathbf{%
n})\equiv V_{c}$, and for the hopping integrals, $T(\mathbf{m})=T(a)$ for $%
l=n=m=a$, and zero otherwise. Hereafter we set the lattice period $a=1$.
There are four nearest neighbours in the ladder, $z=4$. The \textit{%
single-particle} polaronic Hamiltonian takes the form
\begin{eqnarray}
H_{p} &=&-E_{p}\sum_{n}(c_{n}^{\dagger }c_{n}+p_{n}^{\dagger }p_{n})+ \\
&&\sum_{n}[t^{\prime }(c_{n+1}^{\dagger }c_{n}+p_{n+1}^{\dagger
}p_{n})+t(p_{n}^{\dagger }c_{n}+p_{n-1}^{\dagger }c_{n})+H.c.],  \nonumber
\label{ham2}
\end{eqnarray}
where $c_{n}$ and $p_{n}$ are polaron annihilation operators on the lower
and upper legs of the ladder, respectively, Fig.\ref{ladder}. Using Eqs.(\ref%
{polshift},\ref{v},\ref{g2}) one obtains $E_{p} =4g^{2}\omega _{0}$, $%
t^{\prime } =T(a)\exp ( -7E_{p}/8\omega _{0}),$ and $t=T(a)\exp(-3E_{p}/4%
\omega _{0})$.

The Fourier transform of $H_p$ yields two overlapping polaron bands,
\begin{equation}
E_{p}(k)=-E_{p}+2t^{\prime }\cos (k)\pm 2 t\cos (k/2)
\end{equation}
with the effective mass $m^{\ast }=2/|4t^{\prime }\pm t|$ near their edges.

Let us now place two polarons on the ladder. The nearest neighbour
interaction is $v=V_{c}-E_{p}/2,$ if two polarons are on different legs of
the ladder, and $v=V_{c}-E_{p}/4,$ if both polarons are on the same leg. The
attractive interaction is provided via the displacement of the lattice
sites, which are the common nearest neighbours to both polarons. There are
two such nearest neighbours for the intersite bipolaron of type $A$ or $B$,
Fig.\ref{ladder}c, but there is only one common nearest neighbour for
bipolaron $C$, Fig.\ref{ladder}d. When $V_{c}>E_{p}/2$, there are no bound
states and the multi-polaron system is a one-dimensional Luttinger liquid.
However, when $V_{c}<E_{p}/2$ and consequently $v<0$, the two polarons are
bound into an inter-site bipolaron of types $A$ or $B$.

Remarkably, bipolarons tunnel in the ladder already in the first order with
respect to the single-polaron tunnelling amplitude. This case is different
from both onsite bipolarons discussed above, and from intersite chain
bipolarons of \cite{bonc}, where the intersite bipolaron tunnelling appeared
in the second order in $t$ as for the on-site bipolarons. Indeed, the
lowest-energy configurations $A$ and $B$ are degenerate. They are coupled by
$H_{pert}.$ Neglecting all higher-energy configurations, we can project the
Hamiltonian onto the subspace containing $A$, $B$, and empty sites.
\begin{figure}[tbp]
\begin{center}
\includegraphics[angle=-0,width=0.5\textwidth]{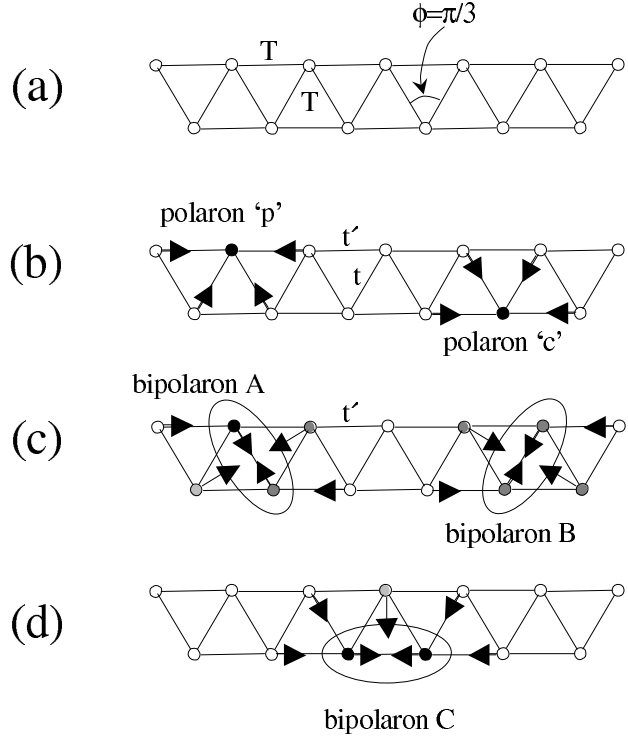}
\end{center}
\caption{One-dimensional zig-zag ladder. (a) Initial ladder with the bare
hopping amplitude $T(a)$. (b) Two types of polarons with their respective
deformations. (c) Two degenerate bipolaron configurations A and B (here $%
t,t^{\prime }$ are renormalised (i.e. polaronic) hopping integrals). (d) A
different bipolaron configuration, C, whose energy is higher than that of A
and B.}
\label{ladder}
\end{figure}
The result of such a projection is the bipolaronic Hamiltonian
\begin{equation}
H_{b}=\left( V_{c}-\frac{5}{2}E_{p}\right) \sum_{n}[A_{n}^{\dagger
}A_{n}+B_{n}^{\dagger }B_{n}]-t^{\prime }\sum_{n}[B_{n}^{\dagger
}A_{n}+B_{n-1}^{\dagger }A_{n}+H.c.],  \label{bipham2}
\end{equation}
where $A_{n}=c_{n}p_{n}$ and $B_{n}=p_{n}c_{n+1}$ are intersite bipolaron
annihilation operators, and the bipolaron-bipolaron interaction is omitted.
The Fourier transform of Eq.(\ref{bipham2}) yields two \textit{bipolaron}
bands,
\begin{equation}
E_{2}(k)=V_{c}-{\frac{5}{{2}}}E_{p}\pm 2t^{\prime }\cos (k/2).
\end{equation}
with a combined width $4|t^{\prime }|$. The bipolaron binding energy in zero
order with respect to $t,t^{\prime }$ is $\Delta \equiv
2E_{1}(0)-E_{2}(0)=E_{p}/2-V_{c}$.

The bipolaron mass near the bottom of the lowest band, $m^{\ast \ast
}=2/t^{\prime }$, is
\begin{equation}
m^{\ast \ast }=4m^{\ast }\left[ 1+0.25\exp \left( \frac{{E_{p}}}{8\omega _{0}%
}\right) \right] .
\end{equation}
The numerical coefficient $1/8$ in the exponent ensures that $m^{\ast \ast }$
remains of the order of $m^{\ast }$ even at sufficiently large $E_{p}$ up to
$E_{p}\approx 10 \omega_0$. This fact combined with a weaker renormalization
of $m^{\ast }$ provides a \emph{superlight} small bipolaron \cite%
{alexandrov:1996,alekor2,jimcrab}.

\subsection{Discrete all-coupling Fr\"ohlich bipolaron}

The CFM model discussed above is analytically solvable in the
strong-coupling nonadibatic ($\omega_0 \gtrsim T(a))$ limit using the
Lang-Firsov transformation of the Hamiltonian, and projecting it on the
inter-site pair Hilbert space \cite{alexandrov:1996,alekor2}. The theory has
been extended to the whole parameter space using CTQMC technique for
bipolarons \cite{jimcrab,jim3}. Refs. \cite{jimcrab,jim3} simulated the CFM
Hamiltonian on a staggered triangular ladder (1D), triangular (2D) and
strongly anisotropic hexagonal (3D) lattices including triplet pairing. On
such lattices, bipolarons are found to move with a crab like motion, Fig.\ref%
{ladder}, which is distinct from the crawler motion found on cubic lattices
\cite{aleran1981}. Such bipolarons are small but very light for a wide range
of electron-phonon couplings and phonon frequencies. EPI has been modeled
using the force function in the site-representation as in Eq.(\ref{force}).
Coulomb repulsion has been screened up to the first nearest neighbors, with
on site repulsion $U$ and nearest-neighbor repulsion $V_c$. The
dimensionless electron-phonon coupling constant $\lambda$ is defined as $%
\lambda=\sum_{\mathbf{m}}f^{2}_{\mathbf{m}}(0)/2M\omega^2 zT(a)$ which is
the ratio of the polaron binding energy to the kinetic energy of the free
electron $zT(a)$, and the lattice constant is taken as $a=1$.

\begin{figure}[tbp]
\includegraphics[height=105mm,angle=270]{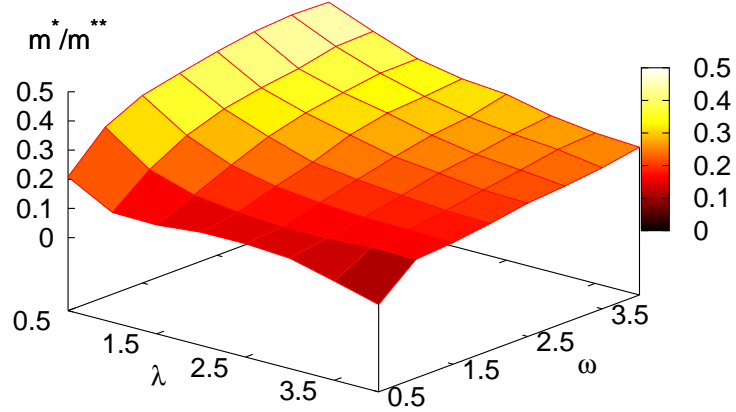}
\caption{Polaron to bipolaron mass ratio for a range of $\bar{\protect\omega}%
=\protect\omega_0/T(a)$ and $\protect\lambda$ on the staggered ladder.
Mobile small bipolarons are seen even in the adiabatic regime $\bar{\protect%
\omega}=0.5$ for couplings $\protect\lambda$ up to 2.5 (Reproduced from J.
P. Hague et al., Phys. Rev. Lett \textbf{98}, 037002 (2007), (c) American
Physical Society, 2007). }
\label{bipjim}
\end{figure}

Extending the CTQMC algorithm to systems of two particles with
strong EPI and Coulomb repulsion solved the bipolaron problem on
different lattices from weak to strong coupling in a realistic
parameter range where usual strong and weak-coupling limiting
approximations fail. Figure \ref{bipjim} shows the ratio of the
polaron to bipolaron masses on the staggered ladder as a function of
effective coupling and phonon frequency for $V_c=0$. The bipolaron
to polaron mass ratio is about 2 in the weak coupling regime
($\lambda\ll1$) as it should be for a large bipolaron, section
\ref{largebip}. In the strong-coupling, large phonon frequency limit
the mass ratio approaches 4, in agreement with strong-coupling
results, section \ref{super}. In a wide region of parameter space,
the bipolaron/polaron mass ratio has been found between 2 and 4 and
a bipolaron radius similar to the lattice spacing,  Figs.
\ref{bipjim}. Thus the bipolaron is small and light at the same
time. Taking into account additional intersite Coulomb repulsion
$V_c$ does not change this conclusion. The bipolaron is stable for
$V_c<4T(a)$. As $V_c$ increases the bipolaron mass decreases but the
radius remains small, at about 2 lattice spacings. Importantly, the
absolute value of the small bipolaron
mass is only about 4 times of the bare electron mass $m_0$, for $%
\lambda=\omega/T(a)=1$ (see Fig. \ref{bipjim}).

Simulations of the bipolaron on an infinite triangular lattice including
exchanges and large on-site Hubbard repulsion $U = 20T(a)$ also lead to the
bipolaron mass of about $6m_{0xy}$ and the bipolaron radius $%
R_{bp}\thickapprox 2a$ for a moderate coupling $\lambda= 0.5$ and a large
phonon frequency $\omega= T(a)$ (for the triangular lattice, $%
m_{0xy}=1/3a^{2}T(a)$). Finally, the bipolaron in a hexagonal lattice with
out-of-plane hopping $T^{\prime }=T(a)/3$ has also a light in-plane inverse
mass, $m_{xy}^{**}\thickapprox 4.5 m_{0xy}$ but a small size, $%
R_{bp}\thickapprox 2.6a$ for experimentally achievable values of the phonon
frequency $\omega=T(a)=200$meV and EPI, $\lambda=0.36$. Out-of-plane $%
m_{z}^{**}\thickapprox70m_{0z}$ is Holstein like, where $m_{0z}=1/2d^2T^{%
\prime }$, ($d$ is the inter-plane spacing). When bipolarons are small and
pairs do not overlap, the pairs can form a BEC at $T_{BEC}=3.31(2n_B/a^2%
\sqrt{3}d)^{2/3}/(m_{xy}^{2/3}m_z^{1/3})$. If we choose realistic values for
the lattice constants of 0.4 nm in the plane and 0.8 nm out of the plane,
and allow the density of bosons to be $n_B$=0.12 per lattice site, which
easily avoids overlap of pairs, then $T_{BEC}\thickapprox$ 300K.

\subsection{Polaronic exciton}

Finally let us briefly mention works on electron-hole bound states coupled
with phonons. Such excitonic polaron states have been analysed by \cite%
{IB1983,IB1984,IB1987}. The binding energies of excitonic states in
interaction with LO phonons were computed using a phonon coherent state and
applying a variational method \cite{IB1983}. Incomplete relaxation of the
lattice is found. The binding energies are larger than those obtained with
static dielectric screening when the polaron radius is comparable to the
exciton radius. Analysing the relative intensities of the one-phonon and
zero-phonon lines for a number of semiconductors, \cite{IB1983} observed
that the zero-phonon exciton states are generally much more probable than
the phonon replicas.

The binding energy of a core exciton, which depends on the interaction of
the conduction electron and of the core hole with the valence electrons via
a Fr\"{o}hlich-type coupling with the electron-hole pairs, in addition to
the Coulomb attraction, was calculated within a functional variational
method \cite{IB1984}. When the exciton radius is comparable to the polaron
radius, the static dielectric screening reduces and the core exciton binding
energy increases. The excitonic-polaron effective mass renormalization was
analysed by \cite{IB1984} using a variational numerical approach. Evidence
for this renormalization is found from the experimental data on polariton
dispersion in hexagonal CdS and in CuCl.

The angular momentum as a constant of the motion of a Fr\"{o}hlich polaron
was introduced by \cite{EKD1970}. Representing the excitonic polaron problem
in angular coordinates and solving it numerically, \cite{IBS1989} showed
that exciton states with various total angular momenta $L$ are differently
affected by the interaction with LO phonons. However, when extending this
approach to the bipolaron problem, the same authors found that the Fr\"{o}%
hlich interaction is not strong enough to guarantee a stable bipolaron state
at least in the case of cubic materials.

Very recently the polaronic exciton problem has been solved  using
the approximation-free DQMC technique \cite{burovski}. Numerically
exact results for the wave function, ground-state energy, binding
energy and effective mass of this quasiparticle were calculated, and
the frequently used instantaneous approximation to the retarded
interaction due to the exchange of phonons was critically analysed.

\section{Current status of polarons and open problems}

At present, the basic properties of single polarons are well understood
theoretically, and to a large extent they are analytically under control at
all coupling. It is remarkable how the Fr\"{o}hlich continuum polaron, one
of the simplest examples of a \textit{Quantum Field Theoretical} problem, as
it basically consists of a single fermion interacting with a scalar Bose
field, has resisted full analytical solution at all coupling since $\sim
1950 $, when its Hamiltonian was first written. Although a mechanism for the
optical absorption of Fr\"{o}hlich polarons was already proposed a long time
ago \cite{KED69,DSG72}, some subtle characteristics were only clarified very
recently \cite{PRL2006} by combining numerical DQMC studies \cite%
{Mishchenko2003} and improved analytical methods \cite{catbook,DK2006} (\ref%
{comparison}). Of special interest are several sum rules derived for the
optical conductivity spectra of arbitrary-coupling Fr\"{o}hlich polarons
\cite{SR,LSD}. A variety of magneto-optical and transport-experiments were
successfully analysed with Fr\"ohlich polaron theory (see e.~g. \cite%
{PD1984,PD86,Hodby1987,Devreese04} and references therein).

The charge carriers in a rich variety of systems of reduced
dimension and dimensionality (submicron- and nanostructures
including heterojunctions, quantum wells, quantum wires, quantum
dots etc.) turn out to be Fr\"ohlich polarons. Several scaling
relations were derived \cite{prb36-4442}, which connect polaron
characteristics (the self-energy, the effective mass, the impedance
and the mobility) in different dimensions.  Quite generally
confinement enhances EPI and the tendency to polaron formation. A
new aspect of the polaron concept has been investigated for
semiconductor {structures at nanoscale}: the exciton-phonon states
are not factorisable into an adiabatic product Ansatz, so that a
\textit{non-adiabatic} treatment is needed \cite{PRB57-2415}.  The
excitonic polarons in nanostructures  lead to the existence of
phonon replicas in the luminescence \cite{verzelen}.  Experimental
evidence of the enhanced phonon-assisted absorption due to effects
of non-adiabaticity has been provided by {the} multi-phonon
photoluminescence (PL)
 spectra observed under
selective excitation in self-assembled InAs/GaAs quantum dots
\cite{Garcia1999} and by the photoluminescence  excitation (PLE)
measurements on single self-assembled InAs/GaAs \cite{lem01} and
InGaAs/GaAs \cite{zre01} quantum dots. The polaron concept was also
invoked for the explanation of the PLE measurements on
self-organised In$_x$Ga$_{1-x}$As/GaAs \cite{Heitz2001} and
CdSe/ZnSe \cite{Woggon} quantum dots.

The Fr\"ohlich polaron has led to many generalisations. The stability region
of the Fr\"{o}hlich large bipolaron is now firmly established \cite%
{verbist1990,verbist1991,bassani1991}(\ref{largebip}). Here the
surprise is double (cf. \cite{verbist1991,devreese1995}): a) only in
a very limited sector of the phase diagram (Coulomb repulsion versus
$\alpha$) the bipolaron is stable, b) most traditional Fr\"{o}hlich
polaron materials (alkali halides and the like) lie completely
outside (and ``far'' from) this bipolaron stability sector, but
several cuprate superconductors lie very close and even inside this
very restricted area of the stability diagram. The stability of a
strong-coupling singlet  bipolaron was studied in two- and
three-dimensional parabolic quantum dots using the Landau-Pekar
variational method \cite{mukhopadhyay}. It was shown that the
confining potential of the quantum dot reduces the stability of the
bipolaron. A theory of bipolaron states in a spherical parabolic
potential well was further developed applying the Feynman
variational principle. The basic parameters of the bipolaron ground
state (the binding energy, the number of phonons in the bipolaron
cloud, and the bipolaron radius were studied as a function of the
radius  of the potential well \cite{pokatilov}. It was found that
confinement can enhance the bipolaron binding energy, when the
radius of a quantum dot is of the same order of magnitude as the
polaron radius. A unified insight into the stability criterion for
bipolaron formation in low-dimensionally confined media was provided
by \cite{senger} using an adiabatic variational method for a pair of
electrons immersed in a reservoir of bulk LO phonons and confined
within an anisotropic parabolic potential box. Bipolaron formation
in a two-dimensional lattice with harmonic confinement, representing
a simplified model for a quantum dot, was investigated by means of
QMC \cite{hohenadler2007a}. This method treats all interactions
exactly and takes into account quantum lattice fluctuations.
Calculations of the bipolaron binding energy reveal that confinement
opposes bipolaron formation for weak electron-phonon coupling but
abets a bound state at intermediate to strong coupling. We also
mention the exciton-polaron formation in nanostructures and
quantum-light sources studied recently by QMC in lattice models with
short- or long-range carrier-phonon interaction (see
\cite{hohenadler2007b}, and references therein)

The richness and profundity of Landau-Pekar's polaron concept is further
illustrated by its extensions to discrete (lattice) polarons. Even the
simplest two-site polaron model by Holstein (\ref{two-site}) proved to be
very useful for a qualitative understanding of nontrivial features of the
polaron problem, and for obtaining some novel analytical and semi-analytical
results (see, for example, \cite{kudfir,wang,berciu}). The ``$1/\lambda
$'' expansion technique based on the Lang-Firsov
transformation (\ref{band}) and unbiased numerical analysis of the finite
and infinite Holstein and Fr\"{o}hlich models combining Lanczos
diagonalisations of clusters, density matrix renormalisation group, cluster
perturbation theory techniques, DMFT and different QMC algorithms, allowed
for a description of properties of a single lattice polaron and a lattice
bipolaron. The Lang-Firsov canonical transformation \cite{lan} was proven
particularly instrumental in calculation of different kinetic and optical
coefficients, which can be represented as expansions in powers of the
unrenormalised hopping integral $t$ (\ref{response}). Sometimes it is
possible to sum the expansion and get results, which are valid for arbitrary
values of parameters providing the understanding of the crossover region
from the Boltzmann kinetics to thermally activated hopping \cite{fir2}.

Recent ED \cite{fehskebook}, CTQMC \cite{kornbook} and DQMC \cite{MN2006}
techniques allow for determination of the ground-state and excited states of
lattice polarons with arbitrary precision in the thermodynamic limit for any
dimension and any type of lattices (\ref{short-range}). The spectral
properties (e.g. photoemission), optical response and thermal transport, as
well as the dynamics of polaron formation in the Holstein model have been
numerically analysed for all EPI strengths and phonon frequencies, including
the intermediate-coupling regime (see \ref{response} and \cite{fehskebook}).
CTQMC methods have proven to be powerful and versatile tools providing
unbiased results for the polaron properties in any lattices for any-range
EPI, including Jahn-Teller polarons (\ref{all}). Combining the Lang-Firsov
transformation and quantum Monte Carlo simulations allows for an exact
sampling without autocorrelations, which proves to be an enormous advantage
for small phonon frequencies or low temperatures \cite{linden2007}.

Importantly, variational and numerical techniques confirmed that the
Fr\"ohlich and Holstein-Lang-Firsov theories are asymptotically exact in the
weak, $\lambda \ll 1$, and strong-coupling, $\lambda \gg 1$, regimes,
respectively, and the polaron formation represents a continuous crossover of
the ground state (\ref{all}). The crossover is related to the exponential
increase of the effective mass, and the band narrowing with a strongly
suppressed electronic quasiparticle residue and the Drude weight,
accompanied by an increase of the incoherent spectral weight (\ref{spectral}%
). These features strongly depend on the phonon dispersion \cite{zoli:1998},
EPI radius \cite{alexandrov:1996,alexandrov:1999,fehske2002,cataudella2005},
lattice geometry \cite{jim} and are more pronounced in higher dimensions
\cite{jim,fehske2002}. Remarkably, the unscreened Fr\"ohlich EPI provides
relatively light lattice polarons (\ref{range}), which are several orders of
magnitude lighter than the Holstein small polarons \cite%
{alexandrov:1996,alexandrov:1999} at strong coupling. This classification of
weak- and strong-coupling regimes still leaves room for continuum Fr\"ohlich
polarons of not only weak but also intermediate and - in the theoretical
analysis - of strong coupling classified with respect to the Fr\"ohlich
electron-phonon coupling constant $\alpha$ (\ref{Ground_Feynman}).

While the single polaron has been actively researched for a long
time and is now well understood, the multi-polaron physics has
gained particular attention in the last two decades. It has been
found--- unexpectedly for many researchers--- that the
Migdal-Eliashberg theory breaks down already at $\lambda \sim 1$ for
any adiabatic ratio $\omega _{0}/E_{F}$. The effective parameter
$\lambda \omega _{0}/E_{F}$ becomes large at $\lambda \gtrsim 1$
since the bandwidth is narrowed and the Fermi energy, $E_{F}$ is
renormalised down exponentially \cite{ale0,alexandrov:2001}.
Extending the BCS theory towards the strong interaction between
electrons and ion vibrations, a charged Bose gas of tightly bound
on-site small bipolarons was predicted \cite{aleran1981}, with a
further conclusion that the highest superconducting transition
temperature is attained in the crossover region of EPI strength
between the BCS-like polaronic and bipolaronic superconductivity
\cite{ale0}. Subsequent studies of the Holstein-Hubbard
model found also two-site bipolarons \cite{aub1995,lamagna,trugman,Macridin}%
. Taking into account that many advanced materials with low density of free
carriers and poor mobility are characterized by poor screening of
high-frequency optical phonons, the Coulomb-Fr\"{o}hlich lattice polaron
model was introduced \cite{alexandrov:1996,alekor2}. The large Hubbard $U$
and intersite Coulomb repulsions and the unscreened Fr\"{o}hlich EPI provide
``superlight''\ but small intersite
bipolarons (\ref{super}). More recent CTQMC simulations of intersite small
bipolarons in the Coulomb-Fr\"{o}hlich model \cite{jimcrab} have found such
quasiparticles in a wide parameter range with achievable phonon frequencies
and couplings. They could have a superconducting transition in excess of
room temperature.

The many-body theory for polarons has been developed for extreme
weak and strong coupling regimes. It became clear how --- in the
weak-coupling limit
--- this problem can be reduced to the study of the structure factor of a
uniform electron gas \cite{LDB1977}. For strong coupling the problem
is reduced to an interacting Bose gas of on-site \cite{aleran1981}
or intersite \cite{alexandrov:1996,alekor2,jimcrab} small bipolarons
in the dilute system.

While correlation effects in transport through metallic quantum dots
with repulsive electron-electron interactions received considerable
attention in the past, and continue to be the focus of intense
investigations,
 much less has been known about a role of attractive
correlations   between small polarons  mediated by EPI in molecular
quantum dots (MQD) and nanowires. In the framework of the negative
$U$ Hubbard  model \cite{alebrawil} it has been found that the
\emph{attractive} electron correlations within the molecule could
lead to a molecular \emph{switching}
 effect where I-V characteristics
have two branches with high and low current at the same bias
voltage. The switching phenomenon
 has been also predicted by a theory of \textit{correlated} polaron
transport  with a full account of both the Coulomb repulsion and EPI
in MQD weakly coupled with electrodes \cite{alebraMQD,erm}.  Ref.
\cite{alebraMQD} has shown that while the phonon side-bands
significantly modify the shape of hysteretic I-V curves in
comparison with the negative-$U$ Hubbard model, switching remains
robust. It shows up at sufficiently low temperatures when the
effective interaction of polarons in MQD is attractive and the
molecular level is multiply degenerate. Importantly, the switching
has not been found in \emph{non-} and \emph{two-fold degenerate}
MQDs  \cite{alebraMQD}.  When the  polaronic energy shift is very
large, the effective charging energy of molecules can become
negative, favoring ground states with even numbers of electrons.
Ref.\cite{koch} has shown that charge transport through such
molecules  is dominated by tunneling of bipolarons which coexists
with  single-electron cotunneling.

The conductance of deformable molecules with a local magnetic moment
has been  studied in  the framework of a two-impurity Anderson model
with positive and negative electron-electron interactions and in the
two-impurity Anderson-Holstein model with a single phonon mode
\cite{bonca2006}. It has been shown that the spin and charge Kondo
effects can occur simultaneously at any coupling strength. At finite
bandwidth and strong coupling the lattice effects lead to a
renormalization of the effective Kondo exchange constants;
nevertheless, universal spin and charge Kondo effects still occur.

The theory of dense polaronic systems in the intermediate regime
remains highly cumbersome, in particular when EPI competes with
strong electron correlations. The corresponding microscopic models
contain (extended) Hubbard, Heisenberg or double-exchange terms, and
maybe also a coupling to orbital degrees of freedom along with
strong EPI, so that even numerical solutions with the same precision
as in the dilute (bi)polaron case are often problematic. A number of
ED, QMC, DMFT and combined numerical results give strong evidence
that the tendency towards lattice polaron formation is enhanced in
strongly correlated electron systems due to a narrowing of the
electron band caused by strong correlations (see recent reviews by
\cite{fehskebook,hohenadler,linden2007,edwards2002} for more
details). Not only antiferromagnetic correlations enhance EPI,
resulting in polaron formation for moderate coupling strength, but
also EPI strongly enhances spin correlations
\cite{macridin2006,capone2002}. Some of these studies show that
increasing carrier density could be accompanied by a dissociation of
polarons leading to normal metallic behavior in the intermediate
coupling adiabatic regime \cite{hohenadler} that is reminiscent of
the ``overcrowding''\ effect hypothesized by Mott (1995). On the
other hand for parameters favoring small polarons no such
density-driven crossover occurs in agreement with simple analytical
arguments \cite{alexandrov:2001}. Thorough investigations of these
models will definitely be a great challenge in the near future. Here
we have focused on the single and two-body problems leaving theories
of strongly-correlated polarons and their applications to
high-temperature superconductors and CMR oxides for future reviews
\cite{aledevbook}.

Finally, exactly solvable models might give a rather limited,
sometimes misleading, description of polarons in real systems.
Qualitative inconsistencies can arise when coupling is assumed to be
just to one phonon mode, often taken as dispersionless, and ad-hoc
approximations for EPI matrix elements are applied. Moreover
electronic nanoscale disorder and long-range strain fields can
interweave with the microscopic mechanisms of polaronic transport
\cite{phillips2003}. Hence ab-initio calculations of the phonon
spectrum, EPI and polaron properties beyond the adiabatic
Born-Oppenheimer approximation are required in many cases for which
the theory and experiment can be compared in detail
\cite{shluger,banacky}.

\section*{Acknowledgments}

We have benefited from discussions with many colleagues. While
writing this review, conversations with Alexander Andreev, Serge
Aubry, Ivan Bozovic, Alex Bratkovsky, Peter Edwards, Janez Bonca,
Holger Fehske, Jim Hague, Yurii Firsov, Martin Hohenadler, Viktor
Kabanov, Pavel Kornilovitch, Wolfgang von der Linden, Peter
Littlewood, Dragan Mihailovic, John Samson, Marshall Stoneham and
other participants of the European Science Foundation workshop
"Mott's Physics in Nanowires and Quantum Dots" (Cambridge, UK, 31
July-2 August, 2006) were especially helpful. We like to thank
Vladimir Fomin for discussions during the preparation of this
review, and to acknowledge discussions with Fons Brosens, Dieter
Bimberg, Vittorio Cataudella, Paolo Calvani, Giulio De Filippis,
Roger Evrard, Vladimir Gladilin, Giuseppe Iadonisi,  Eddy
Kartheuser, Serghei Klimin, Lucien Lemmens, Andrei Mishchenko,
Fran\c{c}ois Peeters and Jacques Tempere. This work was supported in
part by EPSRC under grants No. EP/C518365/1 and No. EP/D07777X/1
(UK), IUAP, FWO-V project G.0435.03, the WOG WO.035.04N (Belgium)
and the European Commission SANDiE Network of Excellence, contract
No. NMP4-CT-2004-500101.

\bibliographystyle{apsrmp}
%\bibliography{BIBMISSING6}

%\end{document}

\end{document}